\documentclass[11pt,a4paper]{article}

\usepackage{amsmath,amsfonts,amssymb,graphicx,theorem,multirow}
\usepackage[usenames]{color}
\usepackage{caption}
\usepackage{lscape}
\usepackage{graphicx}
\usepackage{float}

\usepackage{tikz}
\usetikzlibrary{arrows,arrows.meta,decorations,calc,bending,positioning}
\usetikzlibrary{decorations.pathmorphing,patterns,decorations.pathreplacing,decorations.markings}

\usepackage[backref=false]{hyperref}
\numberwithin{equation}{section}
\usepackage[capitalise,noabbrev]{cleveref} 
\hypersetup{
colorlinks=true,
citecolor=red,
linkcolor=darkblue,
urlcolor=darkblue
}

\usepackage[nobreak]{cite} 

\long\def\ignore#1{}

\definecolor{darkblue}{rgb}{0,0,.8}
\definecolor{red}{rgb}{1,0,0}
\definecolor{purple}{rgb}{1,0,1}
\definecolor{coloroflink}{rgb}{0.7,0,1}
\definecolor{coloroflink}{rgb}{0.180392, 0.545098, 0.341176}
\definecolor{darkpurple}{rgb}{1,.2,1}
\definecolor{newpurple}{rgb}{0.76,0.6,1}
\definecolor{pink}{rgb}{1,.7,.7}
\definecolor{lightblue}{rgb}{.61,.61,1}
\definecolor{midblue}{rgb}{.7,.7,1}
\definecolor{lightmidlightblue}{rgb}{.8,.8,1}
\definecolor{lightlightblue}{rgb}{.9,.9,1}
\definecolor{lightestblue}{rgb}{.96,.96,1}
\definecolor{lightpurple}{rgb}{1,.65,1}
\definecolor{darkgreen}{rgb}{0.180392, 0.545098, 0.341176}
\definecolor{mygray}{rgb}{.75,.75,.75}
\definecolor{lightlightgray}{rgb}{.85,.85,.85}
\definecolor{lightyellow}{rgb}{0.975,0.975,0.65}
\definecolor{lightgreen}{rgb}{1, 1., 1}

\definecolor{amaranth}{rgb}{0.8, 0.17, 0.31}

\theorembodyfont{\itshape} 
\theoremheaderfont{\scshape}
\theoremstyle{plain}  
\newtheorem{Lemme}{Lemma}[section]
\newtheorem{Theoreme}{Theorem}
\newtheorem{Proposition}[Lemme]{Proposition}

\newtheorem{Lemma}[Lemme]{Lemma}
\newtheorem{Corollaire}[Lemme]{Corollary}

\crefname{Proposition}{Proposition}{Propositions}
\crefname{Corollary}{Corollary}{Corollaries}
\crefname{Lemma}{Lemma}{Lemmas}

\newcommand{\nc}{\newcommand}
\nc{\bib}{\bibitem}

\nc{\proof}{{\scshape Proof.\ }} 				
\nc{\eproof}{{\hfill \rule{0.5em}{0.5em}\medskip}}		
\nc{\be}{\begin{equation}}
\nc{\ee}{\end{equation}}
\nc{\chit}{\protect\raisebox{0.25ex}{$\chi$}}

 \tikzset{
  on each segment/.style={
    decorate,
    decoration={
      show path construction,
      moveto code={},
      lineto code={
        \path [#1]
        (\tikzinputsegmentfirst) -- (\tikzinputsegmentlast);
      },
      curveto code={
        \path [#1] (\tikzinputsegmentfirst)
        .. controls
        (\tikzinputsegmentsupporta) and (\tikzinputsegmentsupportb)
        ..
        (\tikzinputsegmentlast);
      },
      closepath code={
        \path [#1]
        (\tikzinputsegmentfirst) -- (\tikzinputsegmentlast);
      },
    },
  },
  mid arrow/.style={postaction={decorate,decoration={
        markings,
        mark=at position .625 with {\arrow[#1]{stealth}}
      }}},
}

\nc{\ir}{\mathrm{i}}
\nc{\eE}{\mathsf{e}} 
\nc{\dd}{\mathrm{d}}   
\nc{\Mod}{\textrm{ mod }}
\nc{\Tr}{\textrm{tr}}
\nc{\tr}{\textrm{tr}}
\nc{\Rcal}{\mathcal{R}}

\renewcommand{\ge}{\geqslant}
\renewcommand{\le}{\leqslant}

\date{}

\begin{document}

\topmargin -15mm
\oddsidemargin 05mm

%
%

\title{\mbox{}\vspace{-.2in}
\bf 
\huge Boundary emptiness formation probabilities\\[0.6cm] in the six-vertex model at $\boldsymbol{\Delta = -\frac12}$
}

\date{}
\maketitle

\begin{center}
{\Large Alexi Morin-Duchesne$^\dagger$, \quad Christian Hagendorf$^\dagger$, \quad Luigi Cantini$^\ddagger$
}
\end{center}\vspace{0cm}

\begin{center}
{\em $^\dagger$Universit\'e catholique de Louvain \\ 
Institut de Recherche en Math\'ematique et Physique\\ 
Chemin du Cyclotron 2, 1348 Louvain-la-Neuve, Belgium
\\[.4cm]
$^\ddagger$CY Cergy Paris Universit\'e\\
Laboratoire de Physique Th\'eorique et Mod\'elisation  \\ 
Avenue Adolphe Chauvin 2, 95300 Pontoise, France
}
\end{center}

\begin{center}
 {\tt alexi.morin-duchesne\,@\,uclouvain.be}
\qquad
{\tt christian.hagendorf\,@\,uclouvain.be}
\qquad
{\tt luigi.cantini\,@\,u-cergy.fr} 
\end{center}
\medskip

%
%
 
\begin{abstract}
We define a new family of overlaps $C_{N,m}$ for the XXZ Hamiltonian on a periodic chain of length~$N$. These are equal to the linear sums of the groundstate components, in the canonical basis, wherein $m$ consecutive spins are fixed to the state~${\uparrow}$. We define the boundary emptiness formation probabilities as the ratios $C_{N,m}/C_{N,0}$ of these overlaps. In the associated six-vertex model, they correspond to correlation functions on a semi-infinite cylinder of perimeter $N$. At the combinatorial point $\Delta = -\frac12$, we obtain closed-form expressions in terms of simple products of ratios of integers. \smallskip

\end{abstract}

\vspace{.5cm}
\noindent\textbf{Keywords:} XXZ Hamiltonian, emptiness formation probability, combinatorial point, quantum Knizhnik-Zamolodchikov equations\\

%
%

\newpage

\tableofcontents
\clearpage

%
\section{Introduction}
\label{sec:Introduction}
%

The \textit{emptiness formation probability} is one of the simplest correlation functions for quantum spin-$\frac12$ chains \cite{KOR97}. It is defined as the groundstate expectation value of a projector onto $m \geqslant 0$ consecutive spins pointing upwards. This expectation value allows one to measure the degree of ferromagnetic order in the groundstate and to detect quantum phase transitions. Therefore, it has been extensively studied in the literature for both finite and infinite spin chains. For the XY spin chain, exact expressions were obtained via free-fermion methods \cite{STN01,AB03,FRA06,AV19}. For the XXZ spin chain, techniques from quantum integrability allowed several authors to exactly compute the emptiness formation probability \cite{KIT02,KOZL08,C12}. In addition to these lattice calculations, several field-theory predictions are available \cite{AB02,KLNS03,S14}. 

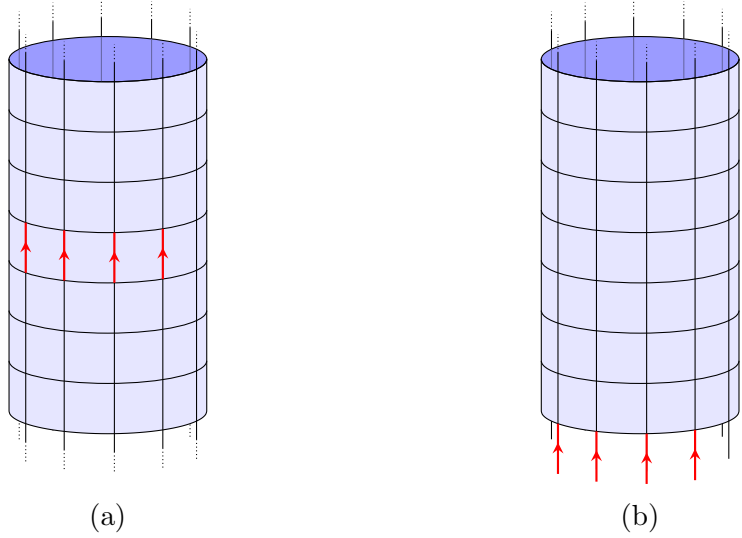
\begin{figure}[h]
\begin{center}
\begin{tikzpicture}

\begin{scope}
\draw (0,-.75) node {(a)};

\foreach \k in {1,2,...,5}
{
  \draw[densely dotted,yshift=.66cm] ( {1.3*cos(pi*6/5+pi*\k/6 r)}, {.3*sin(pi*6/5+pi*\k/6 r)-.5}) -- 
  ( {1.3*cos(pi*6/5+pi*\k/6 r)}, {.3*sin(pi*6/5+pi*\k/6 r)-.25});
  \draw ( {1.3*cos(pi*6/5+pi*\k/6 r)}, {.3*sin(pi*6/5+pi*\k/6 r)+16/3}) --( {1.3*cos(pi*6/5+pi*\k/6 r)}, {.3*sin(pi*6/5+pi*\k/6 r)+16/3+.25});
  \draw[densely dotted] ( {1.3*cos(pi*6/5+pi*\k/6 r)}, {.3*sin(pi*6/5+pi*\k/6 r)+16/3+.25})--( {1.3*cos(pi*6/5+pi*\k/6 r)}, {.3*sin(pi*6/5+pi*\k/6 r)+16/3+.5});
}

\draw [fill=lightlightblue,domain=-pi:0,smooth] (-1.3,16/3) -- plot ( {1.3*cos(\x r)}, {.3*sin(\x r)+2/3}) -- (1.3,16/3) -- cycle;

\begin{scope}
\fill[lightblue] (1.3,16/3) arc [start angle=0, end angle = 360,x radius = 1.3 cm, y radius = .3 cm];
\draw[clip] (1.3,16/3) arc [start angle=0, end angle = 360,x radius = 1.3 cm, y radius = .3 cm];
\foreach \k in {1,2,...,5}
{
  \draw[draw=blue!40!black!60] ( {1.3*cos(pi*6/5+pi*\k/6 r)}, {.3*sin(pi*6/5+pi*\k/6 r)+16/3}) --( {1.3*cos(pi*6/5+pi*\k/6 r)}, {.3*sin(pi*6/5+pi*\k/6 r)+16/3-.75});
}

\end{scope}

\foreach \h in {2,...,8}
{
  \draw[yshift=2/3*\h cm] (-1.3,0) arc [start angle=180, end angle = 360, x radius = 1.3 cm, y radius = .3 cm];
}

\foreach \k in {7,8,...,11}
{
  \draw ( {1.3*cos(pi*6/5+pi*\k/6 r)}, {.3*sin(pi*6/5+pi*\k/6 r)-.25+2/3}) --( {1.3*cos(pi*6/5+pi*\k/6 r)}, {.3*sin(pi*6/5+pi*\k/6 r)+16/3+.25});
  \draw[densely dotted] ( {1.3*cos(pi*6/5+pi*\k/6 r)}, {.3*sin(pi*6/5+pi*\k/6 r)-.5+2/3}) --( {1.3*cos(pi*6/5+pi*\k/6 r)}, {.3*sin(pi*6/5+pi*\k/6 r)-.25+2/3});
   \draw[densely dotted] ( {1.3*cos(pi*6/5+pi*\k/6 r)}, {.3*sin(pi*6/5+pi*\k/6 r)+16/3+.25})--( {1.3*cos(pi*6/5+pi*\k/6 r)}, {.3*sin(pi*6/5+pi*\k/6 r)+16/3+.5});
 }

\foreach \k in {7,8,9,10}
  {\draw [postaction = {on each segment=mid arrow},thick,red] ( {1.3*cos(pi*6/5+pi*\k/6 r)}, {.3*sin(pi*6/5+pi*\k/6 r)+8/3}) --( {1.3*cos(pi*6/5+pi*\k/6 r)}, {.3*sin(pi*6/5+pi*\k/6 r)+10/3});}

\end{scope}

\begin{scope}[xshift = 7cm]
\draw (0,-.75) node {(b)};
\foreach \k in {1,2,...,5}
{
  \draw[yshift=.66cm] ( {1.3*cos(pi*6/5+pi*\k/6 r)}, {.3*sin(pi*6/5+pi*\k/6 r)-.5}) -- 
  ( {1.3*cos(pi*6/5+pi*\k/6 r)}, {.3*sin(pi*6/5+pi*\k/6 r)-.25});
  \draw ( {1.3*cos(pi*6/5+pi*\k/6 r)}, {.3*sin(pi*6/5+pi*\k/6 r)+16/3}) --( {1.3*cos(pi*6/5+pi*\k/6 r)}, {.3*sin(pi*6/5+pi*\k/6 r)+16/3+.25});
  \draw[densely dotted] ( {1.3*cos(pi*6/5+pi*\k/6 r)}, {.3*sin(pi*6/5+pi*\k/6 r)+16/3+.25})--( {1.3*cos(pi*6/5+pi*\k/6 r)}, {.3*sin(pi*6/5+pi*\k/6 r)+16/3+.5});
}
\draw [fill=lightlightblue,domain=-pi:0,smooth] (-1.3,16/3) -- plot ( {1.3*cos(\x r)}, {.3*sin(\x r)+2/3}) -- (1.3,16/3) -- cycle;

\begin{scope}
\fill[lightblue] (1.3,16/3) arc [start angle=0, end angle = 360,x radius = 1.3 cm, y radius = .3 cm];
\draw[clip] (1.3,16/3) arc [start angle=0, end angle = 360,x radius = 1.3 cm, y radius = .3 cm];
\foreach \k in {1,2,...,5}
{
  \draw[draw=blue!40!black!60] ( {1.3*cos(pi*6/5+pi*\k/6 r)}, {.3*sin(pi*6/5+pi*\k/6 r)+16/3}) --( {1.3*cos(pi*6/5+pi*\k/6 r)}, {.3*sin(pi*6/5+pi*\k/6 r)+16/3-.75});
}

\end{scope}

\foreach \h in {2,...,8}
{
  \draw[yshift=2/3*\h cm] (-1.3,0) arc [start angle=180, end angle = 360, x radius = 1.3 cm, y radius = .3 cm];
}
\foreach \k in {7,8,...,11}
{
  \draw ( {1.3*cos(pi*6/5+pi*\k/6 r)}, {.3*sin(pi*6/5+pi*\k/6 r)-.5+2/3}) --( {1.3*cos(pi*6/5+pi*\k/6 r)}, {.3*sin(pi*6/5+pi*\k/6 r)+16/3+.25});
   \draw[densely dotted] ( {1.3*cos(pi*6/5+pi*\k/6 r)}, {.3*sin(pi*6/5+pi*\k/6 r)+16/3+.25})--( {1.3*cos(pi*6/5+pi*\k/6 r)}, {.3*sin(pi*6/5+pi*\k/6 r)+16/3+.5});
 }

\foreach \k in {7,8,9,10}
  {\draw [postaction = {on each segment=mid arrow},thick,red] ( {1.3*cos(pi*6/5+pi*\k/6 r)}, {.3*sin(pi*6/5+pi*\k/6 r)}) --( {1.3*cos(pi*6/5+pi*\k/6 r)}, {.3*sin(pi*6/5+pi*\k/6 r)+2/3});}
 \end{scope}
\end{tikzpicture}
\end{center}
\caption{(a) A square lattice, wrapped around an infinite cylinder, with a sequence of consecutive vertical edges oriented upwards.
(b) A square lattice, wrapped around a semi-infinite cylinder, with a free end at the bottom and a sequence of consecutive boundary edges oriented upwards.
}
\label{fig:Cylinders}
\end{figure}

In this article, we introduce and study a new correlation function for quantum spin-$\frac 12$ chains. We call it the \textit{boundary emptiness formation probability}. To motivate its definition, we recall that for a spin chain of length $N\geqslant 1$, the emptiness formation probability is given by 
\begin{equation}
\label{eqn:DefEFP}
\mathrm{EFP}_{N,m} = \frac{\sum_{\alpha_{m+1},\dots,\alpha_N\in \{\uparrow,\downarrow\}}|\psi_{\uparrow\cdots\uparrow\alpha_{m+1}\cdots\alpha_N}|^2}{\sum_{\alpha_1,\dots,\alpha_N\in \{\uparrow,\downarrow\}}|\psi_{\alpha_1\cdots\alpha_N}|^2},
\end{equation}
where the $\psi_{\alpha_1\cdots\alpha_N}$ are the components of the spin-chain groundstate $|\psi\rangle$ in the canonical basis. Similarly, we define the boundary emptiness formation probability as
\begin{equation}
  \label{eqn:DefBEFP}
  \mathrm{BEFP}_{N,m} = \frac{\sum_{\alpha_{m+1},\dots,\alpha_N\in \{\uparrow,\downarrow\}}\psi_{\uparrow\cdots\uparrow\alpha_{m+1}\cdots\alpha_N}}{\sum_{\alpha_1,\dots,\alpha_N\in \{\uparrow,\downarrow\}}\psi_{\alpha_1\cdots\alpha_N}}.
\end{equation}
We focus on the periodic XXZ spin chain. For this spin chain, $\mathrm{EFP}_{N,m}$ and $\mathrm{BEFP}_{N,m}$ respectively correspond to bulk and boundary correlation functions of the six-vertex model. To see this, we consider the six-vertex model on a square lattice with $N$ vertical lines wrapped around an infinite cylinder, as shown in \cref{fig:Cylinders}(a). On this infinite cylinder, $\mathrm{EFP}_{N,m}$ is the probability that $m$ consecutive vertical edges between two adjacent horizontal lines be aligned upwards. Likewise, for the boundary emptiness formation probability, we consider the six-vertex model on a square lattice wrapped around a semi-infinite cylinder with $N$ vertical lines, as shown in \cref{fig:Cylinders}(b). For free boundary conditions at the cylinder's end, $\mathrm{BEFP}_{N,m}$ is the probability that $m$ consecutive boundary edges be aligned upwards, and is hence a boundary correlation function.

We exploit the relation with the six-vertex model to compute an exact expression for the boundary emptiness formation probability of the periodic XXZ chain with the anisotropy parameter $\Delta =-\frac12$. For certain diagonally-twisted boundary conditions, the groundstates of this spin chain exhibit remarkable connections to the enumerative combinatorics of alternating sign matrices and plane partitions \cite{STROG01,RAZ00,RAZ01,BAT01}. These connections appear if we choose the length $N$ of the spin chain and the twist angle $\phi$ according to
\begin{subequations}
\begin{alignat}{4}
  &(\textrm{i}) \quad&&N=2n+1,\qquad &&\phi=0,
\\
  &(\textrm{ii}) \quad  &&N=2n,\qquad &&\phi=-\frac{\pi}{3},
\end{alignat}
\end{subequations}
where $n\geqslant 1$ is an integer. In the case (i), the groundstate is doubly-degenerate. We choose a basis $|\psi^+_{2n+1}\rangle$ and $|\psi_{2n+1}^-\rangle$ for its eigenspace such that the basis vectors have magnetisation $\mu = +1$ and $\mu = -1$ respectively.\footnote{It is common to include a factor of $\frac12$ in the definition of the magnetisation. We avoid this prefactor for notational convenience, as we will denote these groundstates by $|\psi_N^\mu\rangle$.} In case~(ii), the groundstate is non-degenerate. We write $|\psi^0_{2n}\rangle$ for the corresponding eigenvector. We fix the normalisation of these groundstate vectors by setting
\begin{equation}
  \label{eqn:SpecialComponents}
  (\psi_{2n+1}^{+})_{\underset{\scriptstyle n+1}{\underbrace{\scriptstyle\uparrow\cdots\uparrow}}\underset{\scriptstyle n}{\underbrace{\scriptstyle\downarrow\cdots\downarrow}}}=(\psi_{2n+1}^{-})_{\underset{\scriptstyle n}{\underbrace{\scriptstyle\uparrow\cdots\uparrow}}\underset{\scriptstyle n+1}{\underbrace{\scriptstyle\downarrow\cdots\downarrow}}}=(\psi_{2n}^{0})_{\underset{\scriptstyle n}{\underbrace{\scriptstyle\uparrow\cdots\uparrow}}\underset{\scriptstyle n}{\underbrace{\scriptstyle\downarrow\cdots\downarrow}}}=1.
\end{equation}
For this choice of normalisation, we prove the following result:
\begin{Theoreme}
\label{thm:overlaps.as.products}
For each $n\geqslant 1$, the overlaps
\begin{subequations}
\begin{align}
\label{def-c.mu}
C^{0}_{2n,m} &=
\langle \underbrace{\uparrow\dots \uparrow}_{m}|\otimes \left(\langle {\uparrow}|+\langle {\downarrow}|\right)^{\otimes(2n-m)}|\psi^0_{2n}\rangle,\\
C^{+}_{2n+1,m} &=
\langle \underbrace{\uparrow\dots \uparrow}_{m}|\otimes \left(\langle {\uparrow}|+\langle {\downarrow}|\right)^{\otimes(2n-m+1)}|\psi^+_{2n+1}\rangle,\\
C^{-}_{2n+1,m} &=
\langle \underbrace{\uparrow\dots \uparrow}_{m}|\otimes \left(\langle {\uparrow}|+\langle {\downarrow}|\right)^{\otimes(2n-m+1)}|\psi^-_{2n+1}\rangle,
\end{align}
\end{subequations}
are given by the product formulas
\begin{subequations}
\label{eq:C.results}
\begin{alignat}{2}
C^{0}_{2n,m} &= \eE^{\ir \pi(n-m)/6} \frac{3^{(n-m)/2}}{2^{m(m-1)/2}}\prod_{k=0}^{m-1} \prod_{j=0}^k \frac{(2j+k+2)(2j+n-k)}{(2j+1)(2j+2n-k)} \prod_{\ell=0}^{n-1} \frac{(3\ell+1)!}{(n+\ell)!},\label{eq:C.results.even}\\
C^{+}_{2n+1,m} &= \frac1{3^m 2^{m(m-1)/2}}\prod_{k=0}^{m-1} \prod_{j=0}^k \frac{(2j+k+3)(2j+n+1-k)}{(2j+1)(2j+2n+1-k)} \prod_{\ell=0}^{n} \frac{(3\ell)!}{(n+\ell)!},\\
C^{-}_{2n+1,m} &= \frac1{3^m 2^{m(m-1)/2}}\prod_{k=0}^{m-1} \prod_{j=0}^k \frac{(2j+k+3)(2j+n-k)}{(2j+1)(2j+2n+1-k)} \prod_{\ell=0}^{n} \frac{(3\ell)!}{(n+\ell)!}.
\end{alignat}
\end{subequations}
\end{Theoreme}
Our proof of this theorem relies on the fact that the spin-chain groundstates are eigenstates of the transfer matrix of the six-vertex model \cite{BAXT08}. For $\Delta=-\frac12$, these eigenstates can be obtained from the homogeneous limit of polynomial solutions of the level-one $U_q(\widehat{s\ell_2})$ \textit{quantum Knizhnik-Zamolodchikov system} with $q=\eE^{2\pi \ir/3}$ \cite{RAZ07}. We introduce scalar products for these polynomial solutions that generalise $C^0_{2n,m}$ and $C^\pm_{2n+1,m}$ to the inhomogeneous six-vertex model. Their polynomial and symmetry properties determine them uniquely. This uniqueness allows us to find determinant expressions for the scalar products for $q=\eE^{2\pi \ir/3}$. Finally, we compute the homogeneous limit of these determinants. The idea to compute quantities of interest from determinant expressions for the inhomogeneous six-vertex model was used previously in Kuperberg's proof of the alternating sign matrix conjecture \cite{K96}. The same idea was exploited by Kitanine et al.~\cite{KIT02} to compute the large-$N$ limit of the emptiness formation probability of the XXZ spin chain at the combinatorial point. In the present case, the determinants are more complicated than those of these two earlier papers, and we are able to compute their homogeneous limit
with the help of certain special evaluations of Schur polynomials. This limit yields the expressions \eqref{eq:C.results}.

We note that for $m=0$, the overlaps \eqref{eq:C.results} evaluate to
\begin{subequations}
  \label{eqn:Cm0}
\begin{align}
	C_{2n,0}^0 &= \sum_{\alpha_1,\dots,\alpha_{2n} \in \{\uparrow,\downarrow\}}(\psi^0_{2n})_{\alpha_1\cdots\alpha_{2n}} = \eE^{\ir \pi n/6}\prod_{\ell=0}^{n-1} \frac{(3\ell+1)!}{(n+\ell)!},\\
    C_{2n+1,0}^\pm &= \sum_{\alpha_1,\dots,\alpha_{2n+1} \in \{\uparrow,\downarrow\}}(\psi^\pm_{2n+1})_{\alpha_1\cdots\alpha_{2n+1}} = \prod_{\ell=0}^{n}\frac{(3\ell)!}{(n+\ell)!}.
\end{align}
\end{subequations}
We recognise in these expressions the numbers of alternating sign matrices of size $n$ and diagonally-antidiagonally symmetric alternating sign matrices of size $2n+1$, respectively \cite{BRES99,BEHR17}. Combining \eqref{eq:C.results} and \eqref{eqn:Cm0}, we obtain the boundary emptiness formation probabilities
\begin{subequations}
\label{eqn:BEFP}
\begin{alignat}{3}
\mathrm{BEFP}^{0}_{2n,m} &= \frac{C^0_{2n,m}}{C_{2n,0}^0} = \, \frac{\eE^{-\ir \pi m/6}}{3^{m/2}2^{m(m-1)/2}}\prod_{k=0}^{m-1} \prod_{j=0}^k \frac{(2j+k+2)(2j+n-k)}{(2j+1)(2j+2n-k)},\\
  \mathrm{BEFP}^+_{2n+1,m} &= \frac{C^+_{2n+1,m}}{C_{2n+1,0}^+} = \frac1{3^m 2^{m(m-1)/2}}\prod_{k=0}^{m-1} \prod_{j=0}^k \frac{(2j+k+3)(2j+n+1-k)}{(2j+1)(2j+2n+1-k)}, \\
  \mathrm{BEFP}^-_{2n+1,m} &= \frac{C^-_{2n+1,m}}{C_{2n+1,0}^-} = \frac1{3^m 2^{m(m-1)/2}}\prod_{k=0}^{m-1} \prod_{j=0}^k \frac{(2j+k+3)(2j+n-k)}{(2j+1)(2j+2n+1-k)}.
\end{alignat}
\end{subequations}
Similar factorised formulas exist for the emptiness formation probabilities of the XXZ spin chain at $\Delta=-\frac12$ \cite{RAZ01,C12}.

The plan of this article is as follows. In \cref{sec:Model}, we recall basic facts about the Hamiltonian of the XXZ spin chain and the transfer matrix of the six-vertex model. Furthermore, we establish the relation between the six-vertex model on a semi-infinite cylinder and the boundary emptiness formation probability. In \cref{sec:qKZ}, we discuss the level-one $U_q(\widehat{s\ell_2})$ quantum Knizhnik-Zamolodchikov system, its polynomial solutions and their relations to the groundstates of the XXZ spin chain at $\Delta=-\frac12$. In \cref{sec:DefSP}, we use the polynomial solutions to define the scalar products that generalise the overlaps for the spin-chain groundstates. We discuss their symmetry and polynomial properties in detail. In \cref{sec:Se}, we use these properties to prove a determinant expression for one of the scalar products at $q=\eE^{2\pi \ir/3}$, corresponding to the case where $N$ and $m$ are even. In \cref{sec:homogeneous}, we focus on this scalar product and compute the homogeneous limit of the determinant expression. In \cref{sec:Conclusion}, we present our conclusions and discuss the asymptotic behaviour of the boundary emptiness formation probability along with its possible relations to field theory.

We relegate the technical aspects of our work to several appendices. 
In \cref{sec:Schur.eval}, we prove an evaluation of Schur polynomials that is relevant to the homogeneous limit of our scalar products.
In \cref{app:Barnes}, we provide formulas for the overlaps in terms of the Barnes $G$-function. These are useful for the computation of the asymptotic behaviour of the boundary emptiness formation probability. In \cref{app:five.limits}, we present the computations of the homogeneous limits for the remaining scalar products.

\section{The XXZ spin chain and the six-vertex model}
\label{sec:Model}

In this section, we recall the definition of the XXZ Hamiltonian and the transfer matrix of the six-vertex model. We also establish the relation between the boundary emptiness formation probabilities for the spin-chain groundstates and certain correlation functions for the six-vertex model on a semi-infinite cylinder.

\subsection{The Hamiltonian of the XXZ spin chain}
\label{sec:6VModel}

\paragraph{Notations.} Let us consider a periodic chain of quantum spins $1/2$ with $N$ sites. Its Hilbert space is given by $V^N = V_1\otimes V_2\otimes \cdots \otimes V_N$, where $V_i = \mathbb C^2$ is a copy of the Hilbert space of a single spin, for each $i=1,\dots,N$. A basis of the single spin Hilbert space is given by
\begin{equation}
  |{\uparrow}\rangle =
  \begin{pmatrix}
    1\\ 0
  \end{pmatrix},
  \qquad
  |{\downarrow}\rangle =
  \begin{pmatrix}
    0\\ 1
  \end{pmatrix}.
\end{equation}
A canonical basis of $V^N$ is given by the tensor products $|\alpha_1\alpha_2\cdots \alpha_N\rangle = |\alpha_1\rangle \otimes |\alpha_2\rangle \otimes \cdots \otimes |\alpha_N\rangle$, where $\alpha_i \in \{{\uparrow,\downarrow}\}$ for each $i=1,\dots,N$. For any vector $|\psi\rangle \in V^N$, we may write
\begin{equation}
  |\psi\rangle = \sum_{\alpha_1,\dots,\alpha_N\in \{\uparrow,\downarrow\}}\psi_{\alpha_1\cdots\alpha_N}|\alpha_1\cdots\alpha_N\rangle,
\end{equation}
where the coefficients $\psi_{\alpha_1\cdots\alpha_N}$ are the components of $|\psi\rangle$ in the canonical basis of $V^N$.

For any vector $|\psi\rangle \in V^N$, we define a dual vector (or co-vector) $\langle \psi| \in (V^N)^\ast$ by transposition: $\langle \psi|=|\psi\rangle^t$. The dual pairing between $\langle \psi|\in (V^N)^\ast$ and $|\phi\rangle \in V^N$ is given by
\begin{equation}
  \langle \psi|\phi\rangle = \sum_{\alpha_1,\dots,\alpha_N\in\{{\uparrow,\downarrow}\}} \psi_{\alpha_1\cdots\alpha_N}\phi_{\alpha_1\cdots\alpha_N}.
\end{equation}
We refer to this dual pairing as the scalar product on $V^N$ (even though, strictly speaking, it only defines a scalar product on a real subspace).

\paragraph{The spin-chain Hamiltonian.} The properties and interactions of the quantum spins are described by the Pauli matrices
\begin{equation}
  \sigma^x =
  \begin{pmatrix}
    0 & 1\\
    1 & 0
  \end{pmatrix},
  \quad
  \sigma^y =
  \begin{pmatrix}
    0 & -\ir\\
    \ir & 0
  \end{pmatrix},
  \quad
  \sigma^z =
  \begin{pmatrix}
    1 & 0\\
    0 & -1
  \end{pmatrix}.
\end{equation}
For each $i=1,\dots,N$, we denote by $\sigma_i^\alpha,\,\alpha=x,y,z,$ the Pauli matrix $\sigma^\alpha$ acting on the $i$-th factor of the tensor product $V^N$. In terms of these operators, the Hamiltonian of the XXZ spin chain with the anisotropy parameter $\Delta$ is given by
\begin{equation}
  H = -\frac{1}{2}\sum_{i=1}^N \sigma_i^x\sigma_{i+1}^x + \sigma_i^y\sigma_{i+1}^y + \Delta \sigma_i^z\sigma_{i+1}^z.
  \label{eqn:XXZHamiltonian}
\end{equation}
The boundary conditions for the spin chain are fixed by expressing the spin operators $\sigma_{N+1}^x,\sigma_{N+1}^y,\sigma_{N+1}^z$ in terms of $\sigma_{1}^x,\sigma_{1}^y,\sigma_{1}^z$. In this article, we consider boundary conditions that correspond to a diagonal twist: \begin{equation}
  \sigma_{N+1}^x = \cos (2\phi) \sigma_1^x - \sin (2\phi)\sigma_1^y,\qquad \sigma_{N+1}^y = \sin (2\phi) \sigma_1^x + \cos (2\phi)\sigma_1^y, \qquad \sigma_{N+1}^z = \sigma_1^z.
  \label{eqn:BC}
\end{equation}
Here, the parameter $\phi$ is the twist angle. We will consider that it takes real values. The value $\phi=0$ corresponds to periodic boundary conditions. 

\paragraph{The combinatorial point.} The Hamiltonian \eqref{eqn:XXZHamiltonian} with the boundary conditions \eqref{eqn:BC} is a Hermitian matrix for all twist angles $\phi$. It is therefore diagonalisable. Moreover, it preserves the magnetisation:
\begin{equation}
  \label{eqn:MHComm}
  [M,H]= 0, \qquad  M = \sum_{i=1}^N \sigma_i^z.
\end{equation}
The magnetisation operator is diagonalisable, too. Its eigenvalues are given by $\mu = 2m-N$, where $m=0,\dots,N$. We say that a vector has the magnetisation $\mu$ if it is an eigenvector of $M$ with the eigenvalue $\mu$. The commutation relation \eqref{eqn:MHComm} implies that we may simultaneously diagonalise $H$ and $M$. The corresponding eigenvalue problem is
\begin{equation}
  \label{eqn:EVProblem}
  H|\psi\rangle = E|\psi\rangle, \qquad M|\psi\rangle = \mu|\psi\rangle, \qquad |\psi\rangle \in V^N.
\end{equation}
For finite $N$ and generic values of the anisotropy parameter $\Delta$, it is a challenging problem to compute its solutions. A remarkable exception is however the case $\Delta =-\frac12$, where the groundstate eigenvalues and eigenvectors of the Hamiltonian can be computed for finite $N$ under certain circumstances. This can indeed be achieved for the following special cases: 
\begin{subequations}
\begin{alignat}{4}
  \label{eqn:OddEV}
  &(\textrm{i}) \quad&&N=2n+1,\qquad &&\phi=0, \qquad && \mu = \pm 1,
\\
  \label{eqn:EvenEV}
  &(\textrm{ii}) \quad  &&N=2n,\qquad &&\phi=-\frac{\pi}{3}, \qquad &&\mu = 0.
\end{alignat}
\end{subequations}
In these circumstances, the space of the solutions of \eqref{eqn:EVProblem} is one-dimensional and the groundstate energy is $E=-\frac{3N}4$ \cite{DIFR06,HAG18}. In a suitable normalisation, the components of the corresponding groundstate eigenvectors, and certain correlation functions for these vectors possess remarkable relations with the enumerative combinatorics of alternating sign matrices and plane partitions \cite{RAZ00,RAZ01,DIFR06,RAZ07,C12}. For this reason, the value $\Delta=-\frac12$ is often referred to as the \textit{combinatorial point}.

\subsection{The transfer matrix of the six-vertex model}

\begin{figure}
  \centering
  \begin{tikzpicture}[>=stealth]
    \begin{scope}
    \draw[postaction = {on each segment=mid arrow}] (0,0) -- (.5,0) -- (.5,.5);
    \draw[postaction = {on each segment=mid arrow}] (.5,-.5) -- (.5,0) -- (1,0);
    \draw (.5,-.5) node[below] {$a$};
    \end{scope}
    \begin{scope}[xshift=1.5cm]
    \draw[postaction = {on each segment=mid arrow}] (.5,.5) -- (.5,0) -- (0,0);
    \draw[postaction = {on each segment=mid arrow}] (1,0) -- (.5,0) -- (.5,-.5);
    \draw (.5,-.5) node[below] {$a$};
    \end{scope}
    \begin{scope}[xshift=3cm]
    \draw[postaction = {on each segment=mid arrow}] (0,0) -- (.5,0) -- (1,0);
    \draw[postaction = {on each segment=mid arrow}] (.5,.5) -- (.5,0) -- (.5,-.5);
    \draw (.5,-.5) node[below] {$b$};
    \end{scope}
    \begin{scope}[xshift=4.5cm]
    \draw[postaction = {on each segment=mid arrow}] (1,0) -- (.5,0) -- (0,0);
    \draw[postaction = {on each segment=mid arrow}] (.5,-.5) -- (.5,0) -- (.5,.5);
    \draw (.5,-.5) node[below] {$b$};
    \end{scope}
    \begin{scope}[xshift=6cm]
    \draw[postaction = {on each segment=mid arrow}] (0,0) -- (.5,0) -- (.5,-.5);
    \draw[postaction = {on each segment=mid arrow}] (1,0) -- (.5,0) -- (.5,.5);
    \draw (.5,-.5) node[below] {$c_1$};
    \end{scope}
    \begin{scope}[xshift=7.5cm]
    \draw[postaction = {on each segment=mid arrow}] (.5,-.5) -- (.5,0) -- (0,0);
    \draw[postaction = {on each segment=mid arrow}] (.5,.5) -- (.5,0) -- (1,0);
    \draw (.5,-.5) node[below] {$c_2$};
    \end{scope}
  \end{tikzpicture}
  \caption{The local configurations of the six-vertex model and their Boltzmann weights.}
  \label{fig:Vertices6V}
\end{figure}
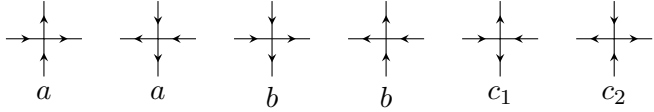

The XXZ spin chain is intimately related to the six-vertex model on the square lattice \cite{BAXT08}. Here, we consider such a lattice wrapped around a cylinder, as in \cref{fig:Cylinders}. A configuration of this model is the assignment of an orientation to each lattice edge, subject to the {\textit{ice rule}}. According to this rule, the number of edges oriented towards any vertex is equal to the number of edges oriented away from it. Hence, there are six possible local configurations around a vertex. They are shown, along with their Boltzmann weights, in \cref{fig:Vertices6V}.
The Boltzmann weight of a configuration is the product of the Boltzmann weights of its vertices. In principle, we may thus compute the model's partition function. In the following, we will be interested in the partition function on the cylinder and its relation to the boundary emptiness formation probability. To compute it, we recall the definitions of the $R$-matrix and of the transfer matrix of the six-vertex model.

\paragraph{The $\boldsymbol{R}$-matrix.} We parameterise the weights $a,b,c_1,c_2$ of the six-vertex model by the spectral parameter $z$ and the crossing parameter $q$. The parameterisation is given by
\begin{equation}
  a(z) = \frac{q z - q^{-1}}{q-q^{-1}z}, \qquad b(z) = \frac{z-1}{q-q^{-1}z}, \qquad c_1(z) = \frac{(q-q^{-1})z}{q-q^{-1}z}, \qquad c_2(z) = \frac{q-q^{-1}}{q-q^{-1}z}.
\end{equation}
The $R$-matrix is an operator $R(z)$ on $\mathbb C^2\otimes \mathbb C^2$. With respect to the canonical basis $|{\uparrow\uparrow}\rangle,|{\uparrow\downarrow}\rangle,|{\downarrow\uparrow}\rangle,|{\downarrow\downarrow}\rangle$ of $\mathbb C^2\otimes \mathbb C^2$, it is given by the matrix
\begin{equation}
  R(z)=
  \begin{pmatrix}
    a(z) & 0 & 0 & 0\\
    0 & b(z) & c_1(z) & 0\\
    0 & c_2(z) & b(z) & 0\\
    0 & 0 & 0 & a(z)
  \end{pmatrix}.
\end{equation}

The parameterisation of the weights leads to several properties of the $R$-matrix that we frequently use in the following. First, the $R$-matrix satisfies the Yang-Baxter equation: On $V_1\otimes  V_2\otimes V_3$, we have
\begin{equation}
  \label{eqn:YBE}
  R_{12}(z_1/z_2)R_{13}(z_1/z_3)R_{23}(z_2/z_3)= R_{23}(z_2/z_3)R_{13}(z_1/z_3)R_{12}(z_1/z_2),
\end{equation}
for all $z_1,z_2,z_3$. Here, we write $R_{ij}(z)$ to denote the operator $R(z)$ acting on the factors $V_i$ and $V_j$ in this tensor product. Second, for $z=1$ the $R$-matrix is
\begin{equation}
  R(z=1) = P
  \label{eqn:RzOne}
\end{equation}
where $P$ is the permutation operator. It is defined through
\begin{equation}
  P(|\mu\rangle \otimes |\nu\rangle)= |\nu\rangle \otimes |\mu\rangle,
  \label{eqn:PermutationOp}
\end{equation}
for all $|\mu\rangle,|\nu\rangle\in \mathbb C^2$. We use this permutation operator to define the $\check R$-matrix: $\check R(z) = P R(z)$. It follows from \eqref{eqn:YBE} that this matrix obeys the braid version of the Yang-Baxter equation: For all $z_1,z_2,z_3$, we have
\begin{equation}
  \check R_{12}(z_2/z_3)\check R_{23}(z_1/z_3) \check R_{12}(z_1/z_2) = \check R_{23}(z_1/z_2)\check R_{12}(z_1/z_3) \check R_{13}(z_2/z_3).
  \label{eqn:BraidYBE}
\end{equation}
Third, the $\check R$-matrix obeys the unitarity relation
\begin{equation}
  \label{eqn:Unitarity}
  \check R(z)\check R(z^{-1}) = 1,
\end{equation}
for all $z$. Finally, let us define the vector
\begin{equation}
  \label{eqn:DefOmega}
  |\omega\rangle = |{\uparrow\downarrow}\rangle - q^{-1}|{\downarrow\uparrow}\rangle.
\end{equation}
One checks that, for all $z$, it is an eigenvector of $\check R(z)$ with eigenvalue one:
\begin{equation}
  \check R(z)|\omega\rangle = |\omega\rangle.
  \label{eqn:OmegaEVOfR}
\end{equation}

\paragraph{The transfer matrix.} The transfer matrix of the six-vertex model on a square lattice with $N\geqslant 1$ vertical lines, and diagonally-twisted periodic boundary conditions along the horizontal line is given by
\begin{equation}
  T(z) = \textup{tr}_0\left(\eE^{\ir \phi \sigma_0^z}R_{0,N}(z)\cdots R_{0,1}(z)\right).
  \label{eqn:TM6V}
\end{equation}
Here,  $R_{0,i}(z)$ is the $R$-matrix acting non-trivially only on the factors $V_0$ and $V_i$ in $V_0\otimes V^N$, where $V_0=\mathbb C^2$ is the {\it auxiliary space}. The trace is taken over this auxiliary space.

The transfer matrix commutes with the magnetisation operator $M$. In addition, the Yang-Baxter equation \eqref{eqn:YBE} and the relation $[R(z), \eE^{\ir \phi \sigma^z} \otimes \eE^{\ir \phi \sigma^z}] = 0$ imply that transfer matrices with different spectral parameters commute \cite{BAXT08}: We have
\begin{equation}
  [T(z),T(w)]=0,
  \label{eqn:TComm}
\end{equation}
for all $z,w$. We now use this commutation relation to recall a result about the diagonalisability of the transfer matrix. We focus on the case $|q|=1$. (A similar result exists for real $q$.)

\begin{Proposition}
  \label[Proposition]{prop:Diagonalisability}
 For $|q|=1$ and $\phi \in \mathbb R$, the transfer matrix $T(z)$ is diagonalisable.
\end{Proposition}
\proof
First, we compute the adjoint matrix $T(z)^\dagger$. To this end, we write $R(z)= R(z;q)$ in order to stress the dependence of the $R$-matrix on the crossing parameter. We have
  \begin{equation}
    T(z)^\dagger = \text{tr}_0 \left(\eE^{-\ir \phi \sigma_0^z}R_{0,N}(z^\ast;q^\ast)^{t_N}\cdots R_{0,1}(z^\ast;q^\ast)^{t_1}\right).
  \end{equation}
  For each $i=1,\dots,N$, the superscript $t_i$ denotes the transpose with respect to the factor $V_i$ of $V^N$. We now use $q^\ast = q^{-1}$ and the relation $R(w;q^{-1}) = R(w^{-1};q)^t$. It leads to
  \begin{equation}
    T(z)^\dagger = \text{tr}_0 \left(\eE^{-\ir \phi \sigma_0^z}R_{0,N}\big((z^*)^{-1};q\big)^{t_0}\cdots R_{0,1}\big((z^*)^{-1};q\big)^{t_0}\right).
  \end{equation}
  We evaluate the transpose with respect to the auxiliary space, indicated by the superscript $t_0$, with the help of the {\it crossing symmetry} of the $R$-matrix:
  \begin{equation}
    R_{0,i}\big((z^*)^{-1};q\big)^{t_0} = -q(z^*)^{-1} U_0 R_{0,i}(q^{-2}z^{*};q) U_0^{-1}.
    \label{eqn:CrossingSymR}
  \end{equation}
  Here, $U_0$ is an operator $U:\mathbb C^2\to \mathbb C^2$ acting on the auxiliary space. In the canonical basis of $\mathbb C^2$, it acts as the antidiagonal matrix
  \begin{equation}
    U =
    \begin{pmatrix}
      0 & -1\\
      q(z^*)^{-1} & 0
    \end{pmatrix}.
  \end{equation}
  Using \eqref{eqn:CrossingSymR}, we obtain
  \begin{equation}
     T(z)^\dagger = \big(\!-q(z^*)^{-1}\big)^N \text{tr}_0 \left(U^{-1}_0\eE^{-\ir \phi \sigma_0^z}U_0R_{0,N}(q^{-2}z^*;q)\cdots R_{0,1}(q^{-2}z^*;q)\right)
  \end{equation}
  We furthermore have $U^{-1}\eE^{-\ir \phi \sigma^z}U = \eE^{\ir \phi \sigma^z}$. This leads to
  \begin{equation}
    T(z)^\dagger = \big(-q(z^*)^{-1}\big)^{N}T(q^{-2}z^*).
  \end{equation}
  
  Second, we use the relation \eqref{eqn:TComm} and find the commutation relation
  \begin{equation}
    [T(z),T(z)^\dagger]= \big(-q(z^*)^{-1}\big)^{N} [T(z),T(q^{-2}z^*)] =0.
  \end{equation}
  We conclude that for $|q|=1$, the transfer matrix commutes with its adjoint: It is a normal matrix. Hence, it is diagonalisable by a unitary transformation \cite{MEY00}.
  \eproof

\Cref{prop:Diagonalisability} implies, in particular, that $T(z)$ is diagonalisable for $z=1$. For this value of the spectral parameter, we use \eqref{eqn:RzOne}
to find the simple expression
\begin{equation}
  \label{eqn:TAndTau}
  T(z=1) = \tau^{-1}\eE^{\ir \phi \sigma_N^z}.
\end{equation}
Here, $\tau$ is the translation operator by one step to the left. It is the linear operator on $V^N$, whose action on the basis vectors is given by
\begin{equation}
  \label{eqn:DefTau}
  \tau |\alpha_1\alpha_2\cdots \alpha_N\rangle = |\alpha_2\cdots \alpha_N \alpha_1\rangle.
\end{equation}
Furthermore, the logarithmic derivative of the transfer matrix at $z=1$ has the simple expression
\begin{equation}
  \left.T(z)^{-1}\frac{\dd}{\dd z}T(z)\right|_{z=1}=-\frac{1}{q-q^{-1}}\left(H-\frac{3N\Delta}{2}\right),
  \label{eqn:HFromT}
\end{equation}
where
\begin{equation}
  \label{eqn:Anisotropy}
  \Delta = \frac12(q+q^{-1})
\end{equation}
and $H$ is the Hamiltonian of the XXZ spin chain \eqref{eqn:XXZHamiltonian} with the twisted boundary conditions \eqref{eqn:BC}.

We conclude from \eqref{eqn:TComm} and \eqref{eqn:HFromT} that the transfer matrix and the spin-chain Hamiltonian commute: We have $[T(z),H]=0$. For $|q|=1$ and $\phi \in \mathbb R$, they can thus be simultaneously diagonalised.\footnote{This is in fact believed to hold for generic values of $q$ and $\phi$.} These values of $q$ correspond to real values of the anisotropy parameter $\Delta$ in the range $[-1,1]$.

\paragraph{The combinatorial point.} Let us now consider $q=\eE^{2\pi\ir/3}$, for which the anisotropy parameter of the spin chain takes the value $\Delta =-\frac12$. At this point, we have the following remarkable result:

\begin{Theoreme}[Simple eigenvalue \cite{DIFR06,RAZ07,HAG18}]
  Let $q=\eE^{2\pi\ir/3}$. The space of solutions of the eigenvalue problem
  \begin{equation}
    T(z)|\psi\rangle = |\psi\rangle, \qquad M|\psi\rangle = \mu |\psi\rangle
  \end{equation}
  is one-dimensional in the following cases:
\begin{subequations}
\begin{alignat}{5}
    \label{eqn:OddSize}
   	&(\mathrm{i}) \qquad &&N=2n+1\qquad &&\phi=0\qquad &&\mu=+1,-1 \qquad &&n\geqslant 0,
	\\[0.15cm]
    \label{eqn:EvenSize}
    	&(\mathrm{ii}) \qquad &&N=2n\qquad &&\phi = -\frac{\pi}{3}\qquad &&\mu=0 \qquad &&n\geqslant 1.
\end{alignat}
\end{subequations}
  In each case, the eigenvalue $1$ is the largest eigenvalue of the transfer matrix.
\end{Theoreme}
Let $|\psi^\pm_{2n+1}\rangle$ and $|\psi_{2n}^0\rangle$ be the basis vectors of the eigenspaces corresponding to the cases \eqref{eqn:OddSize} and \eqref{eqn:EvenSize} respectively. It follows from \eqref{eqn:HFromT} that, for $n\geqslant 1$, they are eigenstates of the XXZ Hamiltonian with the eigenvalue $E = -\frac{3N}4$. We shall see below that the vectors $|\psi^\pm_{2n+1}\rangle$ and $|\psi_{2n}^0\rangle$ can be obtained as specialisations of the solutions of a quantum Knizhnik-Zamolodchikov system.

\subsection{The boundary emptiness formation probability}

In this section, we discuss the connection between partition functions of the six-vertex model on a semi-infinite cylinder and the boundary emptiness formation probability.

Let us consider a cylinder constructed from a square lattice with $N$ vertical and $L$ horizontal lines. The partition function of the vertex model on this cylinder with fixed boundary conditions $\alpha_1\cdots\alpha_N$ and $\beta_1\cdots\beta_N$ at the upper and lower boundary, respectively, is
\begin{equation}
  \label{eqn:ZFixed}
  Z^{\alpha_1\cdots\alpha_N}_{\beta_1\cdots\beta_N}(N,L) = \langle \alpha_1\cdots \alpha_N|T(z)^L|\beta_1\cdots\beta_N\rangle.
\end{equation}
Clearly, this quantity is nonzero only if the spin configurations $\alpha_1\cdots\alpha_N$ and $\beta_1\cdots\beta_N$ have the same magnetisation. We will assume this condition to hold in the following.
The partition function with fixed boundary conditions at the upper and free boundary conditions at the lower boundary is
\begin{equation}
  \label{eqn:ZMixed}
  Z^{\alpha_1\cdots\alpha_N}_f(N,L) = \sum_{\gamma_1,\dots,\gamma_N\in\{\uparrow,\downarrow\}}  Z^{\alpha_1\cdots\alpha_N}_{\gamma_1\cdots\gamma_N}(N,L).
\end{equation}

Let us now make the following three assumptions: (i) In the subsector of $V^N$ with the magnetisation of the spin configuration $\alpha_1\cdots \alpha_N$, the largest eigenvalue of the transfer matrix (in absolute value) is non-degenerate, (ii) the corresponding right-eigenstate $|\varphi\rangle$ has a non-zero overlap with the boundary state in \eqref{eqn:ZFixed}, namely $\langle \alpha_1\cdots \alpha_N|\varphi\rangle \neq 0$, and (iii) the linear sum of all components of $\varphi$ is nonzero. 
We note that the special case $z = 1$ is not covered by assumption (i), as in this case all the eigenvalues have unit norms. With these assumptions,  the ratio of the partition functions \eqref{eqn:ZFixed} and \eqref{eqn:ZMixed} is well-defined and possesses an explicit formula in terms of the components of $|\varphi\rangle$:
\begin{equation}
  \lim_{L\to \infty} \left(\frac{Z^{\alpha_1\cdots\alpha_N}_{\beta_1\cdots\beta_N}(N,L)}{ Z^{\alpha_1\cdots\alpha_N}_f(N,L)}\right) = \frac{\varphi_{\beta_1\cdots\beta_N}}{\sum_{\gamma_1,\dots,\gamma_N\in\{\uparrow,\downarrow\}}\varphi_{\gamma_1\cdots\gamma_N}}.
  \label{eqn:LimitRatio}
\end{equation}
Clearly, this is well-defined only if the sum in the denominator is non-zero, which is ensured by the assumption (iii) above.
We observe that, if all the above assumptions are satisfied, the result is independent of the boundary condition $\alpha_1\cdots\alpha_N$ at the top of the cylinder.

We now recall from the introduction that the boundary emptiness formation probability for the groundstate $|\psi\rangle \in V^N$ of the XXZ spin chain is given by
\begin{equation}
  \textup{BEFP}_{N,m} = \frac{\sum_{\beta_{m+1},\dots,\beta_N\in\{\uparrow,\downarrow\}}\psi_{\uparrow\cdots\uparrow \beta_{m+1}\cdots\beta_N}}{\sum_{\beta_1,\dots,\beta_N\in\{\uparrow,\downarrow\}}\psi_{\beta_1\cdots\beta_N}}.
\end{equation}
Hence, if the transfer-matrix eigenvector $|\varphi\rangle$ is equal to the spin-chain groundstate, $|\varphi\rangle = |\psi\rangle$, then we find, by comparison with \eqref{eqn:LimitRatio}, that
\begin{equation}
 \textup{BEFP}_{N,m}= \lim_{L \to \infty} \left(\frac{\sum_{\beta_{m+1},\dots,\beta_N\in\{\uparrow,\downarrow\}}Z^{\alpha_1\cdots\alpha_N}_{\uparrow\cdots\uparrow\beta_{m+1}\cdots \beta_N}(N,L)}{Z^{\alpha_1\cdots\alpha_N}_f(N,L)}\right).
\end{equation}
The equality $|\varphi\rangle = |\psi\rangle$ holds if the magnetisation of the spin configuration $\alpha_1\cdots\alpha_N$ is equal to the magnetisation of the groundstate sector.

Hence, in terms of the language of statistical mechanics, $\textup{BEFP}_{N,m}$ is an expectation value of an observable on a semi-infinite cylinder with free boundary conditions on one end, and fixed boundary conditions on the other end. The observable is the operator acting on $m$ consecutive aligned spins on the cylinder's free boundary that projects each of these spins to the state ${\uparrow}$. In the case where all the weights of the six-vertex model are real and positive, the boundary emptiness formation probability is indeed a probability.

\section{The level-one quantum Knizhnik-Zamolodchikov system}
\label{sec:qKZ}

In this section, we discuss the level-one $U_q(\widehat{s\ell_2})$ quantum Knizhnik-Zamolodchikov system. We focus on three polynomial solutions of this system and discuss their properties. In the homogeneous limit, these solutions yield eigenvectors of the transfer matrix of the six-vertex model at the combinatorial point.

\subsection{Definition}
\label{sec:DefqKZ}

The level-one $U_q(\widehat{s\ell_2})$ quantum Knizhnik-Zamolodchikov system is a set of equations for a vector $|\Psi\rangle\in V^N$ that depends on $N$ parameters $z_1,\dots,z_N$. This set consists of a set of exchange equations and a covariance property under cyclic shifts. The exchange equations are
\begin{equation}
  \label{eqn:Exchange}
  \check R_{i,i+1}(z_{i+1}/z_i)|\Psi(\dots, z_i,z_{i+1},\dots)\rangle = |\Psi(\dots, z_{i+1},z_i,\dots)\rangle,
\end{equation}
where $i=1,\dots,N-1$. Their consistency follows from the braid Yang-Baxter equation \eqref{eqn:BraidYBE} and the unitarity relation \eqref{eqn:Unitarity} of the $\check R$-matrix of the six-vertex model. The covariance property under cyclic shifts is
\begin{equation}
  \label{eqn:Cyclicity}
  \tau |\Psi(z_1,z_2,\dots,z_N)\rangle =D_N|\Psi(z_2,\dots,z_N,q^6z_1)\rangle.
\end{equation}
Here, $\tau$ denotes the translation operator \eqref{eqn:DefTau} by one step to the left. Furthermore, $D_N$ is an operator $D$ acting on the last factor of the tensor product $V^N$. Its precise form depends on $N$, as we will see below.

\paragraph{Exchange equations for the components.} It is often useful to write the exchange equations for the components $\Psi_{\alpha_1\cdots \alpha_N}$ of the vector $|\Psi\rangle$. To this end, we fix $i=1,\dots,N-1$ and consider the adjacent spins $\alpha_i$ and $\alpha_{i+1}$. We distinguish two cases. First, for adjacent parallel spins, we find
\begin{subequations}
  \label{eqn:EntriesAligned}
\begin{align}
(qz_{i+1}-q^{-1}z_{i})\Psi_{\cdots\uparrow\uparrow\cdots}(\dots,z_i,z_{i+1},\dots) =  (qz_{i}-q^{-1}z_{i+1})\Psi_{\cdots\uparrow\uparrow\cdots}(\dots,z_{i+1},z_{i},\dots),\\[0.1cm]
 (qz_{i+1}-q^{-1}z_{i})\Psi_{\cdots\downarrow\downarrow\cdots}(\dots,z_i,z_{i+1},\dots) =  (qz_{i}-q^{-1}z_{i+1})\Psi_{\cdots\downarrow\downarrow\cdots}(\dots,z_{i+1},z_{i},\dots).
\end{align}
\end{subequations}
Second, for antiparallel spins, we have the equations
\begin{subequations}
  \label{eqn:EntriesDDOps}
\begin{align}
  \Psi_{\cdots \downarrow \uparrow\cdots}(\dots,z_{i},z_{i+1},\dots) &=  \delta_i\Psi_{\cdots \uparrow\downarrow \cdots}(\dots,z_{i},z_{i+1},\dots),\\[0.1cm]
   \Psi_{\cdots \uparrow\downarrow \cdots}(\dots,z_{i},z_{i+1},\dots) &=   \delta_i^{-1}\Psi_{\cdots \downarrow\uparrow \cdots}(\dots,z_{i},z_{i+1},\dots),
\end{align}
\end{subequations}
where $ \delta_i$ and $\delta_i^{-1}$ are the divided-difference operators defined through
\begin{subequations}
\label{eq:divided.diff.ops}
\begin{align}
  \delta_i f(z_i,z_{i+1}) &= \frac{(qz_i-q^{-1}z_{i+1})f(z_{i+1},z_i)-(q-q^{-1})z_if(z_i,z_{i+1})}{z_{i+1}-z_i},\\
  \delta_i^{-1} f(z_i,z_{i+1}) &= \frac{(qz_i-q^{-1}z_{i+1})f(z_{i+1},z_i)-(q-q^{-1})z_{i+1}f(z_i,z_{i+1})}{z_{i+1}-z_i}.
\end{align}
\end{subequations}

We consider solutions to the exchange equations with magnetisation $\mu$. While we shall later focus on the cases $\mu = -1,0,1$, the discussion in this subsection applies more generally to $\mu = -N, -N + 2, \dots, N$. The equations satisfied by the components of $|\Psi\rangle$ hint at a simple strategy to construct these solutions. First, one fixes the component
\begin{align}
  \label{eqn:ReferenceEntry}
  \begin{split}
  \Psi_{\underset{\scriptstyle m}{\underbrace{\scriptstyle\uparrow\cdots\uparrow}}\underset{\scriptstyle N-m}{\underbrace{\scriptstyle\downarrow\cdots\downarrow}}} = \prod_{1\leqslant i<j\leqslant m}\left(\frac{qz_i-q^{-1}z_j}{q-q^{-1}}\right)\prod_{m+1\leqslant i<j\leqslant N}&\left(\frac{qz_i-q^{-1}z_j}{q-q^{-1}}\right) 
  \\&
  \times f(z_1,\dots,z_m;z_{m+1},\dots,z_N),
  \end{split}
\end{align}
where $f(z_1,\dots,z_m;z_{m+1},\dots,z_N)$ is separately symmetric in $z_1,\dots,z_m$ and $z_{m+1},\dots,z_N$. One readily checks that the component obeys \eqref{eqn:EntriesAligned} for each $i=1,\dots,m-1$, and $i=m+1,\dots,N-1$. Second, one \textit{defines} all other components pertaining to the subsector of interest through \eqref{eqn:EntriesDDOps}. By construction, the resulting vector $|\Psi\rangle$ obeys \eqref{eqn:Exchange}.

This strategy illustrates that it is sufficient to know a reference component in order to completely determine a solution to the exchange equation. It leads us to the following result:
\begin{Lemma}
  \label[Lemma]{lem:qKZEqual}
  Let $0\leqslant m \leqslant N$ be an integer and $|\Psi\rangle, |\Phi\rangle\in V^N$ be two solutions to the exchange equations with magnetisation $2m-N$. If
  \begin{equation}
    \Psi_{\underset{m}{\underbrace{\scriptstyle \uparrow\cdots\uparrow}}\underset{N-m}{\underbrace{\scriptstyle \downarrow\cdots \downarrow}}} =\Phi_{\underset{m}{\underbrace{\scriptstyle \uparrow\cdots\uparrow}}\underset{N-m}{\underbrace{\scriptstyle \downarrow\cdots \downarrow}}}\ ,
  \end{equation}
  then $|\Psi\rangle = |\Phi\rangle$.
\end{Lemma}

\paragraph{Cyclic covariance.} We consider solutions $|\Psi\rangle$ to the exchange equations with fixed magnetisation and focus on the case where 
\begin{equation}
  D|\alpha\rangle = \lambda_\alpha|\alpha\rangle, \quad \alpha \in\{\uparrow,\downarrow\}.
\end{equation}
Here, the factor $\lambda_\alpha$ is independent of $z_1,\dots,z_N$. In terms of the components, the cyclic-covariance property reads 
\begin{equation}
  \label{eqn:CyclicCovComps}
  \Psi_{\alpha_N\alpha_1\cdots\alpha_{N-1}}(q^{-6}z_N,z_1,\dots,z_{N-1}) = \lambda_{\alpha_N} \Psi_{\alpha_1\cdots \alpha_N}(z_1,\dots,z_N).
\end{equation}

\subsection{Polynomial solutions}
\label{sec:PolySols}

In this section, we discuss polynomial solutions to the quantum Knizhnik-Zamolodchikov system. To this end, we use the strategy outlined above: We fix a polynomial reference component and define a solution to the exchange equations through the action of the divided-difference operators. That all its components are polynomials is guaranteed by the following lemma:
\begin{Lemma}
  \label[Lemma]{lem:DDOpsPoly}
  Let $f$ be a homogeneous polynomial in $z_1,\dots,z_N$ of total degree $d$ and individual degrees at most $d_j$ for each $j=1,\dots,N$. Then, for each $i=1,\dots,N-1$, $\delta_if$ is a homogeneous polynomial of total degree $d$, and individual degree at most $d_j$ for each $j=1,\dots,N$.
\end{Lemma}
We omit the proof, as it follows directly from the form \eqref{eq:divided.diff.ops} of the divided-difference operators.

Let $n\geqslant 0$ be a non-negative integer and $N=2n+1$. We consider a solution $|\Psi\rangle = |\Psi_{2n+1}^{+}\rangle$ of the exchange relations with magnetisation $\mu=1$, constructed from the reference component \eqref{eqn:ReferenceEntry} for the choice $f(z_1,\dots,z_{n+1};z_{n+2},\dots,z_{2n+1}) = \prod_{i=n+2}^{2n+1}z_i$. We thus have
\begin{equation}
  \label{eqn:SpecialEntryPsiUp}
  (\Psi_{2n+1}^{+})_{\underset{\scriptstyle n+1}{\underbrace{\scriptstyle\uparrow\cdots\uparrow}}\underset{\scriptstyle n}{\underbrace{\scriptstyle\downarrow\cdots\downarrow}}} = \prod_{1\leqslant i<j\leqslant n+1}\left(\frac{qz_i-q^{-1}z_j}{q-q^{-1}}\right)\prod_{n+2\leqslant i<j\leqslant 2n+2}\left(\frac{qz_i-q^{-1}z_j}{q-q^{-1}}\right) \prod_{i=n+2}^{2n+1}z_i.
\end{equation}
The vector $|\Psi_{2n+1}^{+}\rangle$ obeys the exchange equations and satisfies the cyclic-covariance property \eqref{eqn:Cyclicity}. This follows from the work of Razumov, Stroganov, and Zinn-Justin \cite{RAZ07}. Their result is summarised in the following proposition.

\begin{Proposition}[\!\!\cite{RAZ07}] For each $n\geqslant 0$, the vector $|\Psi_{2n+1}^{+}\rangle$ is a solution to the quantum Knizhnik-Zamolodchikov system for $N=2n+1$ and $D=D^{+}$, where
\begin{equation}
  D^{+}= q^{-3(n+1)}q^{3(\sigma^z+1)/2}.
\end{equation}
Its components are homogeneous polynomials of total degree $n(n+1)$ and individual degrees at most $n$.
\end{Proposition}

We use $|\Psi_{2n+1}^{+}\rangle$ to construct two other solutions to the quantum Knizhnik-Zamolochikov system. We obtain a solution $|\Psi_{2n+1}^{-}\rangle$ of magnetisation $\mu=-1$ through the action of a spin-reversal operator. Furthermore, we find a solution $|\Psi_{2n}^{0}\rangle$ of zero magnetisation through the specialisation of a spectral parameter.
\begin{Proposition}
\label[Proposition]{prop:PsiDown}
For each $n\geqslant 0$, the vector
\begin{equation}
\label{eqn:DefPsiDown}
  |\Psi_{2n+1}^{-}\rangle = \bigotimes_{i=1}^{2n+1}
  \begin{pmatrix}
    0 & z_i^{-1}\\
    1 & 0
  \end{pmatrix}
  |\Psi_{2n+1}^{+}\rangle
\end{equation}
is a solution to the quantum Knizhnik-Zamolodchikov system with $N=2n+1$ and $D=D^{-}$, with
\begin{equation}
  D^{-} = q^{-3n} q^{3(\sigma^z+1)/2}.
\end{equation}
Its components are homogeneous polynomials in $z_1,\dots,z_{2n+1}$ of total degree $n^2$ and individual degrees at most $n$.
\end{Proposition}
 \proof
 First, to check the exchange equations, one uses the identity 
 \begin{equation}
   \check R(z/w)
   \left(
   \begin{pmatrix}
     0 & w^{-1}\\
     1 & 0
   \end{pmatrix}
   \otimes 
   \begin{pmatrix}
     0 & z^{-1}\\
     1 & 0
   \end{pmatrix}
   \right)
    = \left(\begin{pmatrix}
     0 & z^{-1}\\
     1 & 0
   \end{pmatrix}
   \otimes 
   \begin{pmatrix}
     0 & w^{-1}\\
     1 & 0
   \end{pmatrix}
   \right)
   \check R(z/w),
 \end{equation}
 and the exchange equations for $|\Psi_{2n+1}^{+}\rangle$. 
 
Second, we compute the component
\begin{align}
  (\Psi_{2n+1}^{-})_{\underset{\scriptstyle n}{\underbrace{\scriptstyle\uparrow\cdots\uparrow}}\underset{\scriptstyle n+1}{\underbrace{\scriptstyle\downarrow\cdots\downarrow}}} &= \Big(\prod_{i=1}^{n}z_i^{-1}\Big)(\Psi_{2n+1}^{+})_{\underset{\scriptstyle n}{\underbrace{\scriptstyle\downarrow\cdots\downarrow}}\underset{\scriptstyle n+1}{\underbrace{\scriptstyle\uparrow\cdots\uparrow}}}\nonumber\\
  &= q^{-3n(n+1)}\Big(\prod_{i=1}^{n}z_i^{-1}\Big)(\Psi_{2n+1}^{+})_{\underset{\scriptstyle n+1}{\underbrace{\scriptstyle\uparrow\cdots\uparrow}}\underset{\scriptstyle n}{\underbrace{\scriptstyle\downarrow\cdots\downarrow}}}(q^{-6}z_{n+1},\dots,q^{-6}z_{2n+1},z_1,\dots,z_n).
\end{align}
From the first to the second line, we used \eqref{eqn:CyclicCovComps} several times. The right-hand side of this equality can be evaluated with the help of \eqref{eqn:SpecialEntryPsiUp}. We obtain  \begin{equation}
  \label{eqn:SpecialEntryPsiDown}
  (\Psi_{2n+1}^{-})_{\underset{\scriptstyle n}{\underbrace{\scriptstyle\uparrow\cdots\uparrow}}\underset{\scriptstyle n+1}{\underbrace{\scriptstyle\downarrow\cdots\downarrow}}} = \prod_{1\leqslant i<j\leqslant n}\frac{qz_i-q^{-1}z_j}{q-q^{-1}}\prod_{n+1\leqslant i<j\leqslant 2n+1}\frac{qz_i-q^{-1}z_j}{q-q^{-1}}.
  \end{equation}
  It is a homogeneous polynomial in $z_1,\dots,z_{2n+1}$ with total degree $n^2$ and individual degrees at most $n$. By \cref{lem:DDOpsPoly}, the same holds for all other components. 
  
  Finally, we check the cyclic-covariance property \eqref{eqn:Cyclicity}. To this end, we write  \begin{align}
    \nonumber \mathcal \tau |\Psi_{2n+1}^{-}(z_1,\dots,z_{2n+1})\rangle &= \bigotimes_{i=2}^{2n+1}
    \begin{pmatrix}
      0 & z_i^{-1}\\
      1 & 0
    \end{pmatrix}
    \otimes 
    \begin{pmatrix}
      0 & z_1^{-1}\\
      1 & 0
    \end{pmatrix}
    D_{2n+1}^{+} |\Psi_{2n+1}^{+}(z_2,\dots,z_{2n+1},q^{6}z_1)\rangle\\
    &
    \nonumber
    = D^{-}_{2n+1}\bigotimes_{i=2}^{2n+1}
    \begin{pmatrix}
      0 & z_i^{-1}\\
      1 & 0
    \end{pmatrix}
    \otimes 
    \begin{pmatrix}
      0 & (q^6z_1)^{-1}\\
      1 & 0
    \end{pmatrix}
    \mathcal |\Psi_{2n+1}^{+}(z_2,\dots,z_{2n+1},q^{6}z_{1})\rangle
    \\
    &= D^{-}_{2n+1}|\Psi_{2n+1}^{-}(z_2,\cdots,z_{2n+1},q^6z_1)\rangle.
  \end{align}
  On the right-hand side of the first line, we used the cyclic-covariance property of $|\Psi_{2n+1}^{+}\rangle$. The second line follows from the first one by commuting the operator $D^{+}_{2n+1}$ to the left, and taking into account additional factors that appear in this operation. They lead to the appearance of $D^{-}_{2n+1}$. From the second to the third line, we used the definition \eqref{eqn:DefPsiDown}.
    \eproof

\begin{Proposition}
  \label[Proposition]{prop:DefPsiE}
  For each $n\geqslant 1$, there is a vector $|\Psi_{2n}^{0}\rangle \in V^{2n}$ of zero magnetisation such that
  \begin{equation}
    \label{eqn:DefPsiE}
      \frac{(q^{-1}-q)^{n}}{\prod_{i=1}^{2n}qz_i}|\Psi_{2n+1}^{+}(z_1,\dots,z_{2n},0)\rangle = |\Psi_{2n}^{0}(z_1,\dots,z_{2n})\rangle \otimes |{\uparrow}\rangle.
     \end{equation}
  This vector is a polynomial solution to the quantum Knizhnik-Zamolochikov system with $N=2n$ and $D=D^{0}$, where
  \begin{equation}
  \label{eq:D0}
    D^{0}=-q^{-3(n-1)}q^{\sigma^z}.
  \end{equation}
  Its components are homogeneous polynomials in $z_1,\dots,z_{2n}$ with total degree $n(n-1)$ and individual degrees at most $n-1$.
\end{Proposition}
 \proof  
    First, we establish the existence of $|\Psi_{2n}^{0}\rangle$. To this end, we note that 
    \begin{equation}
     (\Psi_{2n+1}^{+})_{\underset{\scriptstyle n+1}{\underbrace{\scriptstyle\uparrow\cdots\uparrow}}\underset{\scriptstyle n}{\underbrace{\scriptstyle\downarrow\cdots\downarrow}}}(z_1,\dots,z_{2n},z_{2n+1}=0) = 0.
    \end{equation}
    It follows from the exchange equations that $(\Psi_{2n+1}^{+})_{\alpha_1\cdots\alpha_{2n}\downarrow}(z_1,\dots,z_{2n},z_{2n+1}=0) =0$ for each choice of $\alpha_1,\dots,\alpha_{2n}$. Hence, there is a vector $|\Psi_{2n}^{0}\rangle \in V^{2n}$ such that \eqref{eqn:DefPsiE} holds. Its components are non-vanishing only in the sector of zero magnetisation. Furthermore, the exchange equations for the vector $|\Psi_{2n+1}^{+}\rangle$ straightforwardly imply that $|\Psi_{2n}^{0}\rangle$ obeys the same exchange equations, for $i = 1, \dots, 2n-1$.     
      
      Second, the vector $|\Psi_{2n}^{0}\rangle$ has the component
    \begin{align}
     (\Psi_{2n}^{0})_{\underset{\scriptstyle n}{\underbrace{\scriptstyle\uparrow\cdots\uparrow}}\underset{\scriptstyle n}{\underbrace{\scriptstyle\downarrow\cdots\downarrow}}} &=
      \frac{(q^{-1}-q)^{n}}{\prod_{i=1}^{2n}qz_i}(\Psi_{2n+1}^+)_{\underset{\scriptstyle n}{\underbrace{\scriptstyle\uparrow\cdots\uparrow}}\underset{\scriptstyle n}{\underbrace{\scriptstyle\downarrow\cdots\downarrow}}\uparrow}(z_1,\dots,z_{2n},z_{2n+1}=0).
    \end{align}
     By virtue of \eqref{eqn:CyclicCovComps}, we have
     \begin{equation}
      (\Psi_{2n+1}^+)_{\underset{\scriptstyle n}{\underbrace{\scriptstyle\uparrow\cdots\uparrow}}\underset{\scriptstyle n}{\underbrace{\scriptstyle\downarrow\cdots\downarrow}}\uparrow}(z_1,\dots,z_{2n},z_{2n+1}=0)  = q^{3n}(\Psi_{2n+1}^{+})_{\underset{\scriptstyle n+1}{\underbrace{\scriptstyle\uparrow\cdots\uparrow}}\underset{\scriptstyle n}{\underbrace{\scriptstyle\downarrow\cdots\downarrow}}}(0,z_1,\dots,z_{2n}).
     \end{equation}
    We use \eqref{eqn:SpecialEntryPsiUp} to evaluate the right-hand side of this equality, and we find
    \begin{equation}
    \label{eqn:SpecialEntryPsiE}
     (\Psi_{2n}^{0})_{\underset{\scriptstyle n}{\underbrace{\scriptstyle\uparrow\cdots\uparrow}}\underset{\scriptstyle n}{\underbrace{\scriptstyle\downarrow\cdots\downarrow}}} = \prod_{1\leqslant i<j\leqslant n}\frac{qz_i-q^{-1}z_j}{q-q^{-1}}\prod_{n+1\leqslant i<j\leqslant 2n}\frac{qz_i-q^{-1}z_j}{q-q^{-1}}.
    \end{equation}
    This component is a homogeneous polynomial in $z_1,\dots, z_{2n}$ with total degree $n(n-1)$ and individual degrees at most $n-1$. By virtue of \cref{lem:DDOpsPoly}, the same holds for all other components. 
    
    Finally, we establish the cyclic-covariance property of $|\Psi_{2n}^{0}\rangle$. To this end, we use its exchange equations and the cyclic-covariance property of $|\Psi_{2n+1}^{+}\rangle$. These allow us to write
    \begin{equation}
      \tau |\Psi_{2n+1}^{+}(z_1,\dots,z_{2n+1})\rangle = D_{2n+1}^{+}\check R_{2n,2n+1}(q^{-6}z_{2n+1}/z_1)|\Psi_{2n+1}^{+}(z_2,\dots,z_{2n},q^6z_1,z_{2n+1})\rangle.
    \end{equation}
    We now set $z_{2n+1}=0$ and use \eqref{eqn:DefPsiE}. This leads to
    \begin{equation}
      \tau \left(|\Psi_{2n}^{0}(z_1,\dots,z_{2n})\rangle\otimes |{\uparrow}\rangle\right) = q^{6}D_{2n+1}^{+}\check R_{2n,2n+1}(0)\left(|\Psi_{2n}^{0}(z_2,\dots,z_{2n},q^6z_1)\rangle\otimes |{\uparrow}\rangle\right).
    \end{equation}
    We note that $\tau \left(|\Psi_{2n}^{0}(z_1,\dots,z_{2n})\rangle\otimes |{\uparrow}\rangle\right) = P_{2n,2n+1}\left(\left(\tau|\Psi_{2n}^{0}(z_1,\dots,z_{2n})\rangle\right)\otimes|{\uparrow}\rangle\right)$, where $P_{2n,2n+1}$ is the permutation operator $P$ defined in \eqref{eqn:PermutationOp}, acting on the sites $2n$ and $2n+1$. Hence, it follows that
     \begin{equation}
      \left(\tau|\Psi_{2n}^{0}(z_1,\dots,z_{2n})\rangle\right)\otimes |{\uparrow}\rangle = q^{6}D_{2n}^{+} R_{2n,2n+1}(0)\left(|\Psi_{2n}^{0}(z_2,\dots,z_{2n},q^6z_1)\rangle\otimes |{\uparrow}\rangle\right).
    \end{equation}
    The $R$-matrix has the properties
    \begin{equation}
      R(0)|{\uparrow\uparrow}\rangle = -q^{-2} |{\uparrow\uparrow}\rangle, \qquad R(0)|{\downarrow\uparrow}\rangle = -q^{-1} |{\downarrow\uparrow}\rangle.
    \end{equation}
These relations yield
    \begin{equation}
    \left(\tau|\Psi_{2n}^{0}(z_1,\dots,z_{2n})\rangle\right)\otimes |{\uparrow}\rangle = -q^{9/2}D_{2n}^{+} q^{-\sigma^z_{2n}/2}|\Psi_{2n}^{0}(z_2,\dots,z_{2n},q^6z_1)\rangle\otimes |{\uparrow}\rangle.
    \end{equation}
    We find that $-q^{9/2}D^{+} q^{-\sigma^z/2}=D^{0}$, proving \eqref{eq:D0}.
\eproof

\subsection{Properties of the solutions}

\paragraph{Reduction relations.} The polynomial solutions to the quantum Knizhnik-Zamolochikov system that we discussed in the previous section obey a set of reduction relations. Indeed, for certain specialisations of the parameters $z_1,\dots,z_N$, one may express $|\Psi_N^\mu\rangle$ in terms of $|\Psi_{N-2}^\mu\rangle$. To formulate the reduction relations, we define for each $N\geqslant 1$ and $i=0,\dots,N$ a linear operator 
\begin{equation}
  \Xi_N^{i}:V^N\to V^{N+2},
\end{equation}
through the following action on a basis vector $|\alpha_1\cdots \alpha_N\rangle$:
\begin{equation}
  \Xi_N^{i}|\alpha_1\cdots \alpha_N\rangle =|\alpha_1\cdots \alpha_{i-1}\rangle\otimes |\omega\rangle \otimes |\alpha_{i}\cdots \alpha_N\rangle, \qquad |\omega\rangle =|{\uparrow\downarrow}\rangle - q^{-1}|{\downarrow\uparrow}\rangle.
\end{equation}

\begin{Proposition}[Reduction relations \cite{C12}]
\label[Proposition]{prop:RedPsi}
Let $N=2n$ with $n\ge 1$ and $\mu = 0$ for $N$ even, and $N=2n+1$ with $n\ge 0$ and $\mu =+,-$ for $N$ odd. In each of these cases, we have  
\begin{align}
  \begin{split}
 &  |\Psi^\mu_{N}(\dots,z_i,z_{i+1}=q^2 z_i,\dots)\rangle\\ & \qquad = (-q)^{i-n}(-q z_i)^{\delta(\mu)}\prod_{j=1}^{i-1} \frac{q z_j-q^{-1}z_i}{q-q^{-1}} \prod_{j=i+2}^{N} \frac{q^3 z_i-q^{-1}z_j}{q-q^{-1}} \,\Xi_{N-2}^i |\Psi_{N-2}^\mu(\dots)\rangle,
  \end{split}
\end{align}
  where $\delta(-)=\delta(0)=0$ and $\delta(+)=1$.
\end{Proposition}
\begin{Corollaire}[Wheel condition]
\label[Corollary]{corr:WheelCondition}
Let $N$, $n$ and $\mu$ satisfy the same conditions as in the previous proposition. Then for all triples $(i,j,k)$ satisfying $1\leqslant i < j < k \leqslant N$, we have  
  \begin{equation}
    |\Psi_N^\mu(\dots,z_i,\dots ,z_j=q^2 z_i, \dots, z_{k}=q^4 z_i, \dots)\rangle = 0.
  \end{equation}
\end{Corollaire}

\paragraph{Relations between the vectors.} We have defined both $|\Psi_{2n+1}^{-}\rangle$ and $|\Psi_{2n}^{0}\rangle$ from $|\Psi_{2n+1}^{+}\rangle$. We now show that upon specialisation of a spectral parameter to $0$ or $\infty$, the vectors $|\Psi_{2n-1}^{+}\rangle$ and $|\Psi_{2n-1}^{-}\rangle$ can be obtained from $|\Psi_{2n}^{0}\rangle$.

\begin{Proposition}
\label[Proposition]{prop:BraidRelations}
  For each $n\geqslant 1$ and each $j=1,\dots,2n$, we have
  \begin{subequations}
    \begin{align}
     \lim_{z_j\to 0}\frac{1}{2}(1-\sigma^z_{j})|\Psi_{2n}^{0}\rangle& =
   \left(\prod_{i=1}^{j-1}q^{-\sigma_i^z/2}\right)\frac{\Theta^j_{2n-1}|\Psi_{2n-1}^{+}(\dots,z_{j-1},z_{j+1},\dots)\rangle}{(-1)^jq^{\frac12(6n-3j-1)}(1-q^{-2})^{n-1}},
   \label{eqn:FirstBraidLimit}
   \\[0.15cm]
    \lim_{z_{j}\to \infty}z_{j}^{-(n-1)} \frac12 (1+\sigma^z_{j})|\Psi_{2n}^{0}\rangle &= 
    \left(\prod_{i=1}^{j-1}q^{-\sigma_i^z/2}\right)\frac{\bar\Theta_{2n-1}^j|\Psi_{2n-1}^{-}(\dots,z_{j-1},z_{j+1},\dots)\rangle}{(-1)^{j-1}q^{\frac32(j+1)}(1-q^{-2})^{n-1}},
    \label{eqn:SecondBraidLimit}
    \end{align}
   \end{subequations} 
      where $\Theta_{2n-1}^j: V^{2n-1}\to V^{2n}$ and $\bar \Theta_{2n-1}^j: V^{2n-1}\to V^{2n}$ are linear operators defined by the following actions on a basis vector $|\alpha_1\cdots\alpha_{2n-1}\rangle$:
  \begin{subequations}
\begin{align}
  \Theta_{2n-1}^{j}|\alpha_1\cdots \alpha_{2n-1}\rangle &=|\alpha_1\cdots \alpha_{j-1}\rangle\otimes |{\downarrow}\rangle \otimes |\alpha_{j}\cdots \alpha_{2n-1}\rangle,\\[0.15cm]
  \bar\Theta_{2n-1}^{j}|\alpha_1\cdots \alpha_{2n-1}\rangle &=|\alpha_1\cdots \alpha_{j-1}\rangle\otimes |{\uparrow}\rangle \otimes |\alpha_{j}\cdots \alpha_{2n-1}\rangle.
\end{align}
\end{subequations}
\end{Proposition}
    \proof
      The proofs of \eqref{eqn:FirstBraidLimit} and \eqref{eqn:SecondBraidLimit} are similar. We only present the proof of \eqref{eqn:FirstBraidLimit}, which is in two steps. First, we prove the equality for $j=2n$. Because of the projector on the left-hand side of \eqref{eqn:FirstBraidLimit}, there is a vector $|\Phi\rangle = |\Phi(z_1,\dots,z_{2n-1})\rangle\in V^{2n-1}$ such that 
      \begin{equation}
      \frac12(1-\sigma^z_{2n})|\Psi_{2n}^{0}(z_1,\dots,z_{2n-1},0)\rangle =
   \Theta_{2n-1}^{2n}|\Phi(z_1,\dots,z_{2n-1})\rangle.
      \end{equation}
      Since $|\Psi_{2n}^{0}\rangle$ is a solution to the quantum Knizhnik-Zamolochikov system, the vector $|\Phi\rangle$ obeys the exchange equations. Furthermore, using \eqref{eqn:SpecialEntryPsiE}, we find the component
      \begin{equation}
        (\Phi)_{\underset{\scriptstyle n}{\underbrace{\scriptstyle\uparrow\cdots\uparrow}}\underset{\scriptstyle n-1}{\underbrace{\scriptstyle\downarrow\cdots\downarrow}}} = (1-q^{-2})^{-(n-1)}(\Psi_{2n-1}^{+})_{\underset{\scriptstyle n}{\underbrace{\scriptstyle\uparrow\cdots\uparrow}}\underset{\scriptstyle n-1}{\underbrace{\scriptstyle\downarrow\cdots\downarrow}}}.
      \end{equation}
      Since both vectors, $|\Phi\rangle$ and $|\Psi_{2n-1}^{+}\rangle$, obey the exchange equations, we may use \cref{lem:qKZEqual} to conclude that
      \begin{equation}
        |\Phi(z_1,\dots,z_{2n-1})\rangle = (1-q^{-2})^{-(n-1)}|\Psi_{2n-1}^{+}(z_1,\dots,z_{2n-1})\rangle.
      \end{equation}
      
      Second, the cases where $j=1,\dots,2n-1$ follow from the cyclic covariance property. We have
      \begin{align}
        & \lim_{z_j\to 0}\frac{1}{2}(1-\sigma^z_{j})|\Psi_{2n}^{0}\rangle = \tau^{2n-j}\lim_{z_j\to 0}\frac{1}{2}(1-\sigma^z_{2n})\tau^{-(2n-j)}|\Psi_{2n}^{0}\rangle\nonumber\\
        & = \prod_{k=j+1}^{2n}(D_k^{0})^{-1}\tau^{2n-j}\frac{1}{2}(1-\sigma^z_{2n})|\Psi_{2n}^{0}(q^{-6}z_{j+1},\dots,q^{-6}z_{2n},z_1,\dots,z_{j-1},0)\rangle\nonumber\\
        & =  \prod_{k=j}^{2n}(D_k^{0})^{-1}\frac{\tau^{2n-j}\Theta_{2n-1}^{2n}|\Psi_{2n-1}^+(q^{-6}z_{j+1},\dots,q^{-6}z_{2n},z_1,\dots,z_{j-1})\rangle}{(1-q^{-2})^{n-1}}.
       \end{align}
       We now use $\tau\, \Theta_{2n-1}^i = \Theta_{2n-1}^{i-1}\tau$, for $i=2,\dots,2n-1$, as well as the cyclic covariance property of $|\Psi_{2n-1}^{+}\rangle$. We obtain
       \begin{align}
        &  \lim_{z_j\to 0}\frac{1}{2}(1-\sigma^z_{j})|\Psi_{2n}^{0}\rangle =  \prod_{k=j}^{2n}(D_k^{0})^{-1}\frac{\Theta_{2n}^{j-1}\prod_{k=j}^{2n-1}D_k^{+}|\Psi_{2n-1}^+(\dots,z_{j-1},z_{j+1},\dots)\rangle}{(1-q^{-2})^{n-1}}.
      \end{align}
      Finally, we insert the expressions for $D^{0}$ and $D^{+}$ and use the fact $|\Psi^{+}_{2n-1}\rangle$ has the magnetisation $+1$. This leads to the relation \eqref{eqn:FirstBraidLimit}.
    \eproof
    
\paragraph{Examples.} It is instructive to inspect the components of the polynomial solutions to the quantum Knizhnik-Zamolodchikov system for small system sizes. The non-vanishing components of $|\Psi_3^{+}\rangle$ are
 \begin{equation}
  \label{eqn:EntriesPsiUp3}
  (\Psi_3^{+})_{\uparrow\uparrow\downarrow} = \frac{(qz_1-q^{-1}z_2)z_3}{q-q^{-1}}, \qquad (\Psi_3^{+})_{\uparrow\downarrow\uparrow} = \frac{(q^{-2}z_3-q^{2}z_1)z_2}{q-q^{-1}}, \qquad
  (\Psi_3^{+})_{\downarrow\uparrow\uparrow} = \frac{(qz_2-q^{-1}z_3)z_1}{q-q^{-1}}.
\end{equation}
The definition of $|\Psi_{2n+1}^{-}\rangle$ from $|\Psi_{2n+1}^{+}\rangle$ in \cref{prop:PsiDown} leads to the following non-vanishing components of $|\Psi_3^{-}\rangle$:
 \begin{equation}
 \label{eqn:EntriesPsiDown3}
  (\Psi_3^{-})_{\uparrow\uparrow\downarrow} = \frac{qz_1-q^{-1}z_2}{q-q^{-1}}, \qquad (\Psi_3^{-})_{\uparrow\downarrow\uparrow} = \frac{q^{-2}z_3-q^{2}z_1}{q-q^{-1}}, \qquad
  (\Psi_3^{-})_{\downarrow\uparrow\uparrow} = \frac{qz_2-q^{-1}z_3}{q-q^{-1}}.
\end{equation}
Using \eqref{eqn:DefPsiE}, we obtain $|\Psi_2^{0}\rangle = |\omega\rangle$ from the components \eqref{eqn:EntriesPsiUp3}. Furthermore, the non-vanishing components of $|\Psi_4^{0}\rangle$ are given by
\begin{subequations}
\begin{align}
  (\Psi_{4}^{0})_{\uparrow\uparrow\downarrow\downarrow} &= \frac{(qz_1-q^{-1}z_2)(qz_3-q^{-1}z_4)}{(q-q^{-1})^2},&  &(\Psi_{4}^{0})_{\uparrow\downarrow\downarrow\uparrow} = \frac{(qz_2-q^{-1}z_3)(q^2z_1-q^{-2}z_4)}{q(q-q^{-1})^2},\\
  (\Psi_{4}^{0})_{\downarrow\uparrow\uparrow\downarrow} &= \frac{(qz_2-q^{-1}z_3)(q^2z_1-q^{-2}z_4)}{q(q-q^{-1})^2},&  &(\Psi_{4}^{0})_{\downarrow\downarrow\uparrow\uparrow} = \frac{(qz_1-q^{-1}z_2)(q^2z_3-q^{-2}z_4)}{q^2(q-q^{-1})^2},
\end{align}
and
\begin{align}
(\Psi_4^{0})_{\uparrow\downarrow\uparrow\downarrow}&=\frac{\left(q-q^{-1}\right) z_2 z_4+z_3 \left(q z_2-q^{-1}z_4\right)-q^3 z_1 \left(q^2 z_2-q^{-2}z_4\right)}{q^2 \left(q-q^{-1}\right)^2},\\
(\Psi_4^{0})_{\downarrow\uparrow\downarrow\uparrow}&=\frac{z_3 \left(q^2 z_2-q^{-2}z_4\right)-q^3 z_1 \left(q z_2-q^{-1} z_4\right)-\left(q-q^{-1}\right) q^3 z_1 z_3}{q^3 \left(q-q^{-1}\right)^2}.
\end{align}
\end{subequations}
One can check that these components are consistent with the reduction relation of \cref{prop:RedPsi} and likewise with the relations of \cref{prop:BraidRelations}.

\subsection{Eigenvectors of the inhomogeneous transfer matrix}
\label{sec:qKZandTM}

The transfer matrix of the inhomogeneous twisted six-vertex model is given by
\begin{equation}
   T(z|z_1,\dots,z_N) = \text{tr}_0 \left(\eE^{\ir \phi \sigma^z_0} R_{0N}(z_N/z) \cdots R_{01}(z_1/z)\right).
\end{equation}
Here, $z_i$ is an inhomogeneity parameter associated to the $i$-th vertical line, with $i=1,\dots,N$. If all inhomogeneity parameters are equal, then we recover the transfer matrix of the homogeneous six-vertex model defined in \eqref{eqn:TM6V}: $T(z|w,\dots,w)=T(w/z)$.

\begin{Theoreme}[Transfer-matrix eigenvalue \cite{RAZ07,FON14}]
For $N$ odd, let $N=2n+1$ with $n\geqslant 0$, $\phi=0$ and $\mu=+,-$. For $N$ even, let $N=2n$ with $n\geqslant 1$, $\phi=-\frac{\pi}3$ and $\mu=0$. For $q=\eE^{2\pi\ir/3}$, we have
    \begin{equation}
     T(z|z_1,\dots,z_N)|\Psi_N^\mu\rangle = |\Psi_N^\mu\rangle.
  \end{equation}
\end{Theoreme}

For $q=\eE^{2\pi\ir/3}$, the vectors 
\begin{equation}
\label{eq:homo.psi}
  |\psi^\mu_N\rangle = |\Psi_N^\mu(z_1=1,\dots,z_N=1)\rangle
\end{equation}
span the one-dimensional eigenspaces of the transfer matrix of the homogeneous twisted six-vertex model with the twist angle $\phi$ and magnetisation $\mu$ given in this theorem. They are also the groundstates of the XXZ spin chain at the combinatorial point $\Delta=-\frac12$. Furthermore, it follows from \eqref{eqn:SpecialEntryPsiUp}, \eqref{eqn:SpecialEntryPsiDown} and \eqref{eqn:SpecialEntryPsiE} that these vectors have the components \eqref{eqn:SpecialComponents}.

\section{Scalar products}
\label{sec:DefSP}

In this section, we define and analyse scalar products that involve the polynomial solutions to the quantum Knizhnik-Zamolochikov system. The definition makes use of a state that solves the boundary Yang-Baxter equation. We recall this equation in \cref{sec:bYBE} and focus on a particular solution. In \cref{sec:SP}, we establish a series of properties of the scalar products.

\subsection{Definition}
\label{sec:bYBE}
\paragraph{The boundary Yang-Baxter equation.} Let $\langle\chit(x)|\in (\mathbb C^2)^\ast\otimes  (\mathbb C^2)^\ast$ be a co-vector depending on a parameter $x$. We call it a solution to the boundary Yang-Baxter equation \cite{SKLY88} for the six-vertex model if
\begin{equation}
  \label{eqn:bYBE}
  \left(\langle \chit(x)|\otimes \langle \chit(y)|\right)\check R_{23}(x^{-1}y^{-1})\check R_{12}(xy^{-1})=\left(\langle \chit(y)|\otimes \langle \chit(x)|\right)\check R_{23}(x^{-1}y^{-1})\check R_{34}(xy^{-1}).
\end{equation}
The most general solution to this equation can be found in \cite{VEG93}. In this article, we focus on the special case 
\begin{equation}
   \langle \chit(x)|= \frac{q^{-1}x-q x^{-1}}{q^{-1}-q}\left(\langle{\uparrow\uparrow}|+\langle{\downarrow\downarrow}|\right)+ \frac{x-q}{1-q}\left(\langle{\uparrow\downarrow}|+x^{-1}\langle{\downarrow\uparrow}|\right).
\end{equation}

This solution is factorising in the sense that $\langle \chit(x=1)| = (\langle{\uparrow}|+\langle{\downarrow}|)\otimes (\langle{\uparrow}|+\langle{\downarrow}|)$. Along with $\langle \chit(x)|$, we consider the co-vector
\begin{equation}
  \label{eqn:DefPhi}
  \langle \varphi(x)| = \frac{1-q x}{1-q}\langle{\uparrow}| + \frac{q x-q^{-1}x^{-1}}{q-q^{-1}}\langle{\downarrow}|.
\end{equation}
It has the property $\langle \varphi(x=1)|=\langle{\uparrow}|+\langle{\downarrow}|$. Furthermore, it can be obtained from $\langle \chit(x)|$ through the relation
\begin{equation}
  \langle \chit(x)|\left(\frac{1-\sigma^z_1}{2}\right)=\langle{\downarrow}|\otimes \langle\varphi(x^{-1})|.
\end{equation}
The co-vector $\langle \chit(x)|$ also obeys the fish equation 
\begin{equation}
  \label{eq:Fish} 
  \langle \chit(x)|\check R(x^2) = \langle \chit(x^{-1})|.
\end{equation}

\paragraph{Definition of the scalar products.} We introduce for each $m\geqslant 0$ and $p\geqslant 0$ the vectors
\begin{subequations}
\begin{alignat}{2}
  \langle \gamma_{m,p}| &= \langle\gamma_{m,p}(x_1,\dots,x_p)| = \langle \underset{m}{\underbrace{\uparrow \cdots \uparrow}}|\otimes \bigotimes_{i=1}^p \langle\chit(x_i)|,
\\
    \langle \bar \gamma_{m,p}| &= \langle\bar\gamma_{m,p}(x_1,\dots,x_p)| = \langle \underset{m}{\underbrace{\uparrow \cdots \uparrow}}|\otimes \langle \varphi(x)|\otimes \bigotimes_{i=1}^p \langle\chit(x_i)|.
\end{alignat}
\end{subequations}

We recall that we use the parameterisation $N= 2n+1$ for odd $N$ and $N=2n$ for even $N$, where $n$ is a non-negative integer. In the following, $p$ and $k$ are two non-negative integers such that
\begin{equation}
  n=p+k.
\end{equation}
We define the scalar products
\begin{subequations}
\label{eqn:ScalarProducts}
\begin{align}
  S^{0}_{k,p} &= \langle \gamma_{2k,p}|\Psi_{2n}^{0}(z_1,\dots,z_{2k},x_1,x_1^{-1},\dots,x_p,x_p^{-1})\rangle,\label{eq:Skp0}\\[0.15cm]
   \bar S^{0}_{k,p} &= \langle \bar \gamma_{2k-1,p}|\Psi_{2n}^{0}(z_1,\dots,z_{2k-1},x,x_1,x_1^{-1},\dots,x_p,x_p^{-1})\rangle,
\end{align}
and, for $\mu=+,-$,
\begin{align}
S^\mathrm{\mu}_{k,p} &= \langle \gamma_{2k+1,p}|\Psi_{2n}^\mu(z_1,\dots,z_{2k+1},x_1,x_1^{-1},\dots,x_p,x_p^{-1})\rangle,\\[0.15cm]
   \bar S^\mathrm{\mu}_{k,p} &= \langle \bar \gamma_{2k,p}|\Psi_{2n+1}^\mathrm{\mu}(z_1,\dots,z_{2k},x,x_1,x_1^{-1},\dots,x_p,x_p^{-1})\rangle.
\end{align}
\end{subequations}
When necessary, we explicitly write out their dependence on the parameters, for example as in $S^{0}_{k,p} = S^{0}_{k,p}(z_1,\dots,z_{2k};x_1,\dots,x_p)$ or $\bar S^{0}_{k,p} = \bar S^{0}_{k,p}(z_1,\dots,z_{2k-1};x;x_1,\dots,x_p)$.

\paragraph{Homogeneous limit and relation to the overlaps.} We call {\it homogeneous limit} of the scalar products the limit where all their spectral parameters $z_1,\dots,z_{2k},z_{2k+1}$, $x$ and $x_1,\dots,x_p$ are sent to one. For $q=\eE^{2\ir \pi/3}$, it follows from \eqref{eq:homo.psi} that the homogeneous limit of the scalar products \eqref{eqn:ScalarProducts} produces the overlaps $C_{N,m}$ for the XXZ chain at $\Delta=-\frac12$ defined in \eqref{def-c.mu}. Indeed, we have
\begin{align}
  \label{eqn:HomLimitSe}
  C_{2(p+k),2k}^{0} = \lim_{\substack{z_1,\dots,z_{2k} \to  1\\ x_1,\dots,x_p \to 1}} S^{0}_{k,p},\qquad C_{2(p+k),2k-1}^{0} = \lim_{\substack{z_1,\dots,z_{2k-1} \to  1\\ x,x_1,\dots,x_p \to 1}} \bar S^{0}_{k,p},
\end{align}
and, for $\mu = +,-$,
\begin{align}
  C_{2(p+k)+1,2k+1}^\mu = \lim_{\substack{z_1,\dots,z_{2k+1} \to  1\\ x_1,\dots,x_p \to 1}} S^\mu_{k,p},\qquad C_{2(p+k)+1,2k}^\mu = \lim_{\substack{z_1,\dots,z_{2k} \to  1\\ x,x_1,\dots,x_p \to 1}} \bar S^\mu_{k,p}.
\end{align}
The boundary emptiness formation probabilities are then defined as ratios of these overlaps, as in \eqref{eqn:BEFP}.

\paragraph{Strategy.} We now describe the strategy that allows us to compute the overlaps for the groundstate eigenvectors of the XXZ spin chain at $\Delta=-\frac12$. We proceed in three steps. The first step is to characterise the scalar product $S^{0}_{k,p}$. This is the topic of \cref{sec:SP}. We determine its symmetry and polynomial properties with respect to its parameters $z_1,\dots,z_{2k}$ and $x_1,\dots,x_p$, and establish its reduction relations. We show that these properties uniquely determine the scalar product. Furthermore, we show that the other five scalar products can be obtained from $S_{k,p}^{0}$. The second step is detailed in \cref{sec:Se}. It consists of finding a candidate expression for $S^{0}_{k,p}$ in terms of a determinant and showing that it possesses the same properties as $S_{k,p}^{0}$. By uniqueness, the two quantities are equal. The third and last step is to compute the homogeneous limit. We develop a strategy to compute it in \cref{sec:homogeneous} and apply it to $S_{k,p}^{0}$. As we shall see in \cref{sec:SP}, the homogeneous limit for the five remaining scalar products in \eqref{eqn:ScalarProducts} can be obtained from certain specialisations of $S_{k,p}^{0}$. We relegate their computation to \cref{app:five.limits}.

\subsection{Properties of the scalar products}\label{sec:SP}
In this section, we establish several properties of the scalar products. We derive the polynomial and symmetry properties of  $S_{k,p}^{0}$, as well as several reduction relations that it satisfies. Furthermore, we show that all other scalar products are obtained from $S^{0}_{k,p}$ through a suitable specialisation of its parameters.

To prepare the following discussions, we note that since $|\Psi^{0}_{2n}\rangle$ has zero magnetisation, we have
\begin{equation}
S_{k,p}^{0} = 0 \quad \text{if}\quad k > p.
\end{equation}
Henceforth, we focus on the cases where $k \leqslant p$.

\paragraph{Symmetry and polynomial properties.} In the next two propositions, we characterise $S_{k,p}^{0}$ as a function of $z_1,\dots,z_{2k}$ and $x_1,\dots,x_p$, respectively.  To this end, we recall that the degree width of a Laurent polynomial is the difference of the degrees of its leading and its trailing term.

\begin{Proposition}
  \label[Proposition]{prop:SeZ}
  For $1\leqslant k \leqslant p$, the scalar product $S^{0}_{k,p}$ is given by
  \begin{equation}
    \label{eqn:SeFactorisation}
    S^{0}_{k,p}=\prod_{1\leqslant i < j \leqslant 2k}(qz_i-q^{-1}z_j)\,f_{k,p},
  \end{equation}
  where $f_{k,p}=f_{k,p}(z_1,\dots,z_{2k};x_1,\dots,x_{p})$ is a symmetric polynomial in $z_1,\dots,z_{2k}$ of degree at most $p-k$ with respect to $z_i$ for each $i=1,\dots,2k$. 
  \end{Proposition}
  \proof
    \Cref{prop:DefPsiE} implies that for each $i=1,\dots,2k$, the scalar product $S^0_{k,p}$ is a polynomial in $z_i$ of degree at most $n-1=p+k-1$. The exchange relations for the vector $|\Psi_{2n}^{0}\rangle$ imply that the scalar product obeys the relation
    \begin{equation}
      (qz_{i+1}-q^{-1}z_i)S^{0}_{k,p}(\dots,z_i,z_{i+1},\dots;\dots)=(qz_{i}-q^{-1}z_{i+1})S^{0}_{k,p}(\dots,z_{i+1},z_i,\dots;\dots),
    \end{equation}
    for each $i=1,\dots,2k-1$. It straightforwardly leads to the factorisation \eqref{eqn:SeFactorisation}, as well as the symmetry properties of $f_{k,p}$. The factorisation implies that for each $i=1,\dots,2k$, the function $f_{k,p}$ is a polynomial in $z_i$ of degree at most $p+k-1-(2k-1)=p-k$.
  \eproof

\begin{Proposition}
  \label[Proposition]{prop:SeX}
  For $p\geqslant 1$, the scalar product $S^{0}_{k,p}$ is a Laurent polynomial in each $x_i$ of degree width at most $2n=2(p+k)$. It possesses the property
  \begin{equation}
      \label{eqn:ReflectionX}
    S^{0}_{k,p}(\dots;\dots,x_i^{-1},\dots) = S^{0}_{k,p}(\dots;\dots,x_i,\dots).
  \end{equation}
  Furthermore, for $p\geqslant 2$ the function $S^{0}_{k,p}$ is symmetric with respect to $x_1,\dots,x_{p}$.
  \end{Proposition}
  \proof
    The proof consists of three steps. First, the upper bound $2n=2(k+p)$ on the degree width with respect to $x_i$ follows from \cref{prop:DefPsiE}, and the fact that $\langle\chit(x)|$ is a Laurent polynomial in $x$ of degree width $2$.
  
  Second, we show that $S^{0}_{k,p}$ is invariant under the reversal of $x_i$. To this end, we write
  \begin{align}
    S^{0}_{k,p}(\dots;\dots,x_i^{-1},\dots)
      &=\langle \gamma_{2k,p}(\dots,x_i^{-1},\dots)|\Psi_{2n}^{0}(\dots,x_i^{-1},x_i,\dots)\rangle\nonumber\\
     &=\langle \gamma_{2k,p}(\dots,x_i,\dots)|\check R_{2(k+i)-1,2(k+i)}(x_i^2)|\Psi_{2n}^{0}(\dots,x_i^{-1},x_i,\dots,)\rangle\nonumber\\
     & = S^{0}_{k,p}(\dots;\dots,x_i,\dots) .
  \end{align}
  From the first to the second line, we used \eqref{eq:Fish}, and from the second to the third line the exchange equations.
  
  Third, let $p\geqslant 2$ and $i=1,\dots,p-1$. We use the exchange equations to write
  \begin{align}
    S^{0}_{k,p}(\dots;\dots,x_i,x_{i+1},\dots) =& \langle \gamma_{2k,p}(\dots,x_{i},x_{i+1},\dots)|\check R_{2(k+i),2(k+i)+1}(x_i^{-1}x_{i+1}^{-1})\\
    &\times \check R_{2(k+i)-1,2(k+i)}(x_ix_{i+1}^{-1})|\Psi_{2n}^{0}(\dots,x_{i+1},x_i,x_i^{-1},x_{i+1}^{-1},\dots)\rangle.\nonumber
  \end{align}
  A straightforward application of the boundary Yang-Baxter equation \eqref{eqn:bYBE} leads to
  \begin{align}
    S^{0}_{k,p}(\dots;\dots,x_i,x_{i+1},\dots) =& \langle \gamma_{2k,p}(\dots,x_{i+1},x_{i},\dots)|\check R_{2(k+i),2(k+i)+1}(x_i^{-1}x_{i+1}^{-1})\\
    &\times \check R_{2(k+i)+1,2(k+i)+2}(x_ix_{i+1}^{-1})|\Psi_{2n}^{0}(\dots,x_{i+1},x_i,x_i^{-1},x_{i+1}^{-1},\dots)\rangle. \nonumber
  \end{align}
  We apply once more the exchange equations and obtain
  \begin{equation}
     S^{0}_{k,p}(\dots;\dots,x_i,x_{i+1},\dots) = S^{0}_{k,p}(\dots;\dots,x_{i+1},x_i,\dots).
  \end{equation}
  This relation implies the symmetry of $S^{0}_{k,p}$ with respect to $x_1,\dots,x_p$.
  \eproof

\begin{Corollaire}
  \label[Corollary]{corr:SeTrivialZeroes}
  For $p \ge 1$ and $i=1,\dots,p$, the scalar product $S_{k,p}^{0}$ vanishes for $x_i = q^{\pm 1}$.
\end{Corollaire}
  \proof
    We have $S_{k,p}^{0}=0$ for $x_i = q$ because $\langle \chit(q)|=0$. Furthermore, it follows from \eqref{eqn:ReflectionX} that $S_{k,p}^{0}=0$ for $x_i = q^{-1}$.
  \eproof

\paragraph{Reduction relations.} In the following two propositions, we establish reductions relations for the scalar products. They are the result of the reduction relations for the polynomial solutions to the quantum Knizhnik-Zamolodchikov system, see \cref{prop:RedPsi}.
\begin{Proposition}
  \label[Proposition]{prop:SeRed}
  For $p\geqslant 2$ and $k \ge 0$, we have
  \begin{alignat}{2} 
  S_{k,p}^{0}\big|_{x_p = q^{-2} x_{p-1}} &= q^{2(p+k-2)} \frac{(q x_{p-1}-q^{-1}x_{p-1}^{-1})(x_{p-1}-x_{p-1}^{-1})}{(q-q^{-1})^2}\prod_{j=1}^{2k}\frac{(q z_j - q^{-1} x_{p-1}^{-1})(q z_j-q^{-3}x_{p-1})}{(q-q^{-1})^2}\nonumber\\
&\times \prod_{i=1}^{p-2} \frac{(q x_i-q^{-1}x_{p-1}^{-1})(q x_i^{-1}-q^{-1}x_{p-1}^{-1})(q x_i-q^{-3}x_{p-1})(q x_i^{-1}-q^{-3}x_{p-1})}{(q-q^{-1})^4}\nonumber\\
&\times \frac{(1+q^2)(q-x_{p-1})^2(q^2-x_{p-1})(q^3-x_{p-1})(q^3+x_{p-1})}{q^5 (q^2-1)^2 x_{p-1}^2(1+x_{p-1})} S_{k,p-2}^{0}.\label{eq:first.reduction}
\end{alignat}
\end{Proposition}
\proof
  We give the main steps of the proofs. We use the exchange equations to write
  \begin{align}
  &S_{k,p}^{0}\big|_{x_p = q^{-2} x_{p-1}}\\ 
  & = \langle \gamma_{2k,p}(\dots,x_{p-1},q^{-2}x_{p-1})|\check R_{2(n-1),2n-1}(q^2x_{p-1}^{-2})|\Psi_{2n}^{0}(\dots, x_{p-1},q^{-2}x_{p-1},x_{p-1}^{-1},q^2x_{p-1}^{-1})\rangle. \nonumber
  \end{align}
  The last two arguments $x_{p-1}^{-1}$ and $q^2 x_{p-1}^{-1}$ of the vector $|\Psi_{2n}^{0}\rangle$ prompt us to use \cref{prop:RedPsi}. We obtain
  \begin{align}
  S_{k,p}^{0}\big|_{x_p = q^{-2} x_{p-1}} = &-(-q)^{n-2}\prod_{j=1}^{2k}\frac{q z_j-q^{-1}x_{p-1}^{-1}}{q-q^{-1}}\prod_{i=1}^{p-2}\frac{(q x_i-q^{-1}x_{p-1}^{-1})(q x_i^{-1}-q^{-1}x_{p-1}^{-1})}{(q-q^{-1})^2}\\
  &\times \frac{(q x_{p-1}-q^{-1}x_{p-1}^{-1})(x_{p-1}-x_{p-1}^{-1})}{(q-q^{-1})^2}\langle \gamma_{2k,p}(\dots,x_{p-1},q^2x_{p-1})\nonumber|\\
  & \times \check R_{2(n-1),2n-1}(q^2x_{p-1}^{-2})\check R_{2n-1,2n}(q^{-2})|\Big(|\Psi_{2(n-1)}^{0}(\dots, x_{p-1},q^{-2}x_{p-1})\rangle\otimes |\omega\rangle\Big).\nonumber
  \end{align}
  In the last line of this expression, we used \eqref{eqn:OmegaEVOfR} to write $|\omega\rangle=\check R(q^{-2})|\omega\rangle$. Thanks to the boundary Yang-Baxter equation \eqref{eqn:bYBE}, we have
  \begin{align}
    &\langle \gamma_{2k,p}(\dots,x_{p-1},q^{-2}x_{p-1})\nonumber|\check R_{2(n-1),2n-1}(q^2x_{p-1}^{-2})\check R_{2n-1,2n}(q^{-2})\\[0.1cm]
    & = \langle \gamma_{2k,p}(\dots,q^{-2}x_{p-1},x_{p-1})|\check R_{2(n-1),2n-1}(q^2x_{p-1}^{-2})\check R_{2n-3,2n-2}(q^{-2}).
  \end{align}
  This equality, combined with the exchange equations, leads to 
  \begin{align}
  S_{k,p}^{0}\big|_{x_p = q^{-2} x_{p-1}} = &-(-q)^{n-2}\prod_{j=1}^{2k}\frac{q z_j-q^{-1}x_{p-1}^{-1}}{q-q^{-1}}\prod_{i=1}^{p-2}\frac{(q x_i-q^{-1}x_{p-1}^{-1})(q x_i^{-1}-q^{-1}x_{p-1}^{-1})}{(q-q^{-1})^2}\\
  &\times \frac{(q x_{p-1}-q^{-1}x_{p-1}^{-1})(x_{p-1}-x_{p-1}^{-1})}{(q-q^{-1})^2}\langle \gamma_{2k,p}(\dots,q^{-2}x_{p-1},x_{p-1})\nonumber|\\
  & \times \check R_{2n-2,2n-1}(q^2x_{p-1}^{-2})|\Big(|\Psi_{2(n-1)}^{0}(\dots,q^{-2}x_{p-1}, x_{p-1})\rangle\otimes |\omega\rangle\Big).\nonumber
  \end{align}
  In this expression, the last two arguments $q^{-2}x_{p-1}$ and $x_{p-1}$ of $|\Psi_{2(n-1)}^{0}\rangle$ suggest yet another application of \cref{prop:RedPsi}. After a few simplifications, we find
    \begin{align}
  S_{k,p}^{0}\big|_{x_p = q^{-2} x_{p-1}} = &-q^{2(n-2)}\frac{(q x_{p-1}-q^{-1}x_{p-1}^{-1})(x_{p-1}-x_{p-1}^{-1})}{(q-q^{-1})^2}\prod_{j=1}^{2k}\frac{(q z_j-q^{-1}x_{p-1}^{-1})(q z_j-q^{-3}x_{p-1})}{(q-q^{-1})^2}\nonumber\\
  &\times \prod_{i=1}^{p-2}\frac{(q x_i-q^{-1}x_{p-1}^{-1})(q x_i^{-1}-q^{-1}x_{p-1}^{-1})(q x_i-q^{-3}x_{p-1})(q x_i^{-1}-q^{-3}x_{p-1})}{(q-q^{-1})^4}\nonumber\\
  &\times \left(\langle \chit(q^{-2} x_{p-1})|\otimes \langle \chit(x_{p-1})|\right)\check R_{23}(q^2x_{p-1}^{-2})\left(|\omega\rangle\otimes |\omega\rangle\right)S_{k,p-2}^{0}.
      \label{eqn:SeRedIntermediate}
  \end{align}
 We have the matrix element
 \begin{align}
    \left(\langle \chit(q^{-2} x)|\otimes \langle \chit(x)|\right)\check R_{23}(q^2x^{-2})\left(|\omega\rangle\otimes |\omega\rangle\right)=-\frac{ \left(1+q^2\right) (q-x)^2 \left(q^2-x\right) \left(q^3-x\right) \left(q^3+x\right)}{q^5 (q^2-1)^2  x^2 (x+1)}.
    \nonumber
  \end{align}
  Inserting this expression into \eqref{eqn:SeRedIntermediate} ends the proof.
\eproof

\begin{Proposition}
\label[Proposition]{prop:SSbar}
  For $p\geqslant 1,k\geqslant 1$, we have
  \begin{align}
    \label{eqn:SeSebar}
    \left.S_{k,p}^{0}\right|_{z_{2k}=q^{-2}x_p}
    & = (-q)^{k-p}\left(\frac{q x_p-q^{-1}x_p^{-1}}{q-q^{-1}}\right)\prod_{j=1}^{2k-1}\frac{q z_j-q^{-3}x_p}{q-q^{-1}} \\
    & \quad \times \prod_{j=1}^{p-1}\frac{(q x_p-q^{-1}x_j)(q x_p-q^{-1}x_j^{-1})}{(q-q^{-1})^2}\left.\bar S^{0}_{k,p-1}\right|_{x=x_p^{-1}}.\nonumber
  \end{align}
      For $p\geqslant 0$, $k\geqslant 1$, we have
  \begin{align}
  \label{eqn:SebarSe}
    \left.\bar S^{0}_{k,p}\right|_{z_{2k-1}=q^{-2}x} &= (-q)^{k-p-1}\left(\frac{qx-q^{-1}x^{-1}}{q-q^{-1}}\right)\prod_{j=1}^{2(k-1)}\frac{qz_j-q^{-1}x}{q-q^{-1}}\\
    &\quad \times \prod_{i=1}^p\frac{(qx-q^{-3}x_i)(qx-q^{-1}x_i^{-1})}{(q-q^{-1})^2}\,S_{k-1,p}^{0}\,.\nonumber
  \end{align}
 For $p\geqslant 1,k\geqslant 0$, we have
  \begin{align}
    \label{eqn:SupSupbar}
    \left.S^{+}_{k,p}\right|_{z_{2k+1}=q^{-2}x_p}= (-q)^{k-p}\prod_{j=1}^{2k}\frac{q z_j-q^{-3}x_p}{q-q^{-1}}\prod_{j=1}^{p}\frac{(q x_p-q^{-1}x_j)(q x_p-q^{-1}x_j^{-1})}{(q-q^{-1})^2}\left.\bar S^{+}_{k,p-1}\right|_{x=x_p^{-1}}.  \end{align}
\end{Proposition}
  \proof
      The proofs of \eqref{eqn:SeSebar}, \eqref{eqn:SebarSe} and \eqref{eqn:SupSupbar} are similar. We present the proof of \eqref{eqn:SeSebar}. In order to simplify its presentation, we first consider the specialisation $z_{2k}=q^{-2}x_1$. It allows us to apply \cref{prop:RedPsi}. We obtain
    \begin{align}
    \left.S^{0}_{k,p}\right|_{z_{2k}=q^{-2}x_1}
    & =(-q)^{k-p}\prod_{j=1}^{2k-1}\frac{(q z_j-q^{-3}x_1)(q x_1-q^{-1}x_1^{-1})}{(q-q^{-1})^2} \prod_{j=2}^{p}\frac{(q x_1-q^{-1}x_j)(q x_1-q^{-1}x_j^{-1})}{(q-q^{-1})^2}\nonumber \\
    & \quad \times \langle \underset{2k}{\underbrace{\uparrow \cdots \uparrow}}|\otimes \left(\bigotimes_{i=1}^p\langle \chit(x_i)|\right)\Xi_{2(n-1)}^{2k}|\Psi^{0}_{2(n-1)}(\dots,z_{2k-1},x_1^{-1},x_2,x_2^{-1},\dots)\rangle.\nonumber
  \end{align}
 From the definition \eqref{eqn:DefPhi} of the co-vector $\langle\varphi(x)|$, we obtain $\left(\langle{\uparrow}|\otimes\langle \chit(x)|\right)\left(|\omega\rangle \otimes |\alpha\rangle\right) = \langle \varphi(x^{-1})|\alpha\rangle$ for each $\alpha \in \{\uparrow,\downarrow\}$. We use this observation to derive the identity
    \begin{equation}
      \left(\langle \underset{2k}{\underbrace{\uparrow \cdots \uparrow}}|\otimes \left(\bigotimes_{i=1}^p\langle \chit(x_i)|\right)\right)\Xi_{2(n-1)}^{2k} = \langle \underset{2k-1}{\underbrace{\uparrow \cdots \uparrow}}|\otimes \langle \varphi(x_1^{-1})|\otimes \left(\bigotimes_{i=2}^p\langle \chit(x_i)|\right).
    \end{equation}
    It leads to
     \begin{align}
    \left.S^{0}_{k,p}\right|_{z_{2k}=q^{-2}x_1}
    & =(-q)^{k-p}\frac{q x_1-q^{-1}x_1^{-1}}{q-q^{-1}}\prod_{j=1}^{2k-1}\frac{q z_j-q^{-3}x_1}{q-q^{-1}} \prod_{j=2}^{p}\frac{(q x_1-q^{-1}x_j)(q x_1-q^{-1}x_j^{-1})}{(q-q^{-1})^2}\nonumber\\
    & \quad \times \langle \bar \gamma_{2k,p-1}(\dots,z_{2k};x_1^{-1};x_2,\dots)|\Psi^{0}_{2(n-1)}(\dots,z_{2k-1},x_1^{-1},x_2,x_2^{-1},\dots)\rangle.
  \end{align}
  The relation \eqref{eqn:SeSebar} follows from the symmetry of $S^{0}_{k,p}$ with respect to $x_1,\dots,x_p$, proven in \cref{prop:SeX}.
  \eproof

The combination of \eqref{eqn:SeSebar} and \eqref{eqn:SebarSe} leads to a second closed reduction relation for the scalar product $S^{0}_{k,p}$:
\begin{Corollaire}
For $p\geqslant 1,k\geqslant 1$, we have the reduction relation
\begin{alignat}{2} 
S_{k,p}^{0}\big|_{\substack{z_{2k-1}=q^{-2}x_p^{-1}\\ z_{2k}=q^{-2}x_p}}=&-q^{-2(p-k)}\frac{(q^{-1}x_p^{-1}-q^{-3}x_p)(q x_p - q^{-1}x_p^{-1})(q^{-1}x_p - q x_p^{-1})}{(q-q^{-1})^3}\nonumber\\
&\times \prod_{i=1}^{p-1} \frac{(qx_p-q^{-1}x_i)(qx_p-q^{-1}x_i^{-1})(qx_p^{-1}-q^{-1}x_i)(qx_p^{-1}-q^{-1}x_i^{-1})}{(q-q^{-1})^4}\nonumber\\
&\times\prod_{j=1}^{2k-2} \frac{(q z_j - q^{-3}x_p)(q z_j - q^{-3}x_p^{-1})}{(q-q^{-1})^2}  S_{k-1,p-1}^{0}.\label{eq:second.reduction}
\end{alignat}
\end{Corollaire}

\paragraph{Factorised scalar products.} Certain scalar products take a simple, factorised form. Indeed, it follows from \eqref{eqn:SpecialEntryPsiE} that
\begin{align}
  S^{0}_{k,k} =& \prod_{1\leqslant i < j \leqslant 2k}\frac{q z_i-q^{-1}z_j}{q-q^{-1}} \prod_{i=1}^k \frac{(q x_i - q^{-1}x_i^{-1})(q x_i^{-1}-q^{-1} x_i)}{(q-q^{-1})^2} \nonumber\\
&\times \prod_{1\le i<j \le k} \frac{(q x_i - q^{-1}x_j)(q x_i - q^{-1}x_j^{-1})(q x_i^{-1} - q^{-1}x_j)(q x_i^{-1} - q^{-1}x_j^{-1})}{(q-q^{-1})^4},\label{eq:special.component.even}
\end{align}
where $k\geqslant 1$.
In the following, we prove that $S^{0}_{k,k+1}$ also takes a simple form:
\begin{Proposition}
\label[Proposition]{prop:SimpleCase}
For $k\geqslant 0$, the scalar product $S_{k,k+1}^{0}$ is given by
\begin{alignat}{2}
S_{k,k+1}^{0} &= (1-q^{-1})(1+q^{-1})^2 \prod_{j=1}^{2k}\ (q z_j + q^{-2})\prod_{1\le i<j\le 2k} \frac{q z_i - q^{-1}z_j}{q-q^{-1}}\prod_{i=1}^{k+1} \frac{(1-q x_i)(1-q x_i^{-1})}{(q-q^{-1})^2}\nonumber\\
& \times \prod_{1\le i<j\le k+1} \frac{(q x_i - q^{-1}x_j)(qx_i - q^{-1}x_j^{-1})(qx_i^{-1} - q^{-1}x_j)(qx_i^{-1} - q^{-1}x_j^{-1})}{(q-q^{-1})^4}.\label{eq:Skk+1}
\end{alignat}
\end{Proposition}
The proof of this proposition relies on the following lemma.
\begin{Lemma}\label[Lemma]{lem:Skkq-3}
For $k\geqslant 1$ and $j=1,\dots,2k$, the scalar product $S_{k,k+1}^{0}$ vanishes for $z_j = -q^{-3}$.
\end{Lemma}
\proof
Because of the symmetry of $S_{k,k+1}^{0}$ under exchanges $z_j \leftrightarrow z_{2k}$, it suffices to prove the lemma for $j = 2k$. \Cref{prop:SeZ} will then ensure that it also holds for $j=1, \dots, 2k$. The scalar product $S_{k,k+1}^{0}\big|_{z_{2k}=-q^{-3}}$ is a Laurent polynomial in $x_1$ of degree width $2n =4k+2$. It is uniquely fixed by its value at $4k+3$ points. By \cref{prop:SeZ,corr:SeTrivialZeroes,prop:SeRed}, 
we know that it vanishes for the following $4k+2$ values: $x_1 = q^{\pm 1}, q^{\pm 2} x_j$ and $q^{\pm 2} x_j^{-1}$, with $j = 2, \dots, k+1$. The last specialisation is $x_1 = -q$. At this value, we have $z_{2k} = -q^{-3}$, $z_{2k+1} = x_1 = -q^{-1}$ and $z_{2k+2} = x_1^{-1}=-q$, and hence $|\Psi^{0}_{2n}\rangle =0$ due to the wheel condition of \cref{corr:WheelCondition}. This implies that $S_{k,k+1}^{0}\big|_{z_{2k}=-q^{-3}}= 0$, ending the proof.
\eproof

\noindent {\scshape Proof of \cref{prop:SimpleCase}.\ }
From the properties listed above, we know that $S_{k,k+1}^{0}$ vanishes at the following specialisations: (i) $z_j = - q^{-3}$, (ii) $z_j = q^2 z_i$ for $i<j$, (iii) $x_i = q^{\pm 1}$ and (iv) $x_i = q^{\pm1} x_j, q^{\pm1} x_j^{-1}$. Furthermore, the right-hand side of \eqref{eq:Skk+1} vanishes for all these specialisations. It is a polynomial in each $z_j$ with degree $n-1 = 2k$, and a centered Laurent polynomial in each $x_i$ of degree width $2 n = 4k+2$. These are the polynomial degrees given in \cref{prop:SeZ,prop:SeX}, confirming that the right-hand side equals $S_{k,k+1}^{0}$ up to an overall constant which we denote by $\kappa_k$. One easily checks that \eqref{eq:Skk+1} holds for $k=0$, so that the constant is correctly set for this case: $\kappa_0 = 1$. The expression \eqref{eq:Skk+1} also satisfies the reduction relation \eqref{eq:second.reduction}, implying that $\kappa_k/\kappa_{k-1}=1$ for all $k\geqslant 1$.
\eproof

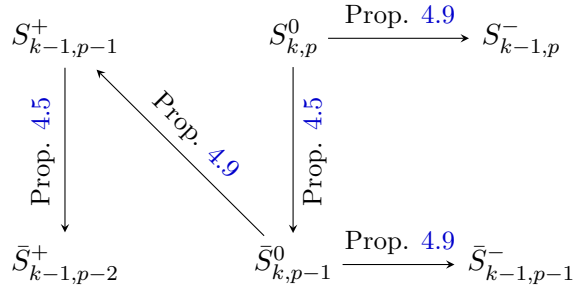
\begin{figure}[h]
  \centering
  \begin{tikzpicture}[
pre/.style={<-,shorten <=1pt,>=stealth},node distance = 2cm]
    \node (Se) at (0,0) {$S^{0}_{k,p}$};
    
    \node (Sebar) at (0,-3)  {$\bar S^{0}_{k,p-1}$}
      edge[pre] node[below,rotate=90] {\small{Prop. \ref{prop:SSbar}}} (Se);
    
    \node (Sdown) at (3,0) {$S^{-}_{k-1,p}$}
      edge[pre] node[auto,swap] {\small{Prop. \ref{prop:SeSdown}}} (Se);
    
    \node (Sdownbar) at (3,-3) {$\bar S^{-}_{k-1,p-1}$}
      edge[pre] node[auto,swap] {\small{Prop. \ref{prop:SeSdown}}} (Sebar);
   
    \node (Sup) at (-3,0) {$S^{+}_{k-1,p-1}$}
      edge[pre] node[above,rotate=-45] {\small{Prop. \ref{prop:SeSdown}}}(Sebar);
   
    \node (Supbar) at (-3,-3) {$\bar S^{+}_{k-1,p-2}$}
      edge[pre] node[above,rotate=90] {\small{Prop. \ref{prop:SSbar}}} (Sup);
   
  \end{tikzpicture}
  \caption{The different scalar products can be obtained from $S^0_{k,p}$ through the specialisations of its parameters. The graph indicates the propositions that establish these specialisations.}
  \label{fig:RelationsSPs}
\end{figure}

\paragraph{Relations between the scalar products.} \Cref{prop:SSbar} establishes relations between different scalar products. We also have the following relations:
\begin{Proposition}
\label[Proposition]{prop:SeSdown}
For $1\leqslant k \leqslant p$, we have 
\begin{subequations}
\begin{alignat}{2}
  \lim_{z_{2k}\to \infty} z_{2k}^{-(n-1)}S^{0}_{k,p} &= -\frac{S^{-}_{k-1,p}}{q^{4k-1}(1-q^{-2})^{n-1}},\\[0.15cm]
  \lim_{z_{2k-1}\to \infty} z_{2k-1}^{-(n-1)}\bar S^{0}_{k,p} &= \frac{\bar S^{-}_{k-1,p}}{q^{4k-1}(1-q^{-2})^{n-1}},
  \end{alignat}
  and
  \be
  \lim_{x\to 0} x \bar S^{0}_{k,p} = -\frac{S^{+}_{k-1,p}}{q^{2p+1}(q-q^{-1})^n}.
  \ee
  \end{subequations}
\proof
  These relations follow from \cref{prop:BraidRelations}.
\eproof
\end{Proposition}
We conclude that all the scalar products and overlaps can be obtained from $S^{0}_{k,p}$ through specialisations or limits of its parameters $z_1, \dots, z_{2k}$ and $x_1, \dots, x_p$. This justifies our focus on the latter in \cref{sec:Se}. For convenience, we summarise the relations between the scalar products in \cref{fig:RelationsSPs}.

%
\section{Determinant expression for the scalar product $\boldsymbol{S_{k,p}^0}$}
%
\label{sec:Se}

From this section onwards, we fix $q = \eE^{2 \pi  \ir/3}$. We write down a candidate expression for the scalar product $S_{k,p}^0$ in terms of a determinant. We show that this expression satisfies the polynomial and reduction properties that uniquely define $S_{k,p}^0$. The result is the following theorem:

\begin{Theoreme}\label{thm:S=R} 
For $q = \eE^{2 \pi \ir/3}$, we have $S_{k,p}^{0} = \Rcal_{k,p}$ with
\begin{alignat}{2}
\Rcal_{k,p} &= \frac{(-1)^{p(p-1)/2} (1-q^2)^{p+2k}}{2^k3^{(p+k)(p+k+1)/2}} \prod_{i=1}^p\frac{(1-q x_i)(1-q^{-1} x_i)}{x_i^{p+k}} \prod_{j=1}^{2k}\, (1+z_j) \prod_{1\le i< j \le 2k}(q z_i - q^{-1}z_j)\nonumber\\
& \times \frac{\det M^{(k,p)}} {C(x_1, \dots, x_p)C(z_1, \dots, z_{2k},x_1, \dots, x_p)},
\label{eq:Rkp}
\end{alignat}
where 
\be 
C(y_1, \dots, y_n) = \prod_{1 \le i<j\le n} (y_j-y_i)(y_iy_j-1)
\ee and $M^{(k,p)}$ is a matrix of size $2(p+k)$ with entries:
\be
M_{ij}^{(k,p)} = 
\left\{\begin{array}{cll}
\displaystyle\frac{x_i^{3(j-1)} + x_i^{3(p+k-j)+1}}{1+x_i} & 1\le i \le p& 1 \le j \le p+k \\[0.3cm]
0& 1\le i \le p& p+k+1 \le j \le 2(p+k) \\[0.3cm]
0& p+1 \le i \le 2p & 1 \le j \le p+k \\[0.2cm]
\displaystyle\frac{x_{i-p}^{3 (j-k-p-1)}-x_{i-p}^{3 (2 (k+p)-j)+1}}{1-x_{i-p}}& p+1 \le i \le 2p & p+k+1 \le j \le 2(p+k) \\[0.4cm]
\displaystyle\frac{z_{i-2 p}^{3 (j-1)}+z_{i-2 p}^{3 (k+p-j)+1}}{1+z_{i-2 p}} & 2p+1 \le i \le 2(p+k) & 1 \le j \le p+k \\[0.4cm]
\displaystyle\frac{z_{i-2 p}^{3 (j-k-p-1)}-z_{i-2 p}^{3 (2 (k+p)-j)+1}}{1-z_{i-2 p}} & 2p+1 \le i \le 2(p+k) & p+k+1 \le j \le 2(p+k).
\end{array}
\right.
\label{eq:Mij}
\ee
\end{Theoreme}
This matrix is divided in six blocks as
\be
M^{(k,p)} = 
\left(\begin{array}{c|c}
m_1 & 0 \\[0.05cm] \hline \\[-0.4cm]
0 & m_2  \\[0.05cm] \hline \\[-0.45cm]
m_3 & m_4
\end{array}\right),
\qquad 
\begin{array}{lll}
(m_1)_{ij} = \eta^+_j(x_i), \quad& 1 \le i \le p, \quad& 1 \le j \le p+k, \\[0.1cm]
(m_2)_{ij} = \eta^-_j(x_i), \quad& 1 \le i \le p, \quad& 1 \le j \le p+k, \\[0.1cm]
(m_3)_{ij} = \eta^+_j(z_i), \quad& 1 \le i \le 2k, \quad& 1 \le j \le p+k, \\[0.1cm]
(m_4)_{ij} = \eta^-_j(z_i), \quad& 1 \le i \le 2k, \quad& 1 \le j \le p+k, \\[0.1cm]
\end{array}
\ee
where
\be
\eta_j^\pm(x) = \frac{x^{3(j-1)} \pm x^{3(p+k-j)+1}}{1\pm x}.
\ee

The remainder of this section is devoted to the proof of this theorem. In \cref{sec:OS.Schurs}, we first discuss properties of two families of polynomials related to even orthogonal and symplectic Schur polynomials. We use them in \cref{sec:reduction.Rkp} to study the reduction properties of the determinant expression $\Rcal_{k,p}$. All the ingredients are put together in \cref{sec:thm.proof} to complete the proof of the theorem.

\subsection{Even orthogonal and symplectic Schur polynomials}\label{sec:OS.Schurs}

We introduce the polynomials
\be
\label{eq:S&O.Schur}
P^{\pm}(y_1, \dots, y_n) = \frac{\det_{i,j=1}^n \big(y_i^{3(j-1)} \pm y_i^{3(n-j)+1}\big)}{C(y_1, \dots, y_n)}, \qquad n \ge 1.
\ee
Up to an overall product of monomials, $P^{+}(y_1, \dots, y_n)$ and $P^{-}(y_1, \dots, y_n)$ are special cases of  even orthogonal Schur and symplectic Schur polynomials, respectively \cite{fulton2013representation}. Their degree in each $y_i$ is $n$. They are invariant under the exchanges $y_i \leftrightarrow y_j$ and satisfy
\be
\label{eq:inverse.yi}
P^{\pm}(y_1, \dots, y_n)\big|_{y_i \to y^{-1}_i} = \pm y_i^{-n} P^{\pm}(y_1, \dots, y_n).
\ee
The following lemma shows that they also satisfy certain reduction properties.

\begin{Lemma}\label[Lemma]{lemma:reductions}
Let $q = \eE^{2 \pi \ir/3}$. The polynomials $P^\pm(y_1, \dots, y_n)$ satisfy the reduction relations
\be
\label{eq:Ppmred}
P^{\pm}(y_1, \dots, y_{n-1},q^{-2} y_{n-1}) = \mp P^{\pm}(y_1, \dots, y_{n-2})\, (1-y^2_{n-1})(1-q^2y^2_{n-1}) \prod_{j=1}^{n-2} (y_j -q^{2}y_{n-1})(q^{2} y_j y_{n-1}-1).
\ee
\end{Lemma}
\proof The polynomial degree of $P^\pm(y_1, \dots, y_{n-1}, q^2 y_{n-1})$ in $y_{n-1}$ is $2n$. From the definition \eqref{eq:S&O.Schur}, it is straightforward to verify that $P^\pm(y_1, \dots, y_{n-1}, q^2 y_{n-1})$ vanishes for $y_{n-1} = \pm 1, \pm q, qy_j, qy_j^{-1}$. For each of these evaluations, there is a simple linear combination of the rows of the matrix in the determinant that vanishes. This gives us a list of $2n$ zeros of $P^\pm(y_1, \dots, y_{n-1}, q^{-2} y_{n-1})$ which exhausts its polynomial degree. To fix the overall constant prefactor, we take the limit $y_{n-1} \to 0$ in the determinant formula and find
\be
\lim_{y_{n-1}\to 0} P^{\pm}(y_1, \dots, y_{n-1},q^{-2} y_{n-1}) = \mp (-1)^n P^{\pm}(y_1, \dots, y_{n-2}) \prod_{j=1}^{n-2} y_j.
\ee
The proposed expression \eqref{eq:Ppmred} has the same limiting behavior, confirming the value of the constant.
\eproof

As an immediate corollary of \cref{lemma:reductions} and \eqref{eq:inverse.yi}, we have
\be
\label{eq:Ppmred2}
P^{\pm}(y_1, \dots, y_{n-1},q^{-2} y^{-1}_{n-1}) = - P^{\pm}(y_1, \dots, y_{n-2})\frac{(1-y^2_{n-1})(q^2-y^2_{n-1})}{y_{n-1}^n} \prod_{j=1}^{n-2} (y_j y_{n-1}-q^{2})(q^{2} y_j-y_{n-1}).
\ee

\subsection{Reduction properties of the determinant formula}\label{sec:reduction.Rkp}

Our proof of \cref{thm:S=R} requires a collection of lemmas and propositions presented in this subsection and the next. A first lemma regards the polynomial properties of $\Rcal_{k,p}$.

\begin{Lemma}\label[Lemma]{lem:poly.degrees}
$\Rcal_{k,p}$ is a polynomial in each $z_j$ of degree $n-1 = p+k-1$ and a centered Laurent polynomial in each $x_i$ of degree width at most $2n = 2 (p+k)$.
\end{Lemma}
\proof Clearly, $\Rcal_{k,p}$ is a rational function of the parameters $z_j$ and $x_i$. By bringing the prefactor $1+z_j$ inside the determinant and multiplying it to the proper row, one obtains a matrix that is regular at $z_j = - 1$. The matrix entries are also regular at $z_j = 1$.  Moreover, at each value of $z_j$ where $C(z_1, \dots, z_{2k},x_1, \dots, x_p) = 0$, the determinant in the numerator vanishes because some of its rows are linearly dependent. As a result, $\Rcal_{k,p}$ is regular for all $z_j \in \mathbb C$, proving that $\Rcal_{k,p}$ is polynomial in $z_j$. With similar arguments, we find that as a function of $x_i$, the only pole of $\Rcal_{k,p}$ is at $x_i = 0$, and observe that its order is finite. The degrees follow from a simple power-counting argument.
\eproof

As a next step, we show that $\Rcal_{k,p}$ satisfies the same two reduction relations \eqref{eq:first.reduction} and \eqref{eq:second.reduction} as $S_{k,p}^{0}$. These are given in the following two propositions.
\begin{Proposition}
\label[Proposition]{prop:first.reduction.R}
$\Rcal_{k,p}$ satisfies the reduction relation\footnote{In \eqref{eq:first.reduction.R} and \eqref{eq:second.reduction.R}, some powers of $q$ could be simplified further using $q^3 = 1$. We have left them as is to make the comparison easier with the corresponding reduction relations \eqref{eq:first.reduction} and \eqref{eq:second.reduction} for $S^{0}_{k,p}$, which hold for generic $q$.}
\begin{alignat}{2} \Rcal_{k,p}\big|_{x_p = q^{-2} x_{p-1}} &= q^{2(p+k-2)} \frac{(q x_{p-1}-q^{-1}x_{p-1}^{-1})(x_{p-1}-x_{p-1}^{-1})}{(q-q^{-1})^2}\prod_{j=1}^{2k}\frac{(q z_j - q^{-1} x_{p-1}^{-1})(q z_j-q^{-3}x_{p-1})}{(q-q^{-1})^2}\nonumber\\
&\times \prod_{i=1}^{p-2} \frac{(q x_i-q^{-1}x_{p-1}^{-1})(q x_i^{-1}-q^{-1}x_{p-1}^{-1})(q x_i-q^{-3}x_{p-1})(q x_i^{-1}-q^{-3}x_{p-1})}{(q-q^{-1})^4}\nonumber\\
&\times \frac{(1+q^2)(q-x_{p-1})^2(q^2-x_{p-1})(q^3-x_{p-1})(q^3+x_{p-1})}{q^5 (q^2-1)^2 x_{p-1}^2(1+x_{p-1})}\Rcal_{k,p-2}.\label{eq:first.reduction.R}
\end{alignat}
\end{Proposition}
\proof After the simplifcation of the prefactors, the proof of \eqref{eq:first.reduction.R} boils down to proving a reduction relation for the determinant of $M^{(k,p)}$,
\begin{alignat}{2}
\frac{\det M^{(k,p)}} {C(\vec x)C(\vec z, \vec x)}
\Big|_{x_p = q^{-2} x_{p-1}}=& - \prod_{i=1}^{p-2} (x_i - q^2 x_{p-1})^2(q^2 x_i x_{p-1}-1)^2 \prod_{j=1}^{2k} (z_j - q^2 x_{p-1})(q^2 z_j x_{p-1}-1)\nonumber\\
&\times (1-q^2 x^2_{p-1})(1-x^2_{p-1})  \frac{\det M^{(k,p-2)}} {C(\vec x\,'')C(\vec z, \vec x\,'')},
\label{eq:det1.red}
\end{alignat}
where $\vec x = \{x_1, \dots, x_p\}$, $\vec x\,'' = \{x_1, \dots, x_{p-2}\}$ and $\vec z = \{z_1, \dots, z_{2k}\}$ are ordered sets. To prove this identity, we split the matrix $M^{(k,p)}$ into two rectangular matrices of size $2(p+k)\times (p+k)$ and use the cofactor expansion formula for determinants to find
\be
\label{eq:cof}
\frac{\det M^{(k,p)}} {C(\vec x)C(\vec z, \vec x)} =\frac{(-1)^{pk}}{\prod_{i=1}^{p}(1-x_i^2)} \sum_{\substack{J \subset \{1,\dots, 2k\}\\|J|=k}}\frac{P^+(\vec x, \vec z_J)P^-(\vec x, \vec z_{J^c})}{\prod_{i\in J,j \in J^c}(z_j-z_i)(z_iz_j-1)}\prod_{j\in J} \frac{1}{(1+z_j)}\prod_{j\in J^c} \frac1{(1-z_j)},
\ee
where $J$, $J^c$, $\vec z_J$, $\vec z_{J^c}$ are ordered sets: $J^c$ is the complement of $J$ in $\{1,\dots, 2k\}$, and the sets $\vec z_J$ and $\vec z_{J^c}$ are defined by $\vec z_J = \{z_j|j \in J\}$ and $\vec z_{J^c} = \{z_j|j \in J^c\}$.
In such cofactor expansions, the summand involves the sign of a permutation, and in the present case it is absorbed in the function in the denominator and in the prefactor $(-1)^{pk}$.

Specialising $x_p$ to $q^{-2}x_{p-1}$, we apply \cref{lemma:reductions} to each of the polynomials in the summand and find
\begin{alignat}{2}
\frac{\det M^{(k,p)}} {C(\vec x)C(\vec z, \vec x)}\Big|_{x_p = q^{-2}x_{p-1}} &= \frac{(-1)^{pk+1}(1-x_{p-1}^2)(1-q^2x_{p-1}^2)}{\prod_{i=1}^{p-2}(1-x_i^2)} \prod_{i=1}^{p-2}(x_i - q^2 x_{p-1})^2(q^2 x_i x_{p-1}-1)^2
\nonumber\\[0.2cm]
&\times\prod_{j=1}^{2k}(z_j-q^2 x_{p-1})(q^2z_jx_{p-1}-1)\\[0.2cm]
&\times \sum_{\substack{J \subset \{1,\dots, 2k\}\\|J|=k}} \frac{P^+(\vec x\,'', \vec z_J)P^-(\vec x\,'', \vec z_{J^c})}{\prod_{i\in J,j \in J^c}(z_j-z_i)(z_iz_j-1)}
\times\prod_{j\in J}\frac{1}{(1+z_j)}\prod_{j\in J^c}\frac{1}{(1-z_j)}.\nonumber
\end{alignat}
Finally, we use \eqref{eq:cof} with $p$ replaced by $p-2$, and obtain the desired reduction relation \eqref{eq:first.reduction.R}.\eproof

\begin{Proposition} $\Rcal_{k,p}$ satisfies the reduction relation
\begin{alignat}{2} 
\Rcal_{k,p}\big|_{\substack{z_{2k-1}=q^{-2}x_p^{-1}\\ z_{2k}=q^{-2}x_p}}&=-q^{-2(p-k)}\frac{(q^{-1}x_p^{-1}-q^{-3}x_p)(q x_p - q^{-1}x_p^{-1})(q^{-1}x_p - q x_p^{-1})}{(q-q^{-1})^3}\nonumber\\
&\times \prod_{i=1}^{p-1} \frac{(qx_p-q^{-1}x_i)(qx_p-q^{-1}x_i^{-1})(qx_p^{-1}-q^{-1}x_i)(qx_p^{-1}-q^{-1}x_i^{-1})}{(q-q^{-1})^4}\nonumber\\
&\times\prod_{j=1}^{2k-2} \frac{(q z_j - q^{-3}x_p)(q z_j - q^{-3}x_p^{-1})}{(q-q^{-1})^2} \times \Rcal_{k-1,p-1}.\label{eq:second.reduction.R}
\end{alignat}
\end{Proposition}
\proof The proof boils down to proving a reduction relation for the determinant of $M^{(k,p)}$:
\be\label{eq:det2red}
\frac{\det M^{(k,p)}}{C(\vec x)C(\vec z, \vec x)} \Big|_{\substack{z_{2k-1}=q^{-2}x_p^{-1}\\ z_{2k}=q^{-2}x_p}} = \frac{2(-1)^p q^{-k}}{x_p^{p-k-1}}  \frac{\det M^{(k-1,p-1)}}{C(\vec x\,')C(\vec z\,'', \vec x\,')} 
\prod_{i=1}^{p-1} (q^2x_i-x_p)(x_i-q^{2}x_p)(q^{2}x_ix_p-1)(x_ix_p-q^{2}),
\ee
where $\vec x\,' = \{x_1, \dots, x_{p-1}\}$ and $\vec z\,'' = \{z_1, \dots, z_{2k-2}\}$ are ordered sets.
To show this relation, we use the cofactor decomposition \eqref{eq:cof}. The terms of the sum split between four cases: (i) $z_{2k-1} \in J$, $z_{2k} \in J^c$, (ii) $z_{2k-1} \in J$, $z_{2k} \in J^c$, (iii) $z_{2k-1}, z_{2k} \in J$, and (iv) $z_{2k-1}, z_{2k} \in J^c$. Upon specialising $z_{2k-1}=q^{-2}x_p^{-1}$ and $z_{2k}=q^{-2}x_p$, the terms of types (iii) and (iv) vanish. This is an immediate consequence of \eqref{eq:Ppmred}, whose right-hand side is zero under this specialisation.

Let us respectively denote by $\Xi_1$ and $\Xi_2$ the sums of terms of types (i) and (ii) specialised to $z_{2k-1}=q^{-2}x_p^{-1}$ and $z_{2k}=q^{-2}x_p$. In each case, to operate the reduction, we apply \eqref{eq:Ppmred} for one of the two polynomials and \eqref{eq:Ppmred2} for the other. We also reduce the products over $J$ and $J'$ in \eqref{eq:cof} to extract the terms involving $z_{2k-1}$ and $z_{2k}$. Many factors simplify, and the resulting expressions for $\Xi_1$ and $\Xi_2$ can be rewritten using \eqref{eq:cof} with $(p,k)$ replaced by $(p-1,k-1)$ as
\begin{subequations}
\begin{alignat}{2}
\Xi_1 &=   \frac{(-1)^p q^{-k}}{x_p^{p-k-1}}  \frac{\det M^{(k-1,p-1)}}{C(\vec x\,')C(\vec z\,'', \vec x\,')} \frac{(q-x_p)(q x_p+1)}{qx_p(q-q^{-1})} \prod_{i=1}^{p-1} (q^2x_i-x_p)(x_i-q^{2}x_p)(q^{2}x_ix_p-1)(x_ix_p-q^{2}),
\\[0.2cm]
\Xi_2 &=   \frac{(-1)^p q^{-k}}{x_p^{p-k-1}}  \frac{\det M^{(k-1,p-1)}}{C(\vec x\,')C(\vec z\,'', \vec x\,')} \frac{(q+x_p)(q x_p-1)} {qx_p(q-q^{-1})} \prod_{i=1}^{p-1} (q^2x_i-x_p)(x_i-q^{2}x_p)(q^{2}x_ix_p-1)(x_ix_p-q^{2}).
\end{alignat}
\end{subequations}
Their sum indeed gives \eqref{eq:det2red}.
\eproof

\subsection{Proof of the theorem}\label{sec:thm.proof}

In this section, we employ the results of the previous section and give a proof of \cref{thm:S=R}. We start our investigation with the case $k=p$, in which case $S_{k,k}^{0}$ is given by \eqref{eq:special.component.even}. We show that the determinant formula \eqref{eq:Rkp} coincides with this expression.

\begin{Proposition}\label[Proposition]{prop:Rkk}
For $q = \eE^{2 \pi \ir/3}$ and $k\ge 1$, we have $S_{k,k}^{0} = \Rcal_{k,k}$.
\end{Proposition}
\proof Because of its prefactors, the expression \eqref{eq:Rkp} for $\Rcal_{k,k}$ vanishes at $z_j = q^2 z_i$ for $i<j$, and likewise at $x_i = \pm q$ and $\pm q^{2}$. As a function of $x_i$, it also vanishes at $x_i = q^{\pm 2} x_j$ and $q^{\pm 2} x_j^{-1}$. This is due to a non trivial vanishing linear combination of the rows of $M^{(k,k)}$. Indeed, let us consider the vector $v$ defined as
\be
\label{eq:v}
v = (a_1, \dots, a_k, a_k, \dots, a_1, -a_1, \dots, -a_k, -a_k, \dots, -a_1).
\ee
It satisfies $(M^{(k,k)} v)_j = 0$ for $j = 2k+1, \dots, 4k$. The equality $(M^{(k,k)} v)_j = 0$ also holds for $j=1, \dots, 2k$ if $\tilde M^{(k,k)} \tilde v = 0$, where $\tilde v$ and $\tilde M^{(k,k)}$ are of size $k$ and are defined by
\be
\label{eq:vM}
\tilde v = (a_1, \dots, a_k), \qquad \tilde M^{(k,k)}_{ij} = x_i^{3(j-1)}+x_i^{3(2k-j)}, \quad 1\le i,j \le k.
\ee
For $x_i = q^{\pm 2} x_j$ or $q^{\pm 2} x_j^{-1}$, two rows of $\tilde M^{(k,k)}$ become linearly dependent, implying that $\det \tilde M^{(k,k)} = 0$ and $\Rcal_{k,k}$ indeed vanishes at this specialisation, as claimed above.

The knowledge of all these zeros tells us that $\Rcal_{k,k}$ can be expressed as
\begin{alignat}{2}
\Rcal_{k,k} &= \kappa_k 
\prod_{1 \le i<j\le 2k} \frac{q z_i - q^{-1}z_j}{q-q^{-1}} \prod_{i=1}^k \frac{(q x_i - q^{-1}x_i^{-1})(q x_i^{-1}-q^{-1} x_i)}{(q-q^{-1})^2} \nonumber\\
&\times \prod_{1\le i<j \le k} \frac{(q x_i - q^{-1}x_j)(q x_i - q^{-1}x_j^{-1})(q x_i^{-1} - q^{-1}x_j)(q x_i^{-1} - q^{-1}x_j^{-1})}{(q-q^{-1})^4},
\label{eq:Rkkpoly}
\end{alignat}
where $\kappa_k$ is possibly a function of the $x_i$ and $z_j$. By counting the powers in \eqref{eq:Rkkpoly}, we find that the products combine into a polynomial in $z_j$ of degree $2k-1$, and a Laurent polynomial in $x_i$ of degree width $4k$. 
This exhausts the degrees of $\Rcal_{k,k}$ in each $x_i$ and $z_j$ given in \cref{lem:poly.degrees}, implying that $\kappa_k$ is a constant. Comparing with \eqref{eq:special.component.even}, we find that $\Rcal_{k,k} = \kappa_k S_{k,k}^{0}$.

An explicit computation for $N=4$ shows that $\kappa_1 = 1$. For $k>1$, the constant $\kappa_k$ is uniquely fixed by imposing the reduction relation \eqref{eq:second.reduction.R}. Using \eqref{eq:second.reduction}, we indeed find $\kappa_k/\kappa_{k-1}=1$, confirming that $\kappa_k = 1$ for all $k\ge 1$. 
\eproof

Next, we investigate the case $p=k+1$, in which case $S_{k,k+1}^{0}$ has the factorised form \eqref{eq:Skk+1}.

\begin{Proposition}
\label[Proposition]{prop:Rkk+1}
For $q = \eE^{2 \pi \ir/3}$, we have $\Rcal_{k,k+1} = S_{k,k+1}^{0}$.
\end{Proposition}
\proof
The expression \eqref{eq:Rkp} for $\Rcal_{k,k+1}$ vanishes for the following specialisations: (i) $z_j = -q^{-3} = -1$, (ii) $z_j = q^2 z_i$ for $i<j$, (iii) $x_i = q^{\pm 1}$ and (iv) $x_i = q^{\pm 2} x_j, q^{\pm 2} x_j^{-1}$. For the case (iv), the proof that $\Rcal_{k,k+1}$ vanishes is similar to the proof given with \eqref{eq:v} and \eqref{eq:vM}. 

The right-hand side of \eqref{eq:Skk+1} vanishes for all of these specialisations and exhausts the degrees in all the variables. This implies that $\Rcal_{k,k+1} = \kappa_k S_{k,k+1}^{0}$ with $\kappa_k$ a constant. An explicit check for $N=2$ shows that $\kappa_0 = 1$. For $k>0$, $\kappa_k$ is uniquely fixed by imposing the reduction relation \eqref{eq:second.reduction.R}. Using \eqref{eq:second.reduction}, we find $\kappa_k/\kappa_{k-1}=1$, confirming that $\kappa_k = 1$ for all $k\ge 1$. 
\eproof

\begin{Proposition}
For $q = \eE^{2 \pi \ir/3}$, we have $\Rcal_{k,p} = S_{k,p}^{0}$.
\end{Proposition}
\proof
For $p<k$, one can always find a vector $v$ of a form similar to \eqref{eq:v} that satisfies $M^{(k,p)}v = 0$, so $\Rcal_{k,p} = S_{k,p}^{0}=0$ in this case. The cases $p=k$ and $p = k+1$ are covered by \cref{prop:Rkk,prop:Rkk+1}. 

For $p\ge k+2$, the proof is inductive in $p$. As a function of $x_p$, $\Rcal_{k,p}$ is a centered Laurent polynomial of width at most $2n$. To prove the equality, we show that it holds for at least $2n+1 = 2p+2k+1$ values of $x_p$. The functions $\Rcal_{k,p}$ and $S_{k,p}^{0}$ both vanish at $x_p = q^{\pm 1}$. They satisfy the same reduction relations at $x_p = q^{-2}x_{p-1}$ and are invariant under the exchanges $x_i \leftrightarrow x_j$ and the inversions $x_i \rightarrow x_i^{-1}$. By the induction hypothesis $\Rcal_{k,p-2} = S_{k,p-2}^{0}$, we find that $\Rcal_{k,p} = S_{k,p}^{0}$ for $x_p = q^{\pm 2} x_i$ and  $q^{\pm 2} x_i^{-1}$, with $i = 1, \dots, p-1$. The equality therefore holds for $4p-2 \ge 2p+2k+2$ values of $x_p$.\eproof

This ends the proof of \cref{thm:S=R}.

%
\section{The homogeneous limit of the scalar product $\boldsymbol{S_{k,p}^{0}}$}\label{sec:homogeneous}
%

In this section, we compute the overlap $C^0_{N,m}$ for even $N$ and even $m$, from the homogeneous limit of the scalar product $S^0_{k,p}$. In \cref{sec:finite.diffs}, we evaluate the limit by applying the method of finite differences, and obtain a expression involving a determinant whose matrix entries are integers. We simplify this determinant in \cref{sec:properties.of.Schurs} using specific evaluations of Schur polynomials that are investigated in \cref{sec:Schur.eval}. The final result is written in terms of a product of Barnes functions.

\subsection{The method of finite differences}\label{sec:finite.diffs}

We start from the scalar product $S^0_{k,p}$ defined in \eqref{eq:Skp0} and its determinant expression \eqref{eq:Rkp}. Applying the limit to the prefactors, we find
\begin{subequations}
\begin{alignat}{2}
&C^{0}_{2(p+k),2k} = \lim_{x,z \to 1} S_{k,p}^{0} = \frac{(-1)^{p(p+1)/2}2^k(q^2-1)^{p-k}}{3^{p(p-1)/2-k(k+1)/2+kp}} \Gamma^{0}_{k,p},
\label{eq:C0Gamma0}\\[0.16cm]
&\Gamma^{0}_{k,p} = \lim_{x,z \to 1} \frac{\det M^{(k,p)}} {C(x_1, \dots, x_p)C(z_1, \dots, z_{2k},x_1, \dots, x_p)},
\end{alignat}
\end{subequations}
where $\lim_{x,z \to 1}$ is a short-hand notation indicating that each $x_i,z_j$ is sent to $1$.
We proceed to evaluate the homogeneous limit of the determinant by using the method of finite differences. The matrix $M^{(k,p)}$ has size $2(p+k)$. As a first step, we define a new matrix $M^{(1)}$ wherein each of the first $2p$ rows is a linear combination of two rows of $M^{(k,p)}$:
\be
M^{(1)} = \begin{pmatrix} 
\mathbb I_p & -\mathbb I_p & 0 \\
\mathbb I_p & \mathbb I_p & 0 \\
0 & 0 & \mathbb I_{2k} \\
\end{pmatrix} \cdot M^{(k,p)},
\ee
where $\mathbb I_n$ is the identity matrix of size $n \times n$. This yields
\be
M^{(1)}_{ij} = 
\left\{\begin{array}{cll}
\displaystyle\frac{x_i^{3(j-1)} + x_i^{3(p+k-j)+1}}{1+x_i} & i \le p& j \le p+k, \\[0.3cm]
\displaystyle-\bigg(\frac{x_i^{3 (j-k-p-1)}-x_i^{3 (2 (k+p)-j)+1}}{1-x_i}\bigg)& i \le p& j > p+k, \\[0.4cm]
\displaystyle\frac{x_{i-p}^{3(j-1)} + x_{i-p}^{3(p+k-j)+1}}{1+x_{i-p}}& p < i \le 2p  & j \le p+k, \\[0.4cm]
\displaystyle\frac{x_{i-p}^{3 (j-k-p-1)}-x_{i-p}^{3 (2 (k+p)-j)+1}}{1-x_{i-p}}& p < i \le 2p & j > p+k,  \\[0.4cm]
\displaystyle\frac{z_{i-2 p}^{3 (j-1)}+z_{i-2 p}^{3 (k+p-j)+1}}{1+z_{i-2 p}} & i>2p & j \le p+k, \\[0.4cm]
\displaystyle\frac{z_{i-2 p}^{3 (j-k-p-1)}-z_{i-2 p}^{3 (2 (k+p)-j)+1}}{1-z_{i-2 p}} & i>2p & j > p+k.
\end{array}
\right.
\label{eq:M1}
\ee
The indices $i$ and $j$ labeling the matrices defined in this section lie in the range $1, \dots, 2(p+k)$. The resulting expression for $\Gamma^{0}_{k,p}$ is
\be
\label{eq:GammaM1}
\Gamma^{0}_{k,p} = \frac1{2^p}\lim_{x,z \to 1} \frac{\det M^{(1)}} {C(x_1, \dots, x_p)C(z_1, \dots, z_{2k},x_1, \dots, x_p)}.
\ee
The limiting procedure that we apply treats separately the first $p$ lines and the last $p+2k$. We rename the variables $x_1, \dots, x_p$ in the first $p$ rows as $y_1^2, \dots, y_p^2$, and the variables $x_1, \dots, x_p, z_1, \dots, z_{2k}$ in the last $p+2k$ rows as $y_{p+1}^2, \dots, y_{2p+2k}^2$. The matrix $M^{(2)}$ is obtained by removing the row-dependent denominators, as well as powers of $y_j$ in such a way that each matrix entry is invariant (up to a sign) under the transformation $y_i \to y_i^{-1}$:
\be
M^{(2)}_{ij} = 
\left\{\begin{array}{cll}
-\Big(y_i^{3(p+k)-6j+4} + y_i^{-3(p+k)+6j-4}\Big)(y_i-y_i^{-1}) \quad& i \le 2(p+k)&  j \le p+k, \\[0.3cm]
\Big(y_i^{9(p+k)-6j+4} - y_i^{-9(p+k)+6j-4}\Big)(y_i+y_i^{-1})\quad& i \le p& j > p+k,   \\[0.3cm]
-\Big(y_i^{9(p+k)-6j+4} - y_i^{-9(p+k)+6j-4}\Big)(y_i+y_i^{-1}) \quad & i > p& j > p+k.
\end{array}
\right.
\label{eq:M2}
\ee
This yields
\be
\label{eq:Gamma.M2}
\Gamma^{0}_{k,p} = \frac1{2^{3p+2k}}\frac{1}{2^{p(p-1)+(p+2k)(p+2k-1)}} \lim_{y \to 1} \frac{1}{\prod_{i=1}^{2p+2k}(y_i-y_i^{-1})}\frac{\det M^{(2)}} {C(y_1, \dots, y_p)C(y_{p+1}, \dots, y_{2p+2k})}.
\ee
We define a new matrix $M^{(3)}$, wherein each column is a linear combination of two columns of $M^{(2)}$:
\begin{subequations}
\be
M^{(3)} = M^{(2)} \cdot 
\frac12\begin{pmatrix}\mathbb I_{p+k} & \mathbb I_{p+k} \\ -\mathbb I_{p+k} & \mathbb I_{p+k}\end{pmatrix},
\ee
\be
M^{(3)}_{ij} = 
\left\{\begin{array}{cll}
-y_i^{3(p+k)-6j+5} + y_i^{-3(p+k)+6j-5} \quad& i \le p\quad&  j \le p+k, \\[0.3cm]
y_i^{9(p+k)-6j+3} - y_i^{-9(p+k)+6j-3}\quad& i \le p\quad& j > p+k,  \\[0.3cm]
y_i^{3(p+k)-6j+3} - y_i^{-3(p+k)+6j-3} \quad & i > p\quad& j \le p+k, \\[0.3cm]
-y_i^{9(p+k)-6j+5} + y_i^{-9(p+k)+6j-5} \quad & i > p\quad& j > p+k.\\[0.3cm]
\end{array}
\right.
\label{eq:M3}
\ee
\end{subequations}
The determinant is given by
\be
\Gamma^{0}_{k,p} = \frac1{2^{2p+k}}\frac{1}{2^{p(p-1)+(p+2k)(p+2k-1)}} \lim_{y \to 1} \frac{1}{\prod_{i=1}^{2p+2k}(y_i-y_i^{-1})}\frac{\det M^{(3)}} {C(y_1, \dots, y_p)C(y_{p+1}, \dots, y_{2p+2k})}.
\ee
Reinserting the factors of $(y_i-y_i^{-1})$ inside the determinant allows us to write the matrix entries in terms of the Chebyshev polynomials $U_\ell(\frac \beta2)$ of the second kind. We use the parameterisation $\beta_j = y_j + y_j^{-1}$ in defining the matrix $M^{(4)}$:
\be
M^{(4)}_{ij} = 
\left\{\begin{array}{lll}
U_{-u_j-1}(\tfrac {\beta_i}2)\quad& i \le p\quad&  j \le p+k, \\[0.3cm]
U_{u_{j-p-k}-3}(\tfrac {\beta_i}2)\quad& i \le p\quad& j > p+k,   \\[0.3cm]
U_{-u_j+1}(\tfrac {\beta_i}2)\quad & i > p\quad& j \le p+k, \\[0.3cm]
U_{u_{j-p-k}-1}(\tfrac {\beta_i}2)\quad & i > p\quad& j > p+k,
\end{array}
\right.
\label{eq:M4}
\ee
with 
\be\label{eq:aj}
u_j = 3(p+k-2j)+5.
\ee
Likewise, we write the functions $C(y_1, \dots, y_p)$ and $C(y_{p+1}, \dots, y_{2p+2k})$ in terms of the variables $\beta_j$, leading to the following expression for $\Gamma^{0}_{k,p}$:
\be
\label{eq:GammaM4}
\Gamma^{0}_{k,p} = \frac{(-1)^p}{2^{2p+k}}\frac{1}{2^{p(p-1)+(p+2k)(p+2k-1)}} \lim_{\beta \to 2} \frac{\det M^{(4)}} {\Delta(\beta_1, \dots, \beta_p)\Delta(\beta_{p+1}, \dots, \beta_{2p+2k})},
\ee
where
\be
\label{eq:Delta}
\Delta(\beta_1, \dots, \beta_n) = \prod_{1 \le i < j \le n} (\beta_j - \beta_i).
\ee
Here, $\lim_{\beta \to 2}$ is a short-hand notation indicating that each $\beta_i$ is sent to $2$.
To evaluate the limit, we use the following lemma, which is a standard result for determinants \cite{DV99}.
\begin{Lemma} \label[Lemma]{sec:finite.diff}
Let $M$ be an $L\times L$ matrix whose rows have the entries $M_{ij} = f_j(x_i)$, for $i = 1, \dots, \ell$ and $\ell \le L$. Then
\be
\lim_{x_1, \dots, x_\ell \to y}\frac{\det M}{\Delta (x_1, \dots, x_\ell)} = \det \tilde M, \qquad \tilde M_{ij} = \left\{ \begin{array}{cl} c_{ij} & i = 1, \dots, \ell,\\[0.1cm]M_{ij}& i>\ell, \end{array}\right.
\ee
where the $c_{kj}$ are the coefficients of $f_j(x)$ in its Taylor expansion around $x = y$:
\be
f_j(x) = \sum_{k=1}^\infty c_{kj}(x-y)^{k-1}.
\ee
\end{Lemma}
\proof
The proof is constructive and obtained by subsequently taking the limits $x_1, \dots, x_\ell \to y$. The limit $x_1 \to y$ fixes the first row of the matrix to $c_{1j}$. To take the limit $x_2 \to y$, we expand $f_j(x_2)$ to order $(x_2-y)^1$ in the second row. Substracting the first row removes the constant term, and the factor of $(x_2-y)$ simplifies with the one in the Vandermonde in the denominator. Taking the limit, we find that the second row is $c_{2j}$. The argument is repeated for the row $i$. We expand in a Taylor series up to order $(x_i-y)^{i-1}$, remove all terms of order lower than $i-1$ by substracting multiples of the first $i-1$ rows, and are left with $c_{ij}(x_i-y)^{i-1}$. Dividing by the factor of $(x_i-y)^{i-1}$ from the Vandermonde in the denominator, we take the limit and obtain $c_{ij}$.
\eproof

Clearly, this lemma also applies in the case where the $\ell$ rows are not the first rows of the matrix. To evaluate \eqref{eq:GammaM4}, we apply it twice: once for the first $p$ rows of the matrix and once for the last $p+2k$ rows. 
We obtain
\be
\label{eq:Gamma.M5}
\Gamma^{0}_{k,p} = \frac{(-1)^p}{2^{2p+k}}\frac{1}{2^{p(p-1)+(p+2k)(p+2k-1)}}  \det M^{(5)},
\ee
where
\be
M^{(5)}_{ij} =
\left\{\begin{array}{lll}
\frac1{(i-1)!}\big(\frac{d}{d\beta}\big)^{i-1}U_{-u_j-1}(\tfrac {\beta}2)\Big|_{\beta=2}\quad& i \le p\quad& j \le p+k, \\[0.3cm]
\frac1{(i-1)!}\big(\frac{d}{d\beta}\big)^{i-1}U_{u_{j-p-k}-3}(\tfrac {\beta}2)\Big|_{\beta=2}\quad& i \le p\quad& j > p+k,   \\[0.3cm]
\frac1{(i-p-1)!}\big(\frac{d}{d\beta}\big)^{i-p-1}U_{-u_j+1}(\tfrac {\beta}2)\Big|_{\beta=2}\quad & i > p\quad& j \le p+k, \\[0.3cm]
\frac1{(i-p-1)!}\big(\frac{d}{d\beta}\big)^{i-p-1}U_{u_{j-p-k}-1}(\tfrac {\beta}2)\Big|_{\beta=2}\quad & i > p\quad& j > p+k.
\end{array}
\right.
\ee

To simplify further the determinant formula \eqref{eq:Gamma.M5}, we note the following identity for the Chebyshev polynomials, 
\be
\label{eq:U.identity}
\frac1{m!}\Big(\frac{d}{d\beta}\Big)^{m}U_{\ell-1}(\tfrac\beta2)\Big|_{\beta =2} =
\frac{1}{(2m+1)!} \prod_{n=-m}^{m} (\ell-n),
\ee
which allows us to rewrite the matrix entries of $M^{(5)}$ in product form. In the resulting expressions, the factorials in the denominators coming from the right side of \eqref{eq:U.identity} depend only on the row label $i$ and are factorised out of the denominator. Because the right side of \eqref{eq:U.identity} is an odd polynomial in $\ell$, we can take linear combinations of the rows to remove the terms which do not have maximal degree in $\ell$, so that a simple power remains. The result is
\be
M^{(6)}_{ij} = 
\left\{\begin{array}{lll}
(-u_j)^{2i-1}\quad& i \le p\quad& j \le p+k, \\[0.3cm]
(u_{j-p-k}-2)^{2i-1}\quad& i \le p\quad& j > p+k,   \\[0.3cm]
(-u_j+2)^{2i-2p-1}\quad & i > p\quad& j \le p+k, \\[0.3cm]
(u_{j-p-k})^{2i-2p-1}\quad & i > p\quad& j > p+k,
\end{array}
\right.
\ee
with
\be
\Gamma^{0}_{k,p} = \frac{(-1)^p}{2^{2p+k}}\frac{1}{2^{p(p-1)+(p+2k)(p+2k-1)}} \prod_{i=1}^p\frac1{(2i-1)!}\prod_{i=1}^{p+2k}\frac1{(2i-1)!} \det M^{(6)}.
\ee
The matrix $\det M^{(6)}$ can equivalently be written as
\be
\label{eq:M6v2}
M^{(6)}_{ij} = 
\left\{\begin{array}{cl}
(v_j-2)^{2i-1}\quad& i \le p, \\[0.2cm]
v_j^{2i-2p-1}\quad & i > p,
\end{array}
\right.\qquad \textrm{with} \qquad
v_j = 
\left\{\begin{array}{cl}
-u_j+2\quad& j \le p+k, \\[0.15cm]
u_{j-p-k}\quad & j > p+k.
\end{array}
\right.
\ee
We expand $(v_j-2)^{2i-1}$ in powers of $v_j$ and take linear combinations of rows to obtain a matrix $M^{(7)}$ with only monomial entries in the $v_j$:
\be
M^{(7)}_{ij} = 
\left\{\begin{array}{cl}
v_j^{2(i-1)}\quad& i \le p, \\[0.2cm]
v_j^{2i-2p-1}\quad & i > p.
\end{array}
\right.
\ee
In terms of $M^{(7)}$, $\Gamma^{0}_{k,p}$ reads
\be
\label{eq:CM7}
\Gamma^{0}_{k,p} = \frac{1}{2^{2p+k}}\frac{\prod_{i=1}^p(4i-2)}{2^{p(p-1)+(p+2k)(p+2k-1)}} \prod_{i=1}^p\frac1{(2i-1)!}\prod_{i=1}^{p+2k}\frac1{(2i-1)!} \det M^{(7)}.
\ee

\subsection{Schur polynomials and a product formula}\label{sec:properties.of.Schurs}

Combining \eqref{eq:C0Gamma0} and \eqref{eq:CM7} yields an expression for $C^0_{2(p+k),2k}$ in terms of the determinant of $M^{(7)}$. This determinant can be written in terms of a Schur polynomial $s_\lambda(\vec x)$. We give some properties of these polynomials in \cref{sec:Schur.eval}. In particular, we recall that $s_\lambda (\vec x)$ admits an expression as a ratio of determinants, see \eqref{eq:Schur.dets}. Up to a reordering of the rows, $M^{(7)}$ has the same form as the matrix appearing in the numerator in this expression, for the staircase partition $\lambda = (2k,2k-1, \dots, 1, 0^{2p})$. Therefore, we have
\be
\label{eq:detM7}
\det M^{(7)} = (-1)^{p(p-1)/2}s_{(2k,2k-1, \dots, 1, 0^{2p})}(\vec v) \hspace{-0.3cm} \prod_{1\le i<j\le2p+2k} \hspace{-0.3cm} (v_j - v_i),
\ee
where the multiplicative factor is obtained from the Vandermonde determinant formula. The overall sign stems from the reordering of the rows.

We need to evaluate this Schur polynomial for the values of $v_j$ given in \eqref{eq:M6v2}. We note that 
\be 
\label{eq:vj.asymm}
v_{p+k+1-j} = -v_j \qquad j = 1, \dots, p+k.
\ee
As a consequence, for $j = 1, \dots, p+k$, either $v_j=0$, or else there is a pair $(v_j,-v_j)$ of opposite sign variables. In both cases, the Schur polynomial reduces to another such polynomial with fewer variables. We have the following Lemma:
\begin{Lemma}
  \label[Lemma]{lem:schur.reduction}
The Schur polynomials $s_{(2k,2k-1, \dots, 1, 0^{2p})}(\vec x)$ satisfy the following reductions:
\begin{subequations}
\begin{alignat}{2}
s_{(2k,2k-1, \dots, 1, 0^{2p})}(\vec x)\big|_{x_1 = 0} &= s_{(2k,2k-1, \dots, 1, 0^{2p-1})}(\vec x'),
\\[0.15cm]
s_{(2k,2k-1, \dots, 1, 0^{2p})}(\vec x)\big|_{x_2 = -x_1} &= s_{(2k,2k-1, \dots, 1, 0^{2p-2})}(\vec x''),\label{eq:schur.reduction.b}
\end{alignat}
\end{subequations}
where $\vec x' = \{x_2, \dots, x_{2(p+k)}\}$ and $\vec x'' = \{x_3, \dots, x_{2(p+k)}\}$.
\end{Lemma}
\proof
In the case $x_1 = 0$, the result follows directly from the determinant form \eqref{eq:Schur.dets}. 
For the case $x_2 = -x_1$, we note that both the numerator and the denominator of $s_{(2k,2k-1, \dots, 1, 0^{2p})}(\vec x)\big|_{x_2 = -x_1}$ in \eqref{eq:Schur.dets} are polynomials in $x_1$. By studying their first two columns, we find that the degree of these polynomials are identical and equal to $4p+4k-3$. Clearly, both are zero for $x_1 = 0$ and $x_1 = \pm x_j$ with $j = 2, \dots, 2p+2k$. This implies that both polynomials have a factor $x_1 \prod_{j=3}^{2(p+k)}(x_1^2-x_j^2)$, which precisely has the degree $4p+4k-3$. Therefore, the ratio of the numerator and denominator is independent of $x_1$. By evaluating the limit $x_1 \to \infty$, we find that this ratio precisely equals the right-hand side of \eqref{eq:schur.reduction.b}.
\eproof

Because the Schur polynomials are symmetric in the variables $x_i$, \cref{lem:schur.reduction} generalises to the case where $x_i = 0$ and $x_i =- x_j$ for other values of $i$ and $j$. Applying this reduction successively for each of the variables $v_1, \dots, v_{p+k}$, we find that
\be
s_{(2k,2k-1, \dots, 1, 0^{2p})}(\vec x)\big|_{x_j = v_j} = s_{(2k,2k-1, \dots, 1, 0^{p-k})} (\vec x)\big|_{x_j = v_{j+p+k}}.
\ee
The polynomial on the right side is therefore symmetric in the variables 
\be
 v_{p+k+j} = u_j, \qquad j = 1, \dots, p+k.
\ee

The final step of the calculation uses the evaluation of the Schur polynomials for equally spaced variables that is given in \cref{sec:Schur.eval}. Indeed, the numbers $u_j$ are equally spaced and of the form \eqref{eq:xj.values} with 
\be
a = 2k,\qquad b = p-k, \qquad \alpha = 3p+3k+5, \qquad \beta=-6.
\ee
\cref{sec:schur.prop} yields
\be\label{eq:sefinal}
s_{(2k,2k-1, \dots, 1,0^{p-k})}(\vec u)  = \frac1{2^{2k}} \prod_{1 \le i \le j\le 2k}\hspace{-0.2cm}(12k-6i-6j+10) \prod_{i=1}^{2k}\prod_{j=1}^{p-k} \frac{2i+j-1}{i+j-1}.
\ee
By combining \eqref{eq:CM7}, \eqref{eq:detM7} and \eqref{eq:sefinal}, we obtain an explicit formula for $C^{0}_{2(p+k),2k}$ in product form. For $q = \eE^{2 \pi \ir/3}$, the result is written in terms of the Barnes $G$-function and reads
\begin{alignat}{2}
C^{0}_{2(p+k),2k} &=  \frac{(\eE^{\ir\pi/6})^{p-k}3^{(3p^2 + p + 9 k^2 + 2k + 6 pk)/2}}{\pi^{k+1/2} 2^{2p^2 + 6k^2-k/3+2pk}} \frac{G(p+k+\frac23)G(p+k+1)G(p+k+\frac43)G(p-k+1)}{G(p+2k+1)G(p+2k+\frac32)G(p+\frac12)G(p+1)}\nonumber\\[0.2cm]
&
\times\frac{G(\frac{p+3k+2}2)G(\frac{p+3k+3}2)}{G(\frac{p-k+2}2)G(\frac{p-k+3}2)}\frac{G(2k+\frac13)}{G(2k+\frac32)}
\frac{G(k+\frac43)G(k+\frac{11}6)}{G(k+\frac16)G(k+\frac23)} \frac{G(\frac16)G(\frac32)^3}{G(\frac13)G(\frac43)^2G(\frac{11}6)}.\label{eq:cefinal}
\end{alignat}
As shown in \cref{app:Barnes}, this is indeed equal to the expression for $C^{0}_{2(p+k),2k}$ given in \eqref{eq:C.results.even}.

\section{Conclusion}
\label{sec:Conclusion}

In this paper, we introduced a new family of correlation functions BEFP$^\mu_{N,m}$ for the XXZ spin chain and the six-vertex model. We computed them at the combinatorial point $\Delta = -\frac 12$ in terms of simple products and ratios of integers. Our computation exploits the tie with the inhomogeneous six-vertex model and the special polynomial solution of the quantum Knizhnik-Zamolodchikov equations. Using the symmetries of this solution, we derived determinant formulas for certain scalar products in the inhomogeneous six-vertex model. The product formulas for BEFP$^\mu_{N,m}$ were obtained from the homogeneous limit of these scalar products using the method of finite differences as well as certain evaluations of Schur polynomials.

We now discuss the asymptotic behavior of the boundary emptiness formation probability. There is a known relation between the emptiness formation probabilities and the counting of alternating sign matrices and plane partitions. Indeed, Kitanine {\it et al.}~\cite{KMST02} computed the emptiness formation probability EFP$_{N,m}$ for $\Delta = -\frac12$ in the limit $N\to \infty$ (with $m$ kept finite) and obtained the simple result $2^{-m^2}A(m)$, where $A(m)$ is the number of alternating sign matrices of size $m$. In \cite{C12}, the same emptiness probability was computed for finite $N$, and it was observed that the corresponding numbers are related to the combinatorics of cyclically symmetric self-complementary plane partition on a $k$-punctured hexagon. For the boundary emptiness formation probabilities, we observe that
\be
\hspace{-0.4cm}\lim_{N \to \infty} \textrm{BEFP}^\mu_{N,m}
= \left\{
\begin{array}{lll}
\displaystyle \frac 1{2^{m(m+1)/2} 3^{(m-1)/2}} \frac{A(m)}{A_V(m)} &\quad \mu = +, -, & \quad m\textrm{\ odd,} \\[0.5cm]
\displaystyle \frac 1{2^{m^2/2} 3^{m/2}} \frac{A(m)}{N_8(m)}\prod_{j=0}^{(m-2)/2}\frac{3j+1}{6j+1} &\quad \mu = +, -,&\quad m\textrm{\ even,}\\[0.5cm]
\displaystyle \frac {\eE^{-\ir \pi m/6}}{2^{(m^2-1)/2} 3^{m/2}} \frac{A(m)}{A_V(m)} \prod_{j=0}^{(m-1)/2}\frac{3j+1}{6j+1} &\quad \mu = 0, &\quad m\textrm{\ odd},\\[0.5cm]
\displaystyle \frac {\eE^{-\ir \pi m/6}}{2^{m(m+1)/2} 3^{m/2}} \frac{A(m)}{N_8(m)}\!\prod_{j=0}^{(m-2)/2}\frac{(3j+1)(6j+5)}{(3j+2)(6j+1)} &\quad \mu = 0, & \quad m\textrm{\ even},
\end{array}\right.
\ee
where 
\begin{subequations}
\begin{alignat}{2}
A(m) &= \prod_{i=0}^{m-1} \frac{(3i+1)!}{(m+i)!},\hspace{2.5cm} A_V(m) = \prod_{i=1}^{m} (3i-1) \frac{(2i-1)!(6i-3)!}{(4i-2)!(4i-1)!}, \\ 
N_8(m) &= \prod_{i=1}^{m-1} (3i+1) \frac{(2i)!(6i)!}{(4i)!(4i+1)!}.
\end{alignat}
\end{subequations}
Here, $A_V(m)$ is the number of vertically-symmetric alternating sign matrices of size $2m+1$ and $N_8(m)$ is the number of cyclically symmetric transpose complement plane partitions of size $2m$. It would be interesting to understand if this coincidence of numbers hides a deeper connection between the spin-chain correlation functions and the combinatorics of alternating sign matrices and plane partitions.

One can consider asymptotic expansions as $N \to \infty$ in two different ways. Let us recall the parameterisation $N=2n$ for even $N$ and $N = 2n+1$ for odd $N$. First, in the scaling limit, the integer $n$ is sent to infinity with the ratio $x = \frac mn$ kept constant in the range $(0,1)$. The logarithm of $C^\mu_{N,nx}$ has a large-$n$ expansion of the form 
\be
\label{eq:C.expansion}
\log |C^\mu_{N,nx}| = n^2 f_{-2} + n \, f^\mu_{-1} + \log n\, f_{\log} + f^\mu_0 + \dots\ \ .
\ee
We compute the coefficients in this expansion using the expressions \eqref{eq:C.Barnes} and the large-$z$ expansion of the Barnes $G$-function:
\begin{subequations}
\begin{alignat}{2}
f_{-2} &= \frac{1}4 \Big(6 \log 3 + x^2(3\log 3 - 4 \log 2) + (1-x)^2 \log(1-x)\nonumber\\ 
&\hspace{0.7cm} - (2-x)^2 \log (2-x) + (1+x)^2 \log(1+x)- (2+x)^2 \log(2 + x)\Big),
\\[0.15cm]
f^\mu_{-1} &= \frac14 \Big((3 \tau^\mu_1 + 5) \log 3 + x (\log 3 - 2 \log 2) +  \tau^\mu_2 (1 - x) \log(1 - x) - \tau^\mu_1 (2 - x) \log(2 - x) \nonumber\\ 
&\hspace{0.7cm}+ (\tau^\mu_2 + 2) (1 + x) \log(1 + x) - (\tau^\mu_1 + 2) (2 + x) \log(2 + x)\Big),
\\[0.15cm]
f^\mu_0 &= \frac{1}{72} \Big(3 + (18 \tau^\mu_1 + 15) \log 3 - 15 \log 2 - 36 \log A - 3 \log(1 - x) + 3 \log(2 - x) 
\nonumber\\ 
&\hspace{0.7cm}
+ (1 - 6 \tau^\mu_1) \log x + (18 \tau^\mu_2 + 15) \log(1 + x) - (18 \tau^\mu_1 + 15) \log(2 + x)\Big),
\\[0.15cm]
f_{\log} &= - \frac 1 {24}, \qquad (\tau_1^+, \tau_1^-, \tau_1^{0}) = (1,1,-1), \qquad (\tau_2^+, \tau_2^-, \tau_2^{0}) = (1,-1,-1).
\end{alignat}
\end{subequations}
Only the surface and constant terms depend on $\mu$, with the dependence simply encoded in the numbers $\tau^\mu_1$ and $\tau^\mu_2$. 

Second, one can also compute an asymptotic expansion for the boundary emptiness formation probability by sending $n \to \infty$ first, and then by considering $m$ large. The corresponding expansion is 
\be
\label{eq:BEFP.g}
\lim_{N\to \infty} \log \big|\textrm{BEFP}^\mu_{N,m}\big| = m^2 g_{-2} + m\, g_{-1} + \log m\, g^\mu_{\log} + g^\mu_0 + \dots\ \ .
\ee
The coefficients in this expansion are given by
\begin{subequations}
\begin{alignat}{2}
g_{-2} &= \frac34(\log 3 - 2 \log 2), \qquad g_{-1} = \frac14 \log 3- \log 2, \qquad 
g^{\mu}_{\textrm{log}} = \left\{\begin{array}{cl}
-\frac{5}{72} \ \ & \mu = +, -,\\[0.15cm]
\frac{7}{72} \ \ & \mu = 0,
\end{array}\right.
\\[0.2cm]
g_0^{+,-} &= -\frac1{24} \log 2 + \frac{11}{72} \log 3 + \frac 16 \log \pi +\frac 1{72} - \frac16 \log A -\frac 13 \log \Gamma(1/3),
\\[0.2cm]
g_0^{0} &= -\frac{13}{24} \log 2 + \frac{11}{72} \log 3 - \frac 16 \log \pi +\frac 1{72} - \frac16 \log A +\frac 23 \log \Gamma(1/3).
\end{alignat}
\end{subequations}
In comparison, a similar expansion for EFP$^\mu_{N,m}$ was computed for $\Delta = 0$ and $\Delta = 1$ in \cite{STN01} and \cite{KMST02} respectively. General formulas for the leading coefficients were conjectured in \cite{KLNS03}. A similar conjecture for BEFP$^\mu_{N,m}$ is currently unknown.

The terms proportional to $n^2$ and $m^2$ in \eqref{eq:C.expansion} and \eqref{eq:BEFP.g} reveal that the presence of a polarised segment on the boundary affects the behaviour of the model in a two-dimensional area in the neighborhood of this segment. In this sense, the coefficients $f_{-2}$ and $g_{-2}$ are bulk free energies per site. The emptiness formation probability has a similar behavior, with the numerical simulations performed in \cite{S14} revealing the presence of a frozen region surrounding the slit. Its exact shape is currently unknown. It would be interesting to perform similar simulations for the boundary emptiness formation probability. Polarised segments or slits are non-conformal boundary conditions in the scaling limit and are responsible for the appearance of a phase separation. For the case at hand, it would be interesting to understand which terms of the expansions \eqref{eq:C.expansion} and \eqref{eq:BEFP.g} are universal and whether the numbers $\tau^\mu_1$ and $\tau^\mu_2$ are related to certain conformal dimensions of an underlying conformal field theory.

\subsection*{Acknowledgments}

AMD was supported by an FNRS Postdoctoral Researcher under the project CR28075116. This work was supported by the Belgian Excellence of Science (EOS) initiative through the project 30889451 PRIMA -- Partners in Research on Integrable Systems and Applications. CH is grateful for support and hospitality from the University of Cergy-Pontoise where early stages of this work were done. LC acknowledges the support and hospitality from the Universit\'e catholique de Louvain where later stages of this work were done.

%
%

\bigskip
\bigskip
\appendix
%

%
\section{Evaluations of Schur polynomials}\label{sec:Schur.eval}
%

In this section, we find simple product expressions for a family of Schur polynomials $s_\lambda(\vec x)$ that come up in our computation of the overlaps $C^\mu_{N,m}$ in \cref{sec:homogeneous} and \cref{app:five.limits}. We consider partitions~$\lambda$ in the form of staircases, $\lambda = (a, a-1, \dots, 1, 0^b)$, where the variables $x_i$ are of the form
\be
\label{eq:xj.values}
x_i = \alpha + \beta i.
\ee
The final result is given in \cref{sec:schur.prop}.

Let us recall that, given a partition $\lambda = (\lambda_1, \lambda_2, \dots \lambda_\ell)$ with $\lambda_1 \ge \lambda_2 \ge \dots \ge \lambda_\ell$, the Schur polynomial $s_\lambda(\vec x)$ in the variables $x_1, \dots, x_\ell$ is the symmetric polynomial defined as 
\be
\label{eq:Schur.dets}
s_\lambda(\vec x) = \det 
\begin{pmatrix}
x_1^{\lambda_1+\ell-1} & x_2^{\lambda_1+\ell-1} & \cdots & x_\ell^{\lambda_1+\ell-1} \\
x_1^{\lambda_2+\ell-2} & x_2^{\lambda_2+\ell-2} & \cdots & x_\ell^{\lambda_2+\ell-2} \\
\vdots & \vdots & \ddots
 & \vdots \\
x_1^{\lambda_\ell} & x_2^{\lambda_\ell} & \cdots & x_\ell^{\lambda_\ell}
\end{pmatrix}
\Bigg/
\det
\begin{pmatrix}
x_1^{\ell-1} & x_2^{\ell-1} & \cdots & x_\ell^{\ell-1} \\
x_1^{\ell-2} & x_2^{\ell-2} & \cdots & x_\ell^{\ell-2} \\
\vdots & \vdots & \ddots
 & \vdots \\
1 & 1 & \cdots & 1 
\end{pmatrix}.
\ee

The combined degree of $s_\lambda(\vec x)$ in the $x_i$ equals the sum of the elements of the partition $\lambda$. For $\lambda = (a, a-1, \dots, 1, 0^b)$, this degree is $a(a+1)/2$. Clearly, for $x_i$ of the form \eqref{eq:xj.values}, $s_\lambda(\vec x)$ is a polynomial in $\alpha$. Because each $x_i$ is linear in $\alpha$, the polynomial degree of $s_{(a, a-1, \dots, 1, 0^b)}(\vec x)$ in $\alpha$ is also $a(a+1)/2$. It has precisely this number of zeros, and we will find all of them explicitly below. Our investigation starts with the following lemma.
\begin{Lemma}\label[Lemma]{sec:order.of.zeros}
Let $M(\alpha)$ be a matrix with polynomial entries in $\alpha$. Suppose that there exists a value $\alpha=\alpha_0$ such that $\det M(\alpha_0)=0$ and that the eigenspace for the eigenvalue zero has dimension $\ell \ge 0$. Then
\be
\det M(\alpha) = (\alpha-\alpha_0)^\ell \times g(\alpha)
\ee
where $g(\alpha)$ is a polynomial in $\alpha$.
\end{Lemma}
\proof
The first step is to construct a complete basis where the first $\ell$ states are the right-eigenstates of $M(\alpha_0)$ of eigenvalue zero. The remaining states can be chosen arbitrarily, as long as they complete the basis and are independent of $\alpha$. Let $W$ be the matrix whose columns are the states of this basis. Clearly $W$ is independent of $\alpha$, and we have
\be
\det M(\alpha) = \frac{\det \big(M(\alpha)W \big)}{\det W}.
\ee
The matrix entries of $M(\alpha)W$ are polynomials in $\alpha$. At $\alpha=\alpha_0$, the first $\ell$ columns vanish. As a result, each of the matrix entries in the first $\ell$ columns of $M(\alpha)W$ has a factor of $(\alpha-\alpha_0)$ that can be factorised from the determinant.
\eproof

For $s_{(a,a-1, \dots, 1,0^{b})}(\vec x)$, we consider the expression \eqref{eq:Schur.dets} and employ the previous lemma for the matrix appearing in the determinant of the numerator, which we denote by $M(\alpha)$. Its entries are indeed polynomial in $\alpha$. If $x_i + x_j = 0$ with $j-i\ge b$, the determinant of $M(\alpha)$ vanishes. Indeed, the vector $w^{i,j}$ with components
\begin{subequations}
\begin{alignat}{3}
[w^{i,j}]_k &= \left\{\begin{array}{cl}
(-1)^k \begin{pmatrix}j-i\\k-i\end{pmatrix} \quad& \textrm{for } k = i, \dots, j,\\[0.35cm]
0 & \textrm{otherwise,}
\end{array}\right.\qquad &j-i-b \textrm{ even,}\\[0.25cm]
[w^{i,j}]_k &= \left\{\begin{array}{cl}
(-1)^k \bigg[\begin{pmatrix}j-i-1\\k-i\end{pmatrix} - \begin{pmatrix}j-i-1\\k-i-1\end{pmatrix}\bigg] \quad&\textrm{for } k = i, \dots, j,\\[0.35cm]
0 & \textrm{otherwise,}
\end{array}\right.\qquad &j-i-b \textrm{ odd,}
\end{alignat}
\end{subequations}
is a right-eigenvector for $M(\alpha)$ of vanishing eigenvalue. The proof of this statement is straightforward. It relies on the fact that the variables $x_i, \dots, x_j$ are equally spaced and on the identity \cite[Corollary~2]{R96}
\be
\sum_{k=0}^j (-1)^k \begin{pmatrix}j \\ k\end{pmatrix} (x-k)^\ell = 0, \qquad \ell = 0, \dots, j-1,
\ee
which holds for arbitrary $x$.

The existence of this eigenvector of vanishing eigenvalue implies that $\det M(\alpha) = (x_i+ x_j) \times g(\alpha)$ for some polynomial $g(\alpha)$. This is true for all $i,j \in \{1, \dots, a+b\}$ satisfying $j-i\ge b$. For two pairs $(i,j)$ and $(i',j')$, the zero is the same if and only if $i' = i - \ell$ and $j' = j + \ell$ for some integer $\ell$. This integer is constrained to the values for which $i',j' \in \{1, \dots, a+b\}$ and $j'-i'\ge b$. For a given pair $(i,j)$, we denote by $L$ the set of such values for $\ell$. The number of non-zero entries of $w^{i-\ell,j+\ell}$ grows linearly with $\ell$, and it is easy to see that these states, for $\ell \in L$, form an independent set. From \cref{sec:order.of.zeros}, we find that the matrix $M(\alpha)$ is proportional to $(x_i+x_j)^{|L|}$, which can alternatively be written as $\prod_{\ell \in L}(x_{i-\ell}+x_{j+\ell})$. This is true for all pairs $(i,j)$ satisfying the above constraints.

In constrast to $M(\alpha)$, the determinant in the denominator of \eqref{eq:Schur.dets} does not depend on $\alpha$. Indeed, this Vandermonde determinant only depends on the differences $x_i - x_j = \beta(i-j)$. We therefore find that 
\be
\label{eq:s.alpha1}
s_{(a,a-1, \dots, 1,0^{b})}(\vec x) = K\, \prod_{i=1}^a\prod_{j = i+b}^{a+b} (x_i + x_j)
\ee
where $K$ is a prefactor that does not depend on $\alpha$. Indeed, each factor $(x_i + x_j)$ is linear in $\alpha$, and the total degree in $\alpha$ of the product in \eqref{eq:s.alpha1} is $a(a+1)/2$, which exhausts the polynomial degree of $s_{(a,a-1, \dots, 1,0^{b})}(\vec x)$. To fix $K$, we note that
\be
\lim_{\alpha\to\infty} \frac{s_{(a,a-1, \dots, 1,0^b)}(\vec x)}{\alpha^{a(a+1)/2}} = s_{(a,a-1, \dots, 1,0^b)}(\vec 1).
\ee
The right-hand side is obtained from the known evaluation of the Schur polynomials for the case where all the variables $x_i$ are set to $1$:
\be 
s_\lambda(\vec 1) = \prod_{1 \le i < j \le \ell} \frac{\lambda_i - \lambda_j + j-i}{j-i}.
\ee
After solving for $K$, we obtain the following proposition.
\begin{Proposition} \label[Proposition]{sec:schur.prop}
Let $x_i = \alpha+\beta i$. For $a,b \ge 0$, the Schur polynomial $s_{(a, a-1, \dots, 1, 0^b)}(\vec x)$ evaluates to
\be
s_{(a, a-1, \dots, 1, 0^b)}(\vec x) = \frac1{2^a} \prod_{i=1}^{a}\prod_{j = i+b}^{a+b} (x_i + x_j) \times \prod_{k = 1}^{a}\prod_{\ell=1}^{b}\frac{2k+\ell-1}{k+\ell-1}.
\ee
\end{Proposition}

\section{Expressions for the overlaps in terms of the Barnes $\boldsymbol{G}$-function}\label{app:Barnes}

The overlaps \eqref{eq:C.results} are expressed in terms of the Barnes $G$-function as follows:
\begin{subequations}
\label{eq:C.Barnes}
\begin{alignat}{2}
C^{0}_{2n,m} &= (\eE^{\ir \pi/6})^{n-m}\frac{2^{m(m+1)+n(m-1)}3^{\frac12(n-m)(3n+3m+1)}}{\pi^{n-m}}\frac{G(n+m+1)G(n+\frac23)G(n+1)G(n+\frac43)}{G(2n+m+1)}\nonumber\\
&\times\frac{G(n+\frac m2 + \frac 12)G(n + \frac m2 + 1)}{G(n-\frac m2 + \frac 12)G(n - \frac m2 + 1)}
\frac{G(\frac n2 - \frac m2 + \frac 12)G(\frac n2 - \frac m2 + 1)}{G(\frac n2 + \frac m2 + \frac 12)G(\frac n2 + \frac m2 + 1)}\label{eq:final.C.even}\\
&\times\frac{1}{G(m+\frac23)G(m+1)G(m+\frac43)G(m+\frac32)}
\frac{G(\frac{3m}2+1)G(\frac{3m}2 + \frac32)}{G(\frac m2 + 1)G(\frac m2 + \frac 32)}\,G(\tfrac32),\nonumber\\[0.5cm]
C^{+}_{2n+1,m} &= \frac{2^{m(m+2)+n(m-1)}3^{\frac12(n-m)(3n+3m+4)}}{\pi^{n-m}}\frac{G(n+m+2)G(n+1)G(n+\frac43)G(n+\frac53)}{G(2n+m+2)}\nonumber\\
&\times\frac{G(n+\frac m2 + 1)G(n + \frac m2 + \frac32)}{G(n-\frac m2 + 1)G(n - \frac m2 + \frac32)}
\frac{G(\frac n2 - \frac m2 + 1)G(\frac n2 - \frac m2 + \frac 32)}{G(\frac n2 + \frac m2 + 1)G(\frac n2 + \frac m2 + \frac 32)}
\\
&\times\frac{1}{G(m+1)G(m+\frac43)G(m+\frac53)G(m+\frac32)}
\frac{G(\frac{3m}2 + \frac32)G(\frac{3m}2+2)}{G(\frac m2 + \frac 32)G(\frac m2 + 2)}\,G(\tfrac32),\nonumber\\[0.5cm]
C^{-}_{2n+1,m} &= \frac{2^{m(m+3)+n(m-1)}3^{\frac12(n-m)(3n+3m+4)}}{\pi^{n-m}}\frac{G(n+m+1)G(n+\frac43)G(n+\frac53)G(n+2)}{G(2n+m+2)}\nonumber\\
&\times\frac{G(n+\frac m2 + 1)G(n + \frac m2 + \frac32)}{G(n-\frac m2 + 1)G(n - \frac m2 + \frac32)}
\frac{G(\frac n2 - \frac m2 + \frac 12)G(\frac n2 - \frac m2 + 1)}{G(\frac n2 + \frac m2 + \frac 12)G(\frac n2 + \frac m2 + 1)}
\label{eq:final.C.down.odd}
\\
&\times\frac{1}{G(m+1)G(m+\frac43)G(m+\frac53)G(m+\frac32)}
\frac{G(\frac{3m}2 + \frac32)G(\frac{3m}2+2)}{G(\frac m2 + \frac 32)G(\frac m2 + 2)}\,G(\tfrac32).\nonumber
\end{alignat}
\end{subequations}

These expressions are different yet equivalent to those obtained via our derivations in \cref{sec:homogeneous} and \cref{app:five.limits}. We prove this in one case, namely for even $N$ and even $m$. The arguments are identical for the other cases. Let us consider the ratio $\mathcal J^{0}_{k,p}$ of the right sides of \eqref{eq:cefinal} and \eqref{eq:final.C.even} for $n= p+k$ and $m = 2k$. We want to show that this ratio equals one. We find
\begin{alignat}{2}
\mathcal J^{0}_{k,p} &=  \frac{3^{9 k^2+\frac{3 k}{2}} \pi ^{p-2 k-\frac12}}{2^{2 p^2 - p + 12 k^2 + \frac{2 k}{3} + 4 k p}}
\frac{G(\frac{1}{6})G(\frac32)^2}{G(\frac{1}{3}) G(\frac{4}{3})^2G(\frac{11}{6})}
\frac{G(2p+4k+1)}{G(p+2k+\frac12)G(p+2k+1)^2 G(p+2k+\frac32)}
\nonumber\\[0.2cm]
&\times \frac{G(p-k+1)}{G(\frac{p-k+1}2)G(\frac{p-k+2}2)^2G(\frac{p-k+3}2)}
\frac{G(\frac{p+3k+1}2)G(\frac{p+3k+2}2)^2G(\frac{p+3k+3}2)}{G(p+3k+1)}
\nonumber\\[0.2cm]
&\times \frac{G(2k+\frac13)G(2k+\frac23)G(2k+1)G(2k+\frac43)}{G(3k+1)G(3k+\frac32)}\frac{G(k+1)G(k+\frac32)G(k+\frac43)G(k+\frac{11}6)}{G(k+\frac16)G(k+\frac23)}.
\end{alignat}
This expression is simplified using the multiplication formulas for the Barnes $G$-function:
\begin{subequations}
\begin{alignat}{2}
G(2 z) &= \frac{2^{2z(z-1)}}{(2 \pi)^{z-1}}\frac{G(z)G(z+\frac12)^2G(z+1)}{G(\frac32)^2}, 
\\ G(3z) &=  \frac{3^{\frac32(3z+1)(z-1)}}{(2 \pi)^{3(z-1)}} \frac{G(z)G(z+\frac13)^2G(z+\frac23)^3G(z+1)^2G(z+\frac43)}{G(\frac43)^2G(\frac53)^3G(\frac73)}.
\end{alignat}
\end{subequations}
Using these formulas, we find that the dependence on $p$ and $k$ disappears:
\be
\mathcal J^{0}_{k,p} = \frac{3^{\frac{31}8}}{2^{\frac{17}6}\pi^{\frac76}} \frac{G(\frac16)G(\frac43)^2 G(\frac53)^6 G(\frac73)^2}{G(\frac13)G(\frac{11}6)G(\frac32)^8} = \frac{2^{\frac{17}6}\pi^{\frac56}}{3^{\frac98}} \frac{1}{\Gamma(\frac 56)}\frac{G(\frac13)^3G(\frac23)^6G(\frac16)}{G(\frac12)^8G(\frac56)}.
\ee
At the last equality, we used the identities
\be
G(z+1) = G(z)\Gamma (z), \qquad \Gamma(z)\Gamma(1-z) = \frac \pi{\sin(\pi z)}.
\ee
The arguments of the remaining $G(z)$ functions are in the range $(0,1)$. There are known expressions for these specialisations \cite{A01}:
\begin{subequations}
\begin{alignat}{2}
\log G(\tfrac 12) &= \frac{\log 2}{24} - \frac{\log \pi}4 - \frac{3 \log A}2 + \frac 18,\\[0.15cm]
\log G(\tfrac 13) &= \frac{\log 3}{72} + \frac{\pi}{18 \sqrt 3} - \frac23 \log \Gamma(\tfrac13) - \frac{4 \log A}3 - \frac{\psi^{(1)}(\frac13)}{12 \pi \sqrt 3}+ \frac 19,\\[0.15cm]
\log G(\tfrac 23) &= \frac{\log 3}{72} + \frac{\pi}{18 \sqrt 3} - \frac13 \log \Gamma(\tfrac23) - \frac{4 \log A}3 - \frac{\psi^{(1)}(\frac23)}{12 \pi \sqrt 3}+ \frac 19,\\[0.15cm]
\log G(\tfrac 16) &= \frac{\log 12}{144} + \frac{\pi}{20 \sqrt 3} - \frac56 \log \Gamma(\tfrac16) - \frac{5 \log A}6 - \frac{\psi^{(1)}(\frac16)}{40 \pi \sqrt 3}+ \frac 5{72},\\[0.15cm]
\log G(\tfrac 56) &= \frac{\log 12}{144} + \frac{\pi}{20 \sqrt 3} - \frac16 \log \Gamma(\tfrac56) - \frac{5 \log A}6 - \frac{\psi^{(1)}(\frac56)}{40 \pi \sqrt 3}+ \frac 5{72},
\end{alignat}
\end{subequations}
where $A$ is the Glaisher-Kinkelin constant and $\psi^{(1)}(z) = \frac{d^2}{dz^2}\ln\Gamma(z)$
 is the polygamma function. This yields
\be\label{eq:Jexp}
\mathcal J^{0}_{k,p} = \exp\bigg[\frac1{40 \sqrt 3 \pi}\Big(20 \pi^2 - \psi^{(1)}(\tfrac16) - 10 \psi^{(1)}(\tfrac13) - 20 \psi^{(1)}(\tfrac23) + \psi^{(1)}(\tfrac56)\Big)\bigg].
\ee
Using two properties of the polygamma function,
\be
\psi^{(1)}(z) + \psi^{(1)}(1-z) = \frac{\pi^2}{\sin^2(\pi z)}, \qquad \psi^{(1)}(2z) = \frac14 \big(\psi^{(1)}(z) + \psi^{(1)}(z+ \tfrac12)\big),
\ee
we find that the argument of the exponential in \eqref{eq:Jexp} vanishes, confirming that $\mathcal J^{0}_{k,p} = 1$.

\section{The homogeneous limit of the scalar products for the other cases}
\label{app:five.limits}

In this section, we evaluate the homogeneous limits of the five remaining cases for $q = \eE^{2 \pi \ir/3}$, starting from the determinant formula \eqref{eq:Rkp} for $S^{0}_{k,p}$. We proceed by taking the limits in \eqref{eqn:HomLimitSe}, using the relations between the scalar products given in \cref{prop:SSbar,prop:SeSdown}, and following the strategy illustrated in \cref{fig:RelationsSPs}. After simplification, we find
\begin{subequations}
\begin{alignat}{2}
C^{0}_{2(p+k-1),2k-1} &= (-q)^{p+k-1} \lim_{x,z\to 1}\lim_{z_{2k}\to q^{-2} x_1} S^{0}_{k,p},
\label{eq:Cekodd}\\[0.15cm]
C^{-}_{2(p+k-1)+1,2k-1} &= -\frac{(q-q^{-1})^{p+k-1}}{q^{p+1}} \lim_{x,z\to 1}\lim_{z_{2k}\to \infty} \frac{S^{0}_{k,p}}{z_{2k}^{p+k-1}}
\label{eq:Cdkodd},\\[0.15cm]
C^{-}_{2(p+k-2)+1,2k-2} &= (q-q^{-1})^{p+k-1}(-1)^{p+k}q^{k+1} \lim_{x,z\to 1}\lim_{z_{2k-1}\to \infty}\lim_{z_{2k}\to q^{-2}x_1} \frac{S^{0}_{k,p}}{z_{2k-1}^{p+k-1}}
\label{eq:Cdkeven},\\[0.15cm]
C^{+}_{2(p+k-2)+1,2k-1} &= (q-q^{-1})^{3(p+k-1)}(-q)^{p-k}\lim_{x,z\to 1}\lim_{x_1\to \infty}\lim_{z_{2k}\to q^{-2}x_1} \frac{S^{0}_{k,p}}{x_{1}^{2p+2k-1}},
\label{eq:Cukodd}\\[0.15cm]
C^{+}_{2(p+k-3)+1,2k-2} &= (q-q^{-1})^{3(p+k-1)}q^{1-p}\lim_{x,z\to 1}\lim_{z_{2k-1}\to q^{-2}x_2}\lim_{x_1\to \infty}\lim_{z_{2k}\to q^{-2}x_1} \frac{S^{0}_{k,p}}{x_{1}^{2p+2k-1}}.
\label{eq:Cukeven}
\end{alignat}
\end{subequations}

\subsection[The homogeneous limit of $S^{-}$]{The homogeneous limit of $\boldsymbol{S^{-}}$}\label{sec:Sd}

We compute $C^{-}_{N,m}$ for $m$ odd, starting from \eqref{eq:Cdkodd} and the determinant expression \eqref{eq:Rkp}. Simplifying the prefactors, we find
\begin{subequations}
\begin{alignat}{2}
&C^{\rm -}_{2(p+k-1)+1,2k-1} = \frac{(-1)^{p(p+1)/2+k}2^{k-1}}{3^{p(p-3)/2-k(k-1)/2+kp}} \Gamma^{-}_{k,p},\\
&\Gamma^{-}_{k,p}  = \lim_{x,z \to 1}\lim_{z_{2k}\to \infty} \frac1{z^{3(p+k-1)}_{2k}}\frac{\det M^{(k,p)}} {C(x_1, \dots, x_p)C(z_1, \dots, z_{2k-1},x_1, \dots, x_p)}.
\end{alignat}
\end{subequations}
We insert the factor of $z^{-3(p+k-1)}_{2k}$ inside the determinant in the last row of the matrix and take the limit $z_{2k} \to \infty$ of each matrix element of this row. The result is zero except on the columns with labels $j=1$ and $j=p+k+1$, for which it is $1$. We perform the same row and column manipulations that were described in \cref{sec:homogeneous} to transition from \eqref{eq:GammaM1} to \eqref{eq:GammaM4}. We replace the columns $j=1$ and $j=p+k+1$ by their sum and their difference. This leaves only one non-zero entry in the last row. The size of the matrix in the determinant is reduced by one, to $2p+2k-1$. The result reads
\begin{subequations}
\be
\Gamma^{-}_{k,p} = \frac{(-1)^{p+k+1}}{2^{2p+k-1}}\frac{1}{2^{p(p-1)+(p+2k-1)(p+2k-2)}} \lim_{\beta \to 2} \frac{\det M^{(4)}} {\Delta(\beta_1, \dots, \beta_p)\Delta(\beta_{p+1}, \dots, \beta_{2p+2k-1})},
\ee
\be
M^{(4)}_{ij} = 
\left\{\begin{array}{lll}
U_{-u_j-1}(\tfrac {\beta_i}2)\quad& i \le p\quad&  j \le p+k, \\[0.3cm]
U_{u_{j-p-k}-9}(\tfrac {\beta_i}2)\quad& i \le p\quad& j > p+k,   \\[0.3cm]
U_{-u_j+1}(\tfrac {\beta_i}2)\quad & i > p\quad& j \le p+k, \\[0.3cm]
U_{u_{j-p-k}-7}(\tfrac {\beta_i}2)\quad & i > p\quad& j > p+k,
\end{array}
\right.
\ee
\end{subequations}
where the indices $i,j$ run from $1$ to $2p+2k-1$. We recall that $u_j$ and $\Delta(\beta_1, \dots, \beta_n)$ are defined in \eqref{eq:aj} and \eqref{eq:Delta}. The next steps again follow those presented in \cref{sec:homogeneous}. We apply \cref{sec:finite.diff} twice and use the formula \eqref{eq:U.identity} for the Taylor coefficients of the Chebyshev polynomials of the second kind. We remove common row factors, take linear combinations of the rows and obtain
\begin{subequations}
\be
\Gamma^{\rm -}_{k,p} = \frac{(-1)^{p+k+1}}{2^{2p+k-1}}\frac{1}{2^{p(p-1)+(p+2k-1)(p+2k-2)}} \prod_{i=1}^p\frac1{(2i-1)!}\prod_{i=1}^{p+2k-1}\frac1{(2i-1)!} \det M^{(6)},
\ee
\be
M^{(6)}_{ij} = 
\left\{\begin{array}{lll}
(-u_j)^{2i-1}\quad& i \le p\quad& j \le p+k, \\[0.3cm]
(u_{j-p-k}-8)^{2i-1}\quad& i \le p\quad& j > p+k,   \\[0.3cm]
(-u_j+2)^{2i-2p-1}\quad & i > p\quad& j \le p+k, \\[0.3cm]
(u_{j-p-k}-6)^{2i-2p-1}\quad & i > p\quad& j > p+k.
\end{array}
\right.
\ee
\end{subequations}
This matrix is equivalently written as 
\be
\label{eq:M6v2.C1}
M^{(6)}_{ij} = 
\left\{\begin{array}{cl}
(v_j-2)^{2i-1}\quad& i \le p, \\[0.2cm]
v_j^{2i-2p-1}\quad & i > p,
\end{array}
\right.\qquad \textrm{with} \qquad
v_j = 
\left\{\begin{array}{cl}
-u_j+2\quad& j \le p+k, \\[0.15cm]
u_{j-p-k}-6\quad & j > p+k.
\end{array}
\right.
\ee
Expanding $(v_j-2)^{2i-1}$ in powers of $v_j$ and taking linear combinations of rows to keep only the leading powers in the $v_j$, we find
\begin{subequations}
\be
\Gamma^{-}_{k,p} = \frac{(-1)^{k+1}}{2^{2p+k-1}}\frac{\prod_{i=1}^{p}(4i-2)}{2^{p(p-1)+(p+2k-1)(p+2k-2)}} \prod_{i=1}^p\frac1{(2i-1)!}\prod_{i=1}^{p+2k-1}\frac1{(2i-1)!} \det M^{(7)},
\ee
\be
M^{(7)}_{ij} = 
\left\{\begin{array}{cl}
v_j^{2(i-1)}\quad& i \le p, \\[0.2cm]
v_j^{2i-2p-1}\quad & i > p.
\end{array}
\right.
\ee
\end{subequations}
The determinant of $M^{(7)}$ is written in terms of a Schur polynomial:
\be
\det M^{(7)} = (-1)^{p(p-1)/2}s_{(2k-1,2k-2, \dots, 1, 0^{2p})}(\vec v) \hspace{-0.1cm} \prod_{1\le i<j\le2p+2k-1} \hspace{-0.3cm} (v_j - v_i).
\ee
Because $v_{p+k-j} = -v_j$ for $j = 1, \dots, p+k$, the Schur polynomial reduces to a lower degree Schur polynomial following the argument detailed below \eqref{eq:vj.asymm}:
\be
s_{(2k-1,2k-2, \dots, 1, 0^{2p})}(\vec x)\big|_{x_j = v_j} = s_{(2k-1,2k-2, \dots, 1, 0^{p-k})} (\vec x)\big|_{x_j = v_{j+p+k}}.
\ee 
The right side is evaluated using \cref{sec:schur.prop}, with 
\be
a = 2k-1,\qquad b = p-k, \qquad \alpha = 3p+3k-1, \qquad \beta=-6.
\ee
We find
\be
s_{(2k,2k-1, \dots, 1,0^{p-k})}(\vec x)\big|_{x_j = v_{j+p+k}}  = \frac1{2^{2k-1}} \prod_{1 \le i \le j\le 2k-1}\hspace{-0.2cm}(12k-6i-6j-2) \prod_{i=1}^{2k-1}\prod_{j=1}^{p-k} \frac{2i+j-1}{i+j-1}.
\ee
By combining these results, we obtain an explicit formula for $C^{-}_{2(p+k-1),2k-1}$ in product form. We write it in terms of the Barnes $G$-function and find
\begin{alignat}{2}
C^{-}_{2(p+k-1)+1,2k-1}&=  \frac{3^{(3p^2 -2 p + 9 k^2 - 7k + 6 pk + 1)/2}}{\pi^{k-1/2} 2^{2p^2 -p + 6k^2-10k/3+2pk-1/3}} \frac{G(p+k+\frac13)G(p+k+\frac23)G(p+k+1)G(p-k+1)}{G(p+2k)G(p+2k+\frac12)G(p+\frac12)G(p+1)}\nonumber\\[0.2cm]
&
\times\frac{G(\frac{p+3k}2)G(\frac{p+3k+1}2)}{G(\frac{p-k+2}2)G(\frac{p-k+3}2)}\frac{G(2k+\frac13)}{G(2k+\frac12)}
\frac{G(k+\frac13)G(k+\frac56)}{G(k+\frac16)G(k+\frac23)} \frac{G(\frac16)G(\frac32)^3}{G(\frac13)G(\frac43)^2G(\frac{11}6)}.
\end{alignat}

\subsection[The homogeneous limit of $\bar S^{0}$]{The homogeneous limit of $\boldsymbol{\bar S^{0}}$}\label{sec:bar.Se}

We compute $C^{0}_{N,m}$ for $m$ odd, starting from \eqref{eq:Cekodd} and the determinant expression \eqref{eq:Rkp}. Simplifying the prefactors, we find
\begin{subequations}
\begin{alignat}{2}
&C^{0}_{2(p+k-1),2k-1} = \frac{(-1)^{p(p+1)/2+1}2^{k-1}q^{p-k}}{3^{p(p-1)/2-k(k-1)/2+kp}(q-q^{-1})^{p+3k-2}}\bar \Gamma^{0}_{k,p},\\[0.15cm]
&\bar \Gamma^{0}_{k,p}  = \lim_{x,z \to 1}\lim_{z_{2k}\to q^{-2}} \frac{\det M^{(k,p)}} {C(x_1, \dots, x_p)C(z_1, \dots, z_{2k-1},x_1, \dots, x_p)}.
\end{alignat}
\end{subequations}
In the limit $z_{2k} \to q^{-2}$, all the entries of the last row of $M^{(k,p)}$ tend to $1$. We repeat the steps that led to \eqref{eq:Gamma.M2} and find 
\begin{subequations}
\begin{alignat}{2}
&\bar \Gamma^{0}_{k,p} = \frac1{2^{3p+2k-1}}\frac{(-1)^{p-1}}{2^{p(p-1)+(p+2k-1)(p+2k-2)}} \lim_{y \to 1} \frac{\det M^{(2)}} {C(y_1, \dots, y_p)C(y_{p+1}, \dots, y_{2p+2k-1})},
\\[0.15cm]
&M^{(2)}_{ij} = 
\left\{\begin{array}{cll}
-\Big(y_i^{3(p+k)-6j+4} + y_i^{-3(p+k)+6j-4}\Big) \quad& 1 \le i < 2(p+k)\quad&  j \le p+k, \\[0.3cm]
\Big(y_i^{9(p+k)-6j+4} - y_i^{-9(p+k)+6j-4}\Big)\frac{(y_i+y_i^{-1})}{(y_i-y_i^{-1})}\quad& 1 \le i \le p& j > p+k,   \\[0.3cm]
-\Big(y_i^{9(p+k)-6j+4} - y_i^{-9(p+k)+6j-4}\Big)\frac{(y_i+y_i^{-1})}{(y_i-y_i^{-1})} \quad & p< i < 2(p+k)& j > p+k, \\[0.3cm]
1 \quad & i =2(p+k).
\end{array}
\right.
\end{alignat}
\end{subequations}
The next step is to multiply $M^{(2)}$ from the right by the matrix  
\be
\label{eq:V}
V = \begin{pmatrix}
\tilde V & 0\\
0 & \tilde V
\end{pmatrix}, \qquad 
\tilde V = \begin{pmatrix}
1 & -1 & 0 & 0 & 0\\
0 & 1 & -1 & 0 & 0 \\
0 & 0 & 1 & \ddots & 0 \\
0 & 0 & 0 & \ddots & -1 \\
0 & 0 & 0 & 0 & 1 
\end{pmatrix},
\ee
where $\tilde V$ has size $p+k$. The matrix $M^{(3)} = M^{(2)}V$ satisfies: 
\be
M^{(3)}_{i=2p+2k,j} = \delta_{j,1} + \delta_{j,p+k+1},
\qquad
\lim_{y_1 \to 1}M^{(3)}_{1,j} + \lim_{y_{p+1} \to 1}M^{(3)}_{p+1,j} = -4\, \delta_{j,1}.
\ee
This allows us reduce the size of the matrix by two, to $2p+2k-2$, and we find
\begin{subequations}
\begin{alignat}{2}
&\bar \Gamma^{0}_{k,p} = \frac1{2^{p-2}} \frac1{2^{2(p+k)-1}}\frac{(-1)^{k-1}}{2^{p(p-1)+(p+2k-1)(p+2k-2)}} \lim_{y \to 1} \frac{\prod_{i=2}^{2p+2k-1}(y_i^3-y_i^{-3})\det M^{(4)}} {C(1,y_2, \dots, y_p)C(y_{p+1}, \dots, y_{2p+2k-1})},
\\[0.15cm]
&M^{(4)}_{ij} = 
\left\{\begin{array}{cll}
\Big(y_{i+1}^{3(p+k)-6j+1} - y_{i+1}^{-3(p+k)+6j-1}\Big) \quad& 1 \le i \le 2(p+k-1)&  j \le p+k-1, \\[0.3cm]
-\Big(y_{i+1}^{9(p+k)-6j-5} + y_{i+1}^{-9(p+k)+6j+5}\Big)\frac{(y_{i+1}+y_{i+1}^{-1})}{(y_{i+1}-y_{i+1}^{-1})}\quad& 1 \le i \le p-1& j \ge p+k,   \\[0.3cm]
\Big(y_{i+1}^{9(p+k)-6j-5} + y_{i+1}^{-9(p+k)+6j+5}\Big)\frac{(y_{i+1}+y_{i+1}^{-1})}{(y_{i+1}-y_{i+1}^{-1})} \quad & p\le i \le 2(p+k-1)& j \ge p+k.
\end{array}
\right.
\end{alignat}
\end{subequations}
The next step is to take linear combinations of the columns of $M^{(4)}_{ij}$, replacing the rows $j$ and $j+p+k-1$ by their sum and their difference. This yields
\begin{subequations}
\begin{alignat}{2}
&\bar \Gamma^{0}_{k,p} =  \frac1{2^{2p+k-2}}\frac{(-1)3^{2(p+k-1)}}{2^{p(p-1)+(p+2k-1)(p+2k-2)}} \lim_{y \to 1} \frac{\det M^{(5)}} {C(1,y_2, \dots, y_p)C(y_{p+1}, \dots, y_{2p+2k-1})},
\\[0.15cm]
&M^{(5)}_{ij} = 
\left\{\begin{array}{cll}
y_{i+1}^{3(p+k)-6j+2} + y_{i+1}^{-3(p+k)+6j-2} \quad& i < p\quad&  j < p+k, \\[0.3cm]
y_{i+1}^{9(p+k)-6j-6} + y_{i+1}^{-9(p+k)+6j+6} \quad& i < p\quad& j \ge p+k,   \\[0.3cm]
y_{i+1}^{3(p+k)-6j} + y_{i+1}^{-3(p+k)+6j} \quad& i \ge p\quad&  j < p+k, \\[0.3cm]
y_{i+1}^{9(p+k)-6j-4} + y_{i+1}^{-9(p+k)+6j+4} \quad & i \ge p\quad& j \ge p+k.
\end{array}
\right.
\end{alignat}
\end{subequations}
With the parameterisation $\beta_j = y_j + y_j^{-1}$, the matrix entries of $M^{(5)}$ are equivalently expressed in terms of the Chebyshev polynomials $T_\ell(\frac\beta2)$ of the first kind:
\be
M^{(5)}_{ij} = 
\left\{\begin{array}{cll}
2\, T_{3-u_j}\big(\frac{\beta_{i+1}}2\big) \quad& i < p\quad&  j < p+k, \\[0.3cm]
2\, T_{u_{j-p-k}-11}\big(\frac{\beta_{i+1}}2\big) \quad& i < p\quad& j \ge p+k,   \\[0.3cm]
2\, T_{5-u_j}\big(\frac{\beta_{i+1}}2\big) \quad& i \ge p\quad&  j < p+k, \\[0.3cm]
2\, T_{u_{j-p-k}-9}\big(\frac{\beta_{i+1}}2\big) \quad & i \ge p\quad& j \ge p+k,
\end{array}
\right.
\ee
where $u_j$ is defined in \eqref{eq:aj}. The next step is to apply \cref{sec:finite.diff} with
\be
\label{eq:T.identity}
\frac1{m!}\Big(\frac{d}{d\beta}\Big)^{m}T_{\ell}(\tfrac\beta2)\Big|_{\beta =2} =
\frac{1}{(2m)!} \prod_{n=0}^{m-1} (\ell^2-n^2).
\ee
Performing row operations to keep only the leading powers in the $v_j$, we find
\begin{subequations}
\begin{alignat}{2}
&\bar \Gamma^{0}_{k,p} =  \frac{(-1)^p2^k3^{2(p+k-1)}}{2^{p(p-1)+(p+2k-1)(p+2k-2)}} \frac{\prod_{i=1}^{p-1}(4i)}{\prod_{i=1}^{p-1} (2i)!\prod_{i=1}^{p+2k-2} (2i)!} \det M^{(6)},
\\[0.15cm]
&M^{(6)}_{ij} = 
\left\{\begin{array}{cl}
v_j^{2i-1}\quad& i < p, \\[0.2cm]
v_j^{2(i-p)}\quad & i \ge p,
\end{array}
\right.\qquad
v_j = 
\left\{\begin{array}{cl}
-u_j+5\quad& j < p+k, \\[0.15cm]
u_{j-p-k}-9\quad & j \ge p+k.
\end{array}
\right.
\end{alignat}
\end{subequations}
The determinant of $M^{(6)}$ is then expressed in terms of a Schur polynomial:
\be
\det M^{(6)} = (-1)^{p(p-1)/2}s_{(2k-1,2k-2, \dots, 1, 0^{2p-1})}(\vec v) \hspace{-0.1cm} \prod_{1\le i<j\le2p+2k-2} \hspace{-0.3cm} (v_j - v_i).
\ee
Because $v_{p+k-j} =-v_j$ for $j = 1, \dots, p+k-1$, this Schur polynomial reduces in degree: 
\begin{alignat}{2}
s_{(2k-1,2k-2, \dots, 1, 0^{2p-1})}(\vec x)\big|_{x_j = v_j} &= s_{(2k-1,2k-2, \dots, 1, 0^{p-k})} (\vec x)\big|_{x_j = v_{j+p+k-1}}
\nonumber\\[0.1cm]
&=\frac{1}{2^{2k-1}} \prod_{i=1}^{2k-1}\prod_{j=i}^{2k-1}(12k-6i-6j+4) \prod_{i=1}^{2k-1}\prod_{j=1}^{p-k} \frac{2i+j-1}{i+j-1},
\end{alignat}
where we used \cref{sec:schur.prop} with
\be
a = 2k-1,\qquad b = p-k, \qquad \alpha = 3p+3k+2, \qquad \beta=-6.
\ee
We have therefore obtained a product expression for $C^{0}_{2(p+k-1),2k-1}$. In terms of the Barnes $G$-function, it reads:
\begin{alignat}{2}
C^{0}_{2(p+k-1),2k-1} &=  \frac{(\eE^{\ir\pi/6})^{p-k}3^{(3p^2 -5 p + 9 k^2 - 10k + 6 pk+3)/2}}{\pi^{k+1} 2^{2p^2 -3p + 6k^2-22k/3+2pk+11/3}} \frac{G(p+k-\frac13)G(p+k)G(p+k+\frac13)G(p-k+1)}{G(p+2k-\frac12)G(p+2k)G(p)G(p+\frac12)}\nonumber\\[0.2cm]
&
\times\frac{G(\frac{p+3k}2)G(\frac{p+3k+1}2)}{G(\frac{p-k+2}2)G(\frac{p-k+3}2)}\frac{G(2k-\frac23)}{G(2k+\frac12)}
\frac{G(k+\frac56)G(k+\frac43)}{G(k-\frac13)G(k+\frac16)} \frac{G(\frac76)G(\frac32)^3}{G(\frac13)G(\frac56)G(\frac43)^2}.
\end{alignat}

\subsection[The homogeneous limit of $\bar S^{-}$]{The homogeneous limit of $\boldsymbol{\bar S^{-}}$}

We compute $C^{-}_{N,m}$ for $m$ even, starting from \eqref{eq:Cdkeven} and the determinant expression \eqref{eq:Rkp}. After simplifying the prefactors, we find
\begin{subequations}
\begin{alignat}{2}
&C^{-}_{2(p+k-2)+1,2k-2} = \frac{(-1)^{p(p-1)/2}2^{k-2}}{3^{p(p-1)/2-k(k-5)/2+kp-2}} \bar\Gamma^{-}_{k,p},\\
&\bar\Gamma^{-}_{k,p}  = \lim_{x,z \to 1}\lim_{z_{2k-1}\to \infty}\lim_{z_{2k}\to q^{-2}} \frac1{z^{3(p+k-1)}_{2k-1}}\frac{\det M^{(k,p)}} {C(x_1, \dots, x_p)C(z_1, \dots, z_{2k-2},x_1, \dots, x_p)}.
\end{alignat}
\end{subequations}
The arguments used for this case combine those presented in \cref{sec:Sd,sec:bar.Se}. In particular, the limits $z_{2k-1}\to \infty$ and $z_{2k}\to q^{-2}$ respectively reduce the size of the matrix by one and two units, and we will be left to compute the determinant of a matrix of size $2p+2k-3$. We thus follow the arguments of \cref{sec:bar.Se}. Taking row combinations and writing the matrix entries in terms of the variables $y_1, \dots, y_{2p+2k-2}$, we find
\begin{subequations}
\begin{alignat}{2}
&\bar \Gamma^{-}_{k,p} = \frac1{2^{3p+2k-2}}\frac{1}{2^{p(p-1)+(p+2k-2)(p+2k-3)}} \lim_{y \to 1} \frac{\det M^{(2)}} {C(y_1, \dots, y_p)C(y_{p+1}, \dots, y_{2p+2k-2})},
\\[0.15cm]
&M^{(2)}_{ij} = 
\left\{\begin{array}{cll}
-\Big(y_i^{3(p+k)-6j+4} + y_i^{-3(p+k)+6j-4}\Big) \quad& 1 \le i \le 2(p+k)-2&  j \le p+k, \\[0.3cm]
\Big(y_i^{9(p+k)-6j+4} - y_i^{-9(p+k)+6j-4}\Big)\frac{(y_i+y_i^{-1})}{(y_i-y_i^{-1})}\quad& 1 \le i \le p& j > p+k,   \\[0.3cm]
-\Big(y_i^{9(p+k)-6j+4} - y_i^{-9(p+k)+6j-4}\Big)\frac{(y_i+y_i^{-1})}{(y_i-y_i^{-1})} \quad & p< i \le 2(p+k)-2& j > p+k, \\[0.3cm]
\delta_{j,1}+ \delta_{j,p+k+1} \quad & i =2(p+k)-1,\\[0.3cm]
1 \quad & i =2(p+k).
\end{array}
\right.
\end{alignat}
\end{subequations}
We right-multiply by the matrix $V$ defined in \eqref{eq:V}. Taking the limit $y_1 \to 1$ allows us to remove the first row and the last two rows, and after column operations we find
\begin{subequations}
\begin{alignat}{2}
&\bar \Gamma^{-}_{k,p} =  \frac1{2^{2p+k-3}}\frac{(-1)^{k+1}3^{2(p+k)-3}}{2^{p(p-1)+(p+2k-2)(p+2k-3)}} \lim_{y \to 1} \frac{\det M^{(5)}} {C(1,y_2, \dots, y_p)C(y_{p+1}, \dots, y_{2p+2k-2})},
\\[0.15cm]
&
M^{(5)}_{ij} = 
\left\{\begin{array}{cll}
y_{i+1}^{3(p+k)-6j+2} + y_{i+1}^{-3(p+k)+6j-2} \quad& i < p\quad&  j < p+k, \\[0.3cm]
y_{i+1}^{9(p+k)-6j-12} + y_{i+1}^{-9(p+k)+6j+12} \quad& i < p\quad& j \ge p+k,   \\[0.3cm]
y_{i+1}^{3(p+k)-6j} + y_{i+1}^{-3(p+k)+6j} \quad& i \ge p\quad&  j < p+k, \\[0.3cm]
y_{i+1}^{9(p+k)-6j-10} + y_{i+1}^{-9(p+k)+6j+10} \quad & i \ge p\quad& j \ge p+k,
\end{array}
\right.
\end{alignat}
\end{subequations}
where the indices $i,j$ are in the set $\{1, \dots, 2p+2k-3\}$. This matrix can be expressed as 
\be
M^{(5)}_{ij} = 
\left\{\begin{array}{cll}
2\, T_{3-u_j}\big(\frac{\beta_{i+1}}2\big) \quad& i < p\quad&  j < p+k, \\[0.3cm]
2\, T_{u_{j-p-k}-17}\big(\frac{\beta_{i+1}}2\big) \quad& i < p\quad& j \ge p+k,   \\[0.3cm]
2\, T_{5-u_j}\big(\frac{\beta_{i+1}}2\big) \quad& i \ge p\quad&  j < p+k, \\[0.3cm]
2\, T_{u_{j-p-k}-15}\big(\frac{\beta_{i+1}}2\big) \quad & i \ge p\quad& j \ge p+k.
\end{array}
\right.
\ee
Applying \cref{sec:finite.diff} and taking linear combinations of the rows, we obtain
\begin{subequations}
\begin{alignat}{2}
&\bar \Gamma^{-}_{k,p} =  \frac{(-1)^{p+k}2^k3^{2(p+k)-3}}{2^{p(p-1)+(p+2k-2)(p+2k-3)}} \frac{\prod_{i=1}^{p-1}(4i)}{\prod_{i=1}^{p-1} (2i)!\prod_{i=1}^{p+2k-3} (2i)!} \det M^{(6)},
\\[0.15cm]
&M^{(6)}_{ij} = 
\left\{\begin{array}{cl}
v_j^{2i-1}\quad& i < p, \\[0.2cm]
v_j^{2(i-p)}\quad & i \ge p,
\end{array}
\right.\qquad
v_j = 
\left\{\begin{array}{cl}
-u_j+5\quad& j < p+k, \\[0.15cm]
u_{j-p-k}-15\quad & j \ge p+k.
\end{array}
\right.
\end{alignat}
\end{subequations}
This determinant is expressed in terms of a Schur polynomial,
\be
\det M^{(6)} = (-1)^{p(p-1)/2}s_{(2k-2,2k-3, \dots, 1, 0^{2p-1})}(\vec v) \hspace{-0.1cm} \prod_{1\le i<j\le2p+2k-3} \hspace{-0.3cm} (v_j - v_i),
\ee
which reduces in degree because $v_{p+k-j} =- v_j$ for $j = 1, \dots, p+k-1$: 
\begin{alignat}{2}
s_{(2k-2,2k-3, \dots, 1, 0^{2p-1})}(\vec x)\big|_{x_j = v_j} &= s_{(2k-2,2k-3, \dots, 1, 0^{p-k})} (\vec x)\big|_{x_j = v_{j+p+k-1}}
\nonumber\\[0.1cm]
&=\frac{1}{2^{2k-2}} \prod_{i=1}^{2k-2}\prod_{j=i}^{2k-2}(12k-6i-6j-8) \prod_{i=1}^{2k-2}\prod_{j=1}^{p-k} \frac{2i+j-1}{i+j-1}.
\end{alignat}
At the last equality, we used \cref{sec:schur.prop} with
\be
a = 2k-2,\qquad b = p-k, \qquad \alpha = 3p+3k-4, \qquad \beta=-6.
\ee
Putting these results together yields a product form for $C^{-}_{2(p+k-2)+1,2k-2}$, which we rewrite in terms of the Barnes $G$-function:
\begin{alignat}{2}
C^{-}_{2(p+k-2)+1,2k-2} &=  \frac{3^{(3p^2 -8 p + 9 k^2 - 19k + 6 pk + 10)/2}}{\pi^{k} 2^{2p^2 -4p + 6k^2-31k/3+2pk+16/3}} \frac{G(p+k-\frac23)G(p+k-\frac13)G(p+k)G(p-k+1)}{G(p+2k-\frac32)G(p+2k-1)G(p)G(p+\frac12)}\nonumber\\[0.2cm]
&
\times\frac{G(\frac{p+3k-2}2)G(\frac{p+3k-1}2)}{G(\frac{p-k+2}2)G(\frac{p-k+3}2)}\frac{G(2k-\frac23)}{G(2k-\frac12)}
\frac{G(k-\frac16)G(k+\frac13)}{G(k-\frac13)G(k+\frac16)} \frac{G(\frac76)G(\frac32)^3}{G(\frac13)G(\frac56)G(\frac43)^2}.\nonumber
\end{alignat}

\subsection[The homogeneous limit of $S^{+}$]{The homogeneous limit of $\boldsymbol{S^{+}}$}\label{sec:Su}

We compute $C^{+}_{N,m}$ for $m$ odd, starting from \eqref{eq:Cukodd} and the determinant expression \eqref{eq:Rkp}. Simplifying the prefactors, we find
\begin{subequations}
\begin{alignat}{2}
&C^{+}_{2(p+k-2)+1,2k-1} = \frac{(-1)^{p(p+1)/2+k+1}2^{k-1}}{3^{p(p-5)/2-k(k+1)/2+kp+2}} \Gamma^{+}_{k,p},\\
&\Gamma^{+}_{k,p}  = \frac{1}{q^2-1}\lim_{x,z \to 1}\lim_{x_1 \to \infty}\lim_{z_{2k}\to q^{-2}x_1} \frac1{x_1^{9(p+k)-10}}\frac{\det M^{(k,p)}} {C(x_2, \dots, x_p)C(z_1, \dots, z_{2k-1},x_2, \dots, x_p)}.
\end{alignat}
\end{subequations}
We insert the factor of $x_1^{9(p+k)-10}$ inside the determinant and evaluate the limit using
\begin{subequations}
\label{eq:row.limits}
\begin{alignat}{2}
&\lim_{x_1 \to \infty} \frac{M^{(k,p)}_{1,j}}{x_1^{3p+3k-3}} = \delta _{j,1},\qquad
\lim_{x_1 \to \infty} \frac{M^{(k,p)}_{p+1,j}}{x_1^{3p+3k-3}} = \delta _{j,p+k+1},\\[0.30cm]
&\lim_{x_1 \to \infty}\lim_{z_{2k}\to q^{-2}x_1} \frac{M^{(k,p)}_{2p+2k,j}-M^{(k,p)}_{1,j}-M^{(k,p)}_{p+1,j}}{x_1^{3p+3k-4}} = (q^2-1)(-\delta _{j,1}+\delta_{j,p+k}+\delta_{j,p+k+1}-\delta_{j,2p+2k}).
\end{alignat}
\end{subequations}
The size of the matrix in the determinant thus reduces by two, and the resulting matrix has only two non-zero elements in its last row. The next steps follow the ideas presented in \cref{sec:homogeneous}. We apply simple row operations, write the results in terms of the variables $y_2, \dots, y_{2p+2k-1}$, and find
\begin{subequations}
\begin{alignat}{2}
&\Gamma^{+}_{k,p} = \frac{(-1)^{p+k}}{2^{3p+2k-4}}\frac{1}{2^{(p-1)(p-2)+(p+2k-2)(p+2k-3)}} \lim_{y \to 1} \frac{\det M^{(2)}\prod_{i=2}^{2p+2k-2}(y_i-y_i^{-1})^{-1}} {C(y_2, \dots, y_p)C(y_{p+1}, \dots, y_{2p+2k-2})},
\\[0.3cm]
&M^{(2)}_{ij} = 
\left\{\begin{array}{cll}
-\Big(y_{i+1}^{3(p+k)-6j-2} + y_{i+1}^{-3(p+k)+6j+2}\Big)(y_{i+1}-y_{i+1}^{-1}) \quad& 1 \le i \le 2(p+k)-3&  j < p+k, \\[0.3cm]
\Big(y_{i+1}^{9(p+k)-6j-8} - y_{i+1}^{-9(p+k)+6j+8}\Big)(y_{i+1}+y_{i+1}^{-1})\quad& 1 \le i < p& j \ge p+k,   \\[0.3cm]
-\Big(y_{i+1}^{9(p+k)-6j-8} - y_{i+1}^{-9(p+k)+6j+8}\Big)(y_{i+1}+y_{i+1}^{-1}) \quad & p \le i \le 2(p+k)-3& j \ge p+k,\\[0.3cm]
\delta_{j,p+k-1} - \delta_{j,2p+2k-2} \quad& i = 2p+2k-2,
\end{array}
\right.
\end{alignat}
\end{subequations}
where $i,j$ are in the range $\{1, \dots, 2p+2k-2\}$. Repeating the steps that allowed us to obtain \eqref{eq:M3}, we apply column operations to replace the columns $j$ and $j+p+k-1$ by their sum and their difference. There subsists a single non-zero element in the last row, so the size of the matrix again reduces by one. This yields
\begin{subequations}
\begin{alignat}{2}
&\Gamma^{+}_{k,p} = \frac1{2^{2p+k-3}}\frac{1}{2^{(p-1)(p-2)+(p+2k-2)(p+2k-3)}} \lim_{y \to 1} \frac{\det M^{(3)}\prod_{i=2}^{2p+2k-2}(y_i-y_i^{-1})^{-1}} {C(y_2, \dots, y_p)C(y_{p+1}, \dots, y_{2p+2k-2})},
\\[0.3cm]
&M^{(3)}_{ij} = 
\left\{\begin{array}{cll}
-y_{i+1}^{3(p+k)-6j-3} + y_{i+1}^{-3(p+k)+6j+3} \quad& i < p\quad& j < p+k, \\[0.3cm]
y_{i+1}^{9(p+k)-6j-7} - y_{i+1}^{-9(p+k)+6j+7}\quad& i < p\quad& j \ge p+k,   \\[0.3cm]
y_{i+1}^{3(p+k)-6j-1} - y_{i+1}^{-3(p+k)+6j+1} \quad & i \ge p\quad& j < p+k, \\[0.3cm]
-y_{i+1}^{9(p+k)-6j-9} + y_{i+1}^{-9(p+k)+6j+9} \quad & i \ge p\quad& j \ge p+k,\\[0.3cm]
\end{array}
\right.
\end{alignat}
\end{subequations}
where $i,j \in \{1, \dots, 2p+2k-3\}$. This is rewritten in terms of Chebyshev polynomials of the second kind, with $\beta_j = y_j+y_j^{-1}$:
\begin{subequations}
\begin{alignat}{2}
&\Gamma^{+}_{k,p} = \frac{1}{2^{2p+k-3}}\frac{1}{2^{(p-1)(p-2)+(p+2k-2)(p+2k-3)}} \lim_{\beta \to 2} \frac{\det M^{(4)}} {\Delta(\beta_2, \dots, \beta_p)\Delta(\beta_{p+1}, \dots, \beta_{2p+2k-2})},
\\[0.3cm]
&M^{(4)}_{ij} = 
\left\{\begin{array}{lll}
U_{-u_j+7}(\tfrac {\beta_{i+1}}2)\quad& i < p\quad&  j < p+k, \\[0.3cm]
U_{u_{j-p-k}-13}(\tfrac {\beta_{i+1}}2)\quad& i < p\quad& j \ge p+k,   \\[0.3cm]
U_{-u_j+5}(\tfrac {\beta_{i+1}}2)\quad & i \ge p\quad& j < p+k, \\[0.3cm]
U_{u_{j-p-k}-15}(\tfrac {\beta_{i+1}}2)\quad & i \ge p\quad& j \ge p+k.
\end{array}
\right.
\end{alignat}
\end{subequations}
We apply \cref{sec:finite.diff} and perform row operations to find
\begin{subequations}
\begin{alignat}{2}
&\Gamma^{+}_{k,p} = \frac{(-1)^p}{2^{2p+k-3}}\frac{\prod_{i=1}^{p-1}(4i-2)}{2^{(p-1)(p-2)+(p+2k-2)(p+2k-3)}} \prod_{i=1}^{p-1}\frac1{(2i-1)!}\prod_{i=1}^{p+2k-2}\frac1{(2i-1)!} \det M^{(7)},
\\[0.3cm]
&M^{(7)}_{ij} = 
\left\{\begin{array}{cl}
v_j^{2(i-1)}\quad& i < p, \\[0.2cm]
v_j^{2i-2p+1}\quad & i \ge p,
\end{array}
\right.
\qquad v_j = 
\left\{\begin{array}{cl}
-u_j+6\quad& j < p+k, \\[0.15cm]
u_{j-p-k}-14\quad & j \ge p+k.
\end{array}
\right.
\end{alignat}
\end{subequations}
The determinant of $M^{(7)}$ is written in terms of a Schur polynomial:
\be
\det M^{(7)} = (-1)^{p(p-1)/2+1}s_{(2k-1,2k-2, \dots, 1, 0^{2p-2})}(\vec v) \hspace{-0.1cm} \prod_{1\le i<j\le2p+2k-3} \hspace{-0.3cm} (v_j - v_i).
\ee
In contrast to the previous cases, it is now the $v_j$ with $j \ge p+k$ which factor out, due to the relation
\be
v_{2p+2k-2-i} = - v_{p+k-1+i}, \qquad i = 1, \dots, p+k-2.
\ee
As a result, we find
\begin{alignat}{2}
s_{(2k-1,2k-2, \dots, 1, 0^{2p-2})}(\vec x)\big|_{x_j = v_j} &= s_{(2k-1,2k-2, \dots, 1, 0^{p-k})} (\vec x)\big|_{x_j = v_j}
\nonumber\\[0.1cm]
&=\frac{1}{2^{2k-1}} \prod_{i=1}^{2k-1}\prod_{j=i}^{2k-1}(-12k+6i+6j+2)\prod_{i=1}^{2k-1}\prod_{j=1}^{p-k} \frac{2i+j-1}{i+j-1},
\end{alignat}
where we used \cref{sec:schur.prop} with
\be
a = 2k-1,\qquad b = p-k, \qquad \alpha = -3p-3k+1, \qquad \beta=6.
\ee
The resulting product expression for $C^{+}_{2(p+k-2)+1,2k-1}$ is expressed in terms of the Barnes $G$-function:
\begin{alignat}{2}
C^{+}_{2(p+k-2)+1,2k-1} & = \frac{3^{(3p^2 -8 p + 9 k^2 - 13k + 6 pk + 5)/2}}{\pi^{k-3/2} 2^{2p^2 -5p + 6k^2-22k/3+2pk+8/3}} \frac{G(p+k-1)G(p+k-\frac23)G(p+k-\frac13)G(p-k+1)}{G(p+2k-1)G(p+2k-\frac12)G(p-\frac12)G(p)}\nonumber\\[0.2cm]
&
\times\frac{G(\frac{p+3k}2)G(\frac{p+3k+1}2)}{G(\frac{p-k+2}2)G(\frac{p-k+3}2)}\frac{G(2k+\frac13)}{G(2k+\frac12)}
\frac{G(k+\frac13)G(k+\frac56)}{G(k+\frac16)G(k+\frac23)} \frac{G(\frac16)G(\frac23)G(\frac32)^3}{G(\frac{11}6)G(\frac43)^3G(\frac53)}.
\end{alignat}

\subsection[The homogeneous limit of $\bar {S}^{+}$]{The homogeneous limit of $\boldsymbol{\bar S^{+}}$}

We compute $C^{+}_{N,m}$ for $m$ even, starting from \eqref{eq:Cukeven} and the determinant expression \eqref{eq:Rkp}. Simplifying the prefactors, we find
\begin{subequations}
\begin{alignat}{2}
&C^{+}_{2(p+k-3)+1,2k-2} = \frac{(-1)^{p(p-1)/2}2^{k-2}}{3^{p(p-3)/2-k(k-3)/2+kp-1}} \bar\Gamma^{+}_{k,p},\\
&\bar\Gamma^{+}_{k,p}  = \frac{1}{q^2-1}\lim_{x,z \to 1}\lim_{z_{2k-1}\to q^{-2}}\lim_{x_1 \to \infty}\lim_{z_{2k}\to q^{-2}x_1} \frac1{x_1^{9(p+k)-10}}\frac{\det M^{(k,p)}} {C(x_2, \dots, x_p)C(z_1, \dots, z_{2k-2},x_2, \dots, x_p)}.
\end{alignat}
\end{subequations}
The argument for this case combines those presented in \cref{sec:bar.Se,sec:Su}. In the limit ${z_{2k-1} \to q^{-2}}$, the entries of the row with label $j = 2p+2k-1$ are equal to one. The relations \eqref{eq:row.limits} then allow us to reduce the size of the matrix by two. Writing this in terms of the variables $y_2, \dots, y_{2p+2k-3}$, we find
\begin{subequations}
\begin{alignat}{2}
&\bar\Gamma^{+}_{k,p} = \frac{(-1)^{k}}{2^{3p+2k-5}}\frac{1}{2^{(p-1)(p-2)+(p+2k-2)(p+2k-3)}} \lim_{y \to 1} \frac{\det M^{(2)}} {C(y_2, \dots, y_p)C(y_{p+1}, \dots, y_{2p+2k-3})},
\\[0.3cm]
&M^{(2)}_{ij} = 
\left\{\begin{array}{cll}
-\Big(y_{i+1}^{3(p+k)-6j-2} + y_{i+1}^{-3(p+k)+6j+2}\Big) \quad& 1 \le i \le 2(p+k)-4&  j < p+k, \\[0.3cm]
\Big(y_{i+1}^{9(p+k)-6j-8} - y_{i+1}^{-9(p+k)+6j+8}\Big)\frac{(y_{i+1}+y_{i+1}^{-1})}{(y_{i+1}-y_{i+1}^{-1})}\quad& 1 \le i < p& j \ge p+k,   \\[0.3cm]
-\Big(y_{i+1}^{9(p+k)-6j-8} - y_{i+1}^{-9(p+k)+6j+8}\Big)\frac{(y_{i+1}+y_{i+1}^{-1})}{(y_{i+1}-y_{i+1}^{-1})} \quad & p \le i \le 2(p+k)-4& j \ge p+k,\\[0.3cm]
1 \quad& i = 2p+2k-3,\\[0.25cm]
\delta_{j,p+k-1} - \delta_{j,2p+2k-2} \quad& i = 2p+2k-2,
\end{array}
\right.
\end{alignat}
\end{subequations}
where $i,j \in \{1, \dots, 2p+2k-2\}$.
We multiply $M^{(2)}$ from the right by the matrix $V$ defined in \eqref{eq:V}, of size $2(p+k-1)$. The entries of $M^{(3)} = M^{(2)} V$ satisfy
\be
M^{(3)}_{i=2p+2k-3,j} = \delta_{j,1} + \delta_{j,p+k},
\qquad
\lim_{y_2 \to 1}M^{(3)}_{1,j} + \lim_{y_{p+1} \to 1}M^{(3)}_{p,j} = -4\, \delta_{j,1},
\ee
allowing us to reduce the matrix size by two more units. This yields
\begin{subequations}
\begin{alignat}{2}
&\bar \Gamma^{+}_{k,p} = \frac{(-1)^p}{2^{3p+2k-7}} \frac{1}{2^{(p-1)(p-2)+(p+2k-3)(p+2k-4)}} \lim_{y \to 1} \frac{\prod_{i=3}^{2p+2k-3}(y_i^3-y_i^{-3})\det M^{(4)}} {C(1,y_3, \dots, y_p)C(y_{p+1}, \dots, y_{2p+2k-3})},
\\[0.15cm]
&M^{(4)}_{ij} = 
\left\{\begin{array}{cll}
\Big(y_{i+2}^{3(p+k)-6j-5} - y_{i+2}^{-3(p+k)+6j+5}\Big) \quad& 1 \le i \le 2(p+k)-5&  j \le p+k-2, \\[0.3cm]
-\Big(y_{i+2}^{9(p+k)-6j-17} + y_{i+2}^{-9(p+k)+6j+17}\Big)\frac{(y_{i+2}+y_{i+2}^{-1})}{(y_{i+2}-y_{i+2}^{-1})}\quad& 1 \le i \le p-2& j \ge p+k-1,   \\[0.3cm]
\Big(y_{i+2}^{9(p+k)-6j-17} + y_{i+2}^{-9(p+k)+6j+17}\Big)\frac{(y_{i+2}+y_{i+2}^{-1})}{(y_{i+2}-y_{i+2}^{-1})} \quad & p-1\le i \le 2(p+k)-5& j \ge p+k-1,\\[0.3cm]
\delta_{j,p+k-2}+\delta_{j,2p+2k-4}\quad&i=2p+2k-4,
\end{array}
\right.
\end{alignat}
\end{subequations}
where  $i,j \in \{1,\dots,2p+2k-4\}$. We perform column operations, replacing the columns $j$ and $j+p+k-2$ by their sum and their difference. In doing so, the two non-zero entries of the last row combine and give rise to a single non-zero element, so the matrix reduces by another unit to $2p+2k-5$. We obtain
\begin{subequations}
\begin{alignat}{2}
&\bar \Gamma^{+}_{k,p} =  \frac1{2^{2p+k-5}}\frac{(-1)3^{2(p+k)-5}}{2^{(p-1)(p-2)+(p+2k-3)(p+2k-4)}} \lim_{y \to 1} \frac{\det M^{(5)}} {C(1,y_3, \dots, y_p)C(y_{p+1}, \dots, y_{2p+2k-3})},
\\[0.15cm]
&
M^{(5)}_{ij} = 
\left\{\begin{array}{cll}
y_{i+2}^{3(p+k)-6j-6} + y_{i+2}^{-3(p+k)+6j+6} \quad& i < p-1\quad&  j < p+k-1, \\[0.3cm]
y_{i+2}^{9(p+k)-6j-16} + y_{i+2}^{-9(p+k)+6j+16} \quad& i < p-1\quad& j \ge p+k-1,   \\[0.3cm]
y_{i+2}^{3(p+k)-6j-4} + y_{i+2}^{-3(p+k)+6j+4} \quad& i \ge p-1\quad&  j < p+k-1, \\[0.3cm]
y_{i+2}^{9(p+k)-6j-18} + y_{i+2}^{-9(p+k)+6j+18} \quad & i \ge p-1\quad& j \ge p+k-1,
\end{array}
\right.
\end{alignat}
\end{subequations}
where $i,j \in \{1, \dots, 2p+2k-5\}$. With the parameterisation $\beta_i = y_i + y_i^{-1}$, this matrix is written as
\be
M^{(5)}_{ij} = 
\left\{\begin{array}{cll}
2\, T_{11-u_j}\big(\frac{\beta_{i+2}}2\big)\quad& i < p-1\quad&  j < p+k-1, \\[0.3cm]
2\, T_{u_{j-p-k}-21}\big(\frac{\beta_{i+2}}2\big) \quad& i < p-1\quad& j \ge p+k-1,   \\[0.3cm]
2\, T_{9-u_j}\big(\frac{\beta_{i+2}}2\big)\quad& i \ge p-1\quad&  j < p+k-1, \\[0.3cm]
2\, T_{u_{j-p-k}-23}\big(\frac{\beta_{i+2}}2\big)\quad& i \ge p-1\quad& j \ge p+k-1.
\end{array}
\right.
\ee
We can now apply \cref{sec:finite.diff} to take the limit $\beta_i \to 2$. After performing row operations, we find
\begin{subequations}
\begin{alignat}{2}
&\bar \Gamma^{+}_{k,p} =  \frac{(-1)2^k3^{2(p+k)-5}}{2^{(p-1)(p-2)+(p+2k-3)(p+2k-4)}} \frac{\prod_{i=1}^{p-2}(4i)}{\prod_{i=1}^{p-2} (2i)!\prod_{i=1}^{p+2k-4} (2i)!} \det M^{(6)},
\\[0.15cm]
&M^{(6)}_{ij} = 
\left\{\begin{array}{cl}
v_j^{2i-1}\quad& i < p-1, \\[0.2cm]
v_j^{2(i-p+1)}\quad & i \ge p-1,
\end{array}
\right.\qquad
v_j = 
\left\{\begin{array}{cl}
-u_j+9\quad& j < p+k-1, \\[0.15cm]
u_{j-p-k}-23\quad & j \ge p+k-1.
\end{array}
\right.
\end{alignat}
\end{subequations}
This determinant is expressed in terms of a Schur polynomial,
\be
\det M^{(6)} = (-1)^{p(p+1)/2+1}s_{(2k-2,2k-3, \dots, 1, 0^{2p-3})}(\vec v) \hspace{-0.1cm} \prod_{1\le i<j\le2p+2k-5} \hspace{-0.3cm} (v_j - v_i),
\ee
which reduces in degree because $v_{2p+2k-4-i}=-v_{i+p+k-2}$ for $i = 1, \dots, p+k-3$: 
\begin{alignat}{2}
s_{(2k-2,2k-3, \dots, 1, 0^{2p-3})}(\vec x)\big|_{x_j = v_j} &= s_{(2k-2,2k-3, \dots, 1, 0^{p-k})} (\vec x)\big|_{x_j = v_{j}}
\\[0.1cm]
&=\frac{1}{2^{2k-2}} \prod_{i=1}^{2k-2}\prod_{j=i}^{2k-2}(-12k+6i+6j+8) \prod_{i=1}^{2k-2}\prod_{j=1}^{p-k} \frac{2i+j-1}{i+j-1}.\nonumber
\end{alignat}
At the last equality, we used \cref{sec:schur.prop} with
\be
a = 2k-2,\qquad b = p-k, \qquad \alpha = -3p-3k+4, \qquad \beta=6.
\ee
Putting everything together yields a product form for $C^{+}_{2(p+k-3)+1,2k-2}$. In terms of the Barnes $G$-function, it reads
\begin{alignat}{2}
C^{+}_{2(p+k-3)+1,2k-2} &=  \frac{3^{(3p^2 -14 p + 9 k^2 - 25k + 6 pk + 20)/2}}{\pi^{k-1} 2^{2p^2 -8p + 6k^2-43k/3+2pk+37/3}} \frac{G(p+k-2)G(p+k-\frac53)G(p+k-\frac43)G(p-k+1)}{G(p+2k-\frac52)G(p+2k-2)G(p-1)G(p-\frac12)}\nonumber\\[0.2cm]
&
\times\frac{G(\frac{p+3k-2}2)G(\frac{p+3k-1}2)}{G(\frac{p-k+2}2)G(\frac{p-k+3}2)}\frac{G(2k-\frac23)}{G(2k-\frac12)}
\frac{G(k-\frac16)G(k+\frac13)}{G(k-\frac13)G(k+\frac16)} \frac{G(\frac23)G(\frac76)G(\frac32)^3}{G(\frac56)G(\frac43)^3G(\frac53)}.
\end{alignat}


\end{document}